\documentclass[reprint,nofootinbib,amsmath,amssymb,fontenc aps,superscriptaddress]{revtex4-2}
\usepackage{tabularray}
\usepackage{graphicx}
\usepackage{bm}
\usepackage{soul}
\usepackage{placeins}
\usepackage{xspace}
\usepackage{bm,amsmath,amssymb,amsfonts,gensymb,bbold}
\usepackage{multirow}
\usepackage{comment}
\usepackage{hyperref}
\hypersetup{
    colorlinks=true,
    linkcolor=blue,
    citecolor=blue,
    filecolor=blue,      
    urlcolor=blue,
}
\usepackage{nicefrac}
\usepackage{cleveref}
\usepackage{color}

\newcommand{\eq}[1]{\begin{align} #1 \end{align}}

\newcommand{\im}{{\rm Im}}
\newcommand{\re}{{\rm Re}}
\usepackage[english]{babel}
\usepackage{verbatim}
\usepackage{makecell}

\usepackage{ amssymb }
\usepackage{bbold,dsfont}

\begin{document}

\title{Perturbations of Charged Black Holes with Higher-Order Interactions}

\author{Oleksandr S.~Stashko}
\email{alexander.stashko@gmail.com}
\affiliation{Institute of Cosmology, Department of Physics and Astronomy, Tufts University, Medford, MA 02155, USA}
\affiliation{Black Hole Initiative, Department of Physics, Harvard University, Cambridge, MA 02138, USA}
\author{Roman A.~Konoplya}
\email{roman.konoplya@gmail.com}
\affiliation{Research Centre for Theoretical Physics and Astrophysics, Institute of Physics, Silesian University in Opava, Bezručovo náměstí 13, CZ-74601 Opava, Czech Republic}
\author{Mark P.~Hertzberg}
\email{mark.hertzberg@tufts.edu}
\affiliation{Institute of Cosmology, Department of Physics and Astronomy, Tufts University, Medford, MA 02155, USA}

\begin{abstract}
We study linear gravitational perturbations and the corresponding quasinormal-mode spectra of electrically and magnetically charged black holes in the presence of higher-order corrections to the Einstein–Hilbert action. In particular, we consider a four-derivative interaction that  couples the Riemann tensor to the electromagnetic field strength in the unique (Horndeski) combination that preserves second-order equations of motion. We provide a comprehensive analysis of the stability conditions. We numerically confirm the absence of exponentially growing  modes throughout the physically admissible region of parameter space where the theory remains free of pathologies. We then investigate in detail the properties of the quasinormal-mode spectra and identify several distinctive features induced by the higher-order interaction.
\end{abstract}

\maketitle

\section{Introduction}

Recent developments in gravitational-wave multi-messenger astronomy \cite{LIGOScientific:2016aoc, LIGOScientific:2017vwq, LIGOScientific:2020zkf,KAGRA:2013rdx} and Event Horizon Telescope (EHT) black hole imaging \cite{Akiyama2019,EHT2022} have made the strong- field regime observationally accessible.  This motivates tests of general relativity (GR) in regimes where deviations may be  conceivable.

Since GR is the unique two-derivative theory of massless spin-2, deviations from GR can arise from the presence of higher-derivative operators. The next lowest order possibility (for bosonic fields and after performing field redefinitions) is four derivatives. At this order, one such possibility is to non-minimally couple the Riemann tensor (and its contractions) to two factors of the electromagnetic field strength.
 This is a kind of leading correction to GR in the sense of effective field theory (EFT). When the interactions are organized in a special way to avoid higher-derivative equations of motion, these are  known as ``vector Horndeski" theories \cite{Horndeski:1976gi}.

These non-minimal couplings between the electromagnetic field and curvature arise in several contexts. In quantum electrodynamics in curved spacetime, one-loop vacuum polarization leads to these kinds of effective interactions   
\cite{Drummond:1980}. 
It also appears in certain Kaluza–Klein compactifications of higher-dimensional gravity theories \cite{Feng:2015sbw,PhysRevD.101.064055}.

Considerable attention has been devoted to black holes in such theories.
The presence of a new interaction  between the electromagnetic field and curvature modifies both the background geometry and the dynamics of perturbations. 
Recent studies have shown that these couplings  can significantly modify photon propagation near black holes, leading to polarization-dependent photon rings and lensing features that may leave observable imprints in black-hole imaging \cite{Carballo-Rubio2025zwz}.

The full phase space of the existence and stability of such black hole solutions remains an open question.
A recent study has derived the full set of linear perturbation equations for such black holes and performed an analysis of their linear stability, focusing on  the conditions under which ghost and gradient instabilities are absent \cite{Chen:2024hkm}. 

Although such conditions are necessary for consistency, they do not exhaust all possible instability channels.
When a black hole is slightly disturbed, its response is governed by a discrete set of damped oscillations known as quasinormal modes (QNMs) \cite{Kokkotas:1999bd,Berti:2009kk,Konoplya:2011qq,Bolokhov:2025rng}, whose complex frequencies depend only on the parameters of the background spacetime. As such, QNMs provide a characteristic “fingerprint” of a given black hole solution and are of direct relevance for gravitational-wave observations in the ringdown phase of compact-object mergers. 
There is a significant amount of related work, including on lensing \cite{Carballo-Rubio2025zwz}, general stability \cite{BeltranJimenez2013btb, Chen:2024hkm}, QNMs in generalized Proca theories \cite{Chiang:2025gpa,Garcia-Saenz:2022wsl,Garcia-Saenz:2021uyv},
EFT corrections \cite{Boyce_2025, DiRusso:2025qpf}.
Structural aspects of charged black holes connects to many important subjects, including Love numbers \cite{barbosa2026runninglovenumberscharged,Noumi:2026shc}, weak gravity conjecture \cite{Noumi:2026shc}, magnetic black holes \cite{Pereniguez:2025jxq}.

The QNM spectra in presence of non-minimal coupling was initially studied in \cite{Garcia-Saenz:2022wsl} for Schwarzschild black holes. In this paper, we extend the analysis to charged black holes. 
Charged black holes are of significant conceptual interest. In astrophysical settings, electric charges may be neutralized via different physical mechanisms, while magnetic charges may not (see ahead to conclusions, for further discussion of this). The corrections from this new interaction to the properties and stability of charged black holes will be carefully analyzed in this work.

The paper is organized as follows: 
In Section \ref{TheorySolutions} we give the relevant action, the equations of motion, and the basic structure of black hole solutions.
In Section \ref{Perturbations} we describe linear perturbations around the black hole solutions.
In Section \ref{sec:QNMs} we compute the quasinormal mode spectra.
In Section \ref{Conclusions} we conclude.
Finally, in Appendix \ref{AppExtra} we provide additional figures  and in Appendix \ref{AppPrecise} we provide additional precise values.

\section{Theory and Solutions}\label{TheorySolutions}

We consider the special class of theories in which the Maxwell field is coupled to gravity as
\eq{\label{eq:action}
S = \int d^4x \sqrt{-g} \left( \frac{1}{2\kappa^2} R - \frac{1}{4} F_{\mu\nu}F^{\mu\nu} + \xi L^{\mu\nu\alpha\beta} F_{\mu\nu} F_{\alpha\beta} \right).}
Here $\kappa^2=8\pi G$ is the gravitational coupling and $L^{\mu\nu\alpha\beta}$ is the double-dual Riemann tensor 
\eq{
L^{\mu\nu\alpha\beta}=\frac{1}{4}\epsilon^{\mu\nu\sigma\delta}\epsilon^{\alpha\beta\gamma\eta} R_{\sigma\delta\gamma\eta},
}
where $\epsilon^{\mu\nu\sigma\delta}$ is Levi-Civita tensor. Writing out this 4-derivative term in detail gives
\eq{
L^{\mu\nu\alpha\beta} F_{\mu\nu} F_{\alpha\beta}=4 R_{\mu\nu}F_{\sigma}{}^{\nu} F^{\sigma\mu}-R_{\mu\nu\alpha\beta}F^{\mu\nu}F^{\alpha\beta}-R F^2.
}
In the above action the coefficient $\xi$ is a single free parameter, whose value remains unmeasured (some work on constraining its value includes Refs.~\cite{Prasanna:2003ix,Hertzberg:2025heq}). The above term is the unique combination of terms that couple the Riemann tensor to the electromagnetic field strength that still generates only second-order equations of motion. One can also include terms $\sim F^4$ (which can be related to some of the above terms by field redefinitions at leading order in a small field expansion), but that will not be our focus here.

The variation of  the action (\ref{eq:action}) with respect to the metric $g_{\mu\nu}$ yields
\eq{\label{eq:EoM_gen}
R_{\mu\nu}-\frac{1}{2}g_{\mu\nu}R=-\kappa^2(T^{M}_{\mu\nu}+\xi T^{H}_{\mu\nu}),}
where 
\eq{
T^{M}_{\mu\nu}=\frac{1}{4}g_{\mu\nu}F_{\alpha\beta}F^{\alpha\beta}- F_{\mu\rho}F^{\rho}_{\nu},
}
is the Maxwell energy-momentum tensor and $T_{\mu\nu}^H$ is a new piece from  the non-minimal curvature coupling  
\begin{align}
T^{H}_{\mu\nu}
={}&
g_{\mu\nu}R_{\rho\sigma\gamma\delta}\tilde{F}^{\rho\sigma}\tilde{F}^{\gamma\delta}
-4\nabla^{\rho}\tilde{F}_{\lambda\mu}\nabla^{\lambda}\tilde{F}_{\rho\nu}
\nonumber\\
&-8R^{\rho\lambda}\tilde{F}_{\mu\rho}\tilde{F}_{\lambda\nu}-4\tilde{F}_{\lambda(\nu|}
R^{\lambda}_{\sigma\rho|\mu)}\tilde{F}^{\rho\sigma},
\end{align}
where $\tilde F^{\mu\nu}=\tfrac{1}{2}\epsilon^{\mu\nu\alpha\beta}F_{\alpha\beta}$.

Variation with respect to the vector field $A_{\mu}$ gives equations of motion for the electromagnetic field
\eq{
\nabla_{\mu}\left(F^{\mu\nu}-4\xi L^{\mu\nu\alpha\beta}F_{\alpha\beta}\right)=0.
}

\subsection{Black Hole Solutions}
The metric of a static, spherically symmetric spacetime in Schwarzschild-like coordinates can be written as
\eq{
ds^{2}=e^{-2\alpha(r)}f(r)dt^{2}-\frac{dr^{2}}{f(r)}-r^{2}\left(d\theta^{2}+\sin^2\!\theta\, d\varphi^{2}\right),
}
and for the vector potential $A_{\mu}$ one can choose the following ansatz
\eq{
A_{\mu}=(A_0(r),0,0,-Q_m\cos\theta),
}
where $A_{0}(r)$ denotes the electric component of the vector potential and $Q_{m}$ is the magnetic charge, respectively.

The nontrivial
Einstein equations  are (from here on, we adopt natural units $\kappa^2=8\pi G=c=1$)
\eq{\label{eq:bkgEE1}
\left[
\alpha
+\frac{3}{4}\ln\left(
1+\frac{4\xi Q_m^2}{r^4}
\right)
\right]'
=
-\frac{4\xi Q_e^2 r^3}
{E^2\left(r^4+4\xi Q_m^2\right)}.
}
\eq{\label{eq:bkgEE2}
\left[r(1-f)\right]'=\frac{Q_e^2r^4+Q_m^2E\left(E-48\xi f\right)}{2E\left(r^4+4\xi Q_m^2\right)},
}
with 
\eq{
\label{eq:denom_f}
E(r)=r^2+8\xi(1-f),
}
and  the vector-field component satisfies 
\eq{\label{eq:bkgA0}
A_0'=\frac{Q_e e^{-\alpha}}{E},
}
where $Q_e$ plays the role of the electric charge. It is convenient to redefine $Q_e=\sqrt{2}q_e$ and $Q_m=\sqrt{2}q_m$.

In the purely magnetic  case ($A_0=0$, $Q_e=0$), the system (\ref{eq:bkgEE1}--\ref{eq:bkgEE2}) admits the exact solution \cite{Verbin_2022}
\eq{\label{eq:sol_anl1}
\alpha(r)=-\frac{3}{4}\ln\left[1+\frac{8\xi q_m^2}{r^4}\right],
}
\eq{\label{eq:sol_anl2}
f(r)=\frac{r^6}{\left(r^4+8\xi q_m^2\right)^{7/4}}\int\limits_{r_h}^{r}\left(1-\frac{q_m^2}{x^2}\right)\left(1+\frac{8\xi q_m^2}{x^4}\right)^{3/4}dx,
}
where the integral can expressed in terms Gauss hypergeometric functions (see \cite{Verbin_2022}). 

For  $\xi<0$, the space-time contains a spherical-like curvature singularity at $r_s^4=8|\xi|q_m^2$, so solutions (\ref{eq:sol_anl1}, \ref{eq:sol_anl2}) are defined only for $r>r_s$. In the BH case   this singularity must
be hidden behind the event horizon ($r_h>r_s$).

For $\xi>0$, all magnetic BHs have an RN-like horizon structure, i.e. have  inner and outer horizons. In the extremal case horizon coincide at $r_h=q_m$. For $\xi<0$, the presence of the spherical-like  curvature
singularity allows also to  have BHs with a single horizon depending on $(\xi,q_m)$ values.

For the purely electric case ($Q_m=0$) no closed-form analytical solution can be obtained. Similarly to the magnetic case, the solutions describe BHs with one or two horizons, as well as naked singularities, including spherical-like singularities.

From (\ref{eq:bkgEE1}--\ref{eq:bkgA0}), one can expect some pathologies when the expression in the denominator vanishes, $E(r)=0$, or diverges, $E\to\infty$, at some $r$. To analyze the behaviour of $E$, it is convenient to rewrite it in the form of the differential equation
\eq{\label{eq:elden}
(rE)'=3r^2+\frac{8\xi q_e^2}{E}.
}
One can show that $E$ cannot diverges at $r=r_s>0$. Indeed, for $E\to\infty$, this yields
\eq{
E(r)\simeq \frac{C}{r_s}+r_s^2,
}
which is finite for all $r_s>0$ and diverge when $r\to0$.

Let us now assume that $E(r_s)=0$ at some $r_s>0$. From (\ref{eq:elden}), this yields
\eq{\label{eq:elden_s}
E(r)=4|q_e|\sqrt{\frac{\xi}{r_s}}\sqrt{r-r_s}+O(r-r_s).
}
We need to distinguish between two cases. For $\xi>0$, the right-hand side of (\ref{eq:elden_s}) is real and can be reached from $r>r_s$. We then obtain the following asymptotic behaviour near $r_s$
\eq{
f(r)&=1+\frac{r_s^2}{8\xi}-\frac{|q_e|}{2\sqrt{|\xi|r_s}}\sqrt{r-r_s}+O(r-r_s),\\
\alpha(r)&=-\frac{1}{2}\ln(r-r_s)+O(1).
}
The Kretschmann and electromagnetic scalars behave as
\eq{
R_{\mu\nu\alpha\beta}R^{\mu\nu\alpha\beta}\simeq\frac{(r_s^2+8\xi)^2}{256\xi^2 (r-r_s)^4},~
F_{\mu\nu}F^{\mu\nu}\simeq-\frac{r_s}{4\xi(r-r_s)}.
}
Thus, $E(r_s)=0$ corresponds to a finite radius curvature singularity. The corresponding BH solutions mostly have a single horizon, except in a small region of parameter space where they resemble RN BHs.

For $\xi<0$, $E(r)$ is real only for $r<r_s$. Hence, global solutions (for $r>r_s)$ with such behavior do not exist.
In the case where $E\to\infty$ as $r\to0$, we have
\eq{
E(r)\simeq\frac{8|\xi|f_{-1}}{r},~
f(r)\simeq\frac{f_{-1}}{r},~
\alpha(r)\simeq\alpha_0,
}
where $f_{-1}$ and $\alpha_0$ are asymptotic constants. This behaviour corresponds to a point-like naked singularity at the centre. Taking into account that $E\sim r^2$ as $r\to\infty$, we obtain $f_{-1}>0$, i.e., $f\to+\infty$ as $r\to0$. Hence, for $\xi<0$, $f(r)$ may have two roots corresponding to the inner and outer horizons.

In the limit $\xi\to0$, in both cases  the Reissner-Nordstrom solution with electric and magnetic charges is recovered.

At spatial infinity the asymptotic behavior is qualitatively similar for electric and magnetic black holes
\eq{\label{eq:AssInf1}
f(r)=1-\frac{2M}{r}+\frac{q_e^2+q_m^2}{r^2}-\frac{16\xi q_m^2}{r^4}+O\left(\frac{1}{r^5}\right),
}
\eq{\label{eq:AssInf2}
\alpha(r)=\frac{2\xi(q_e^2-3q_m^2)}{r^4}+O\left(\frac{1}{r^5}\right),
}

\begin{figure*}
    \centering
    \includegraphics[width=0.49\linewidth]{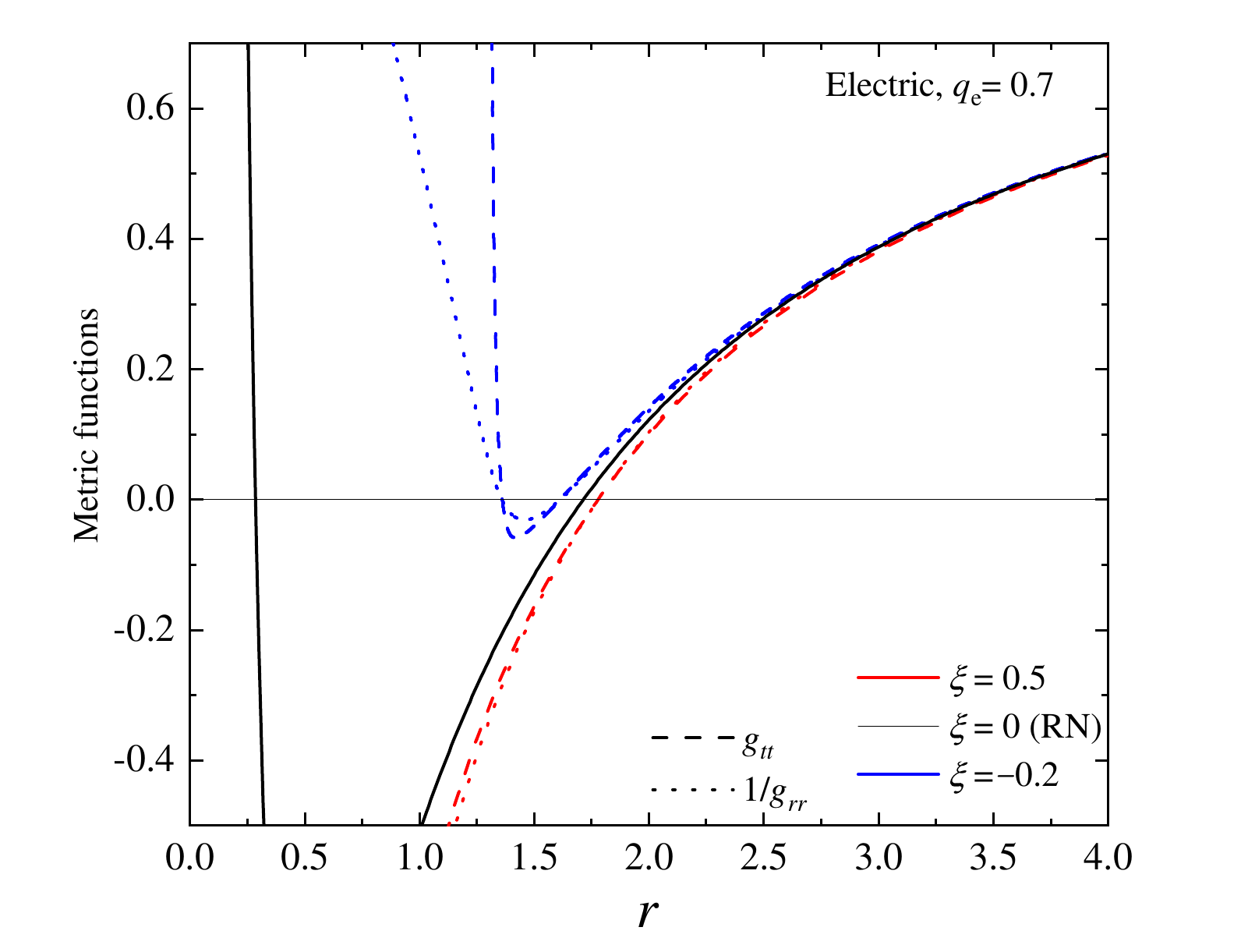}
    \includegraphics[width=0.49\linewidth]{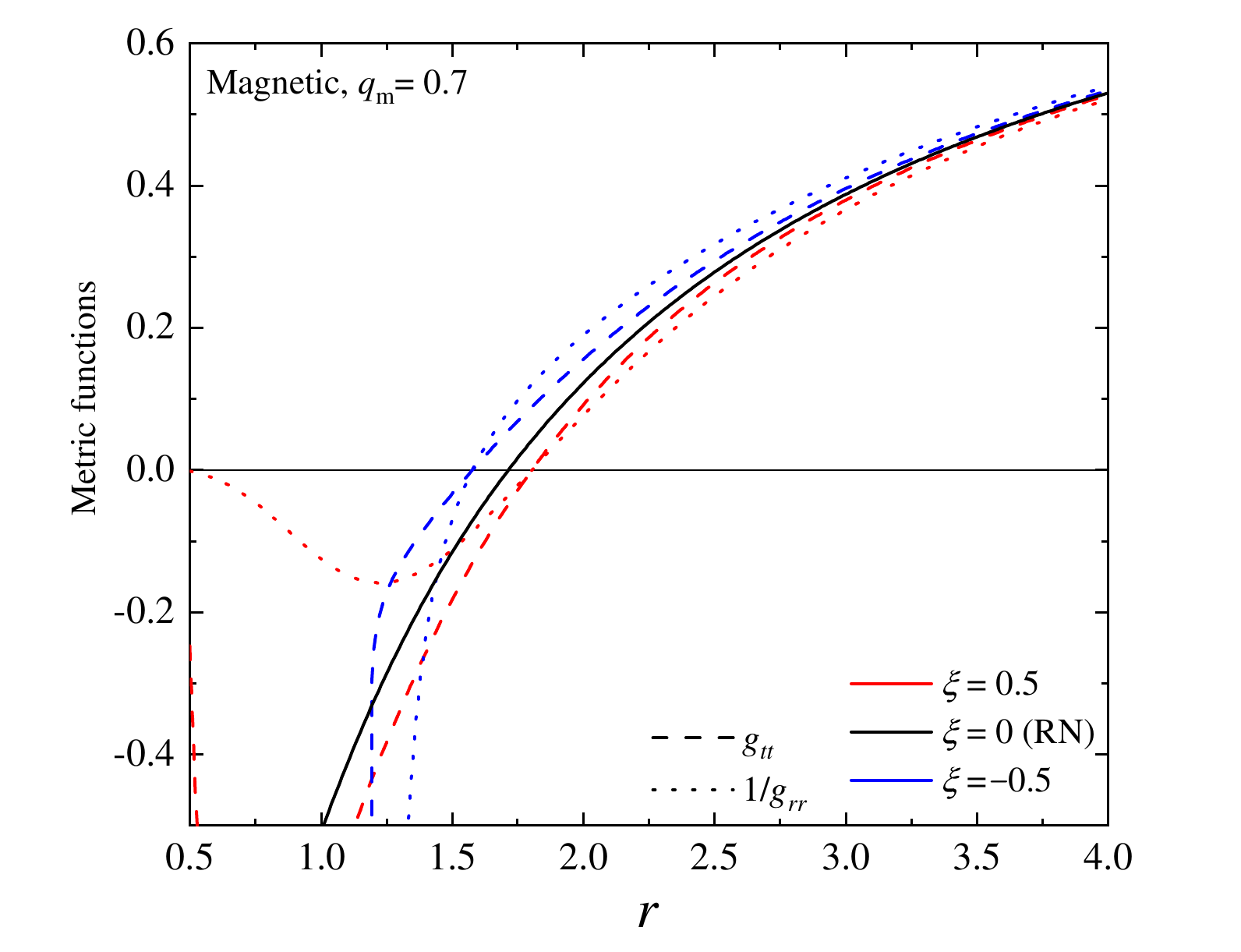}
    \caption{The typical  near horizon behavior of hairy electric (left panel) and magnetic (right panel) black hole solutions in comparison to the RN black hole. }
    \label{fig:sols_example}
\end{figure*}

\begin{figure*}
    \centering
    \includegraphics[width=0.49\linewidth]{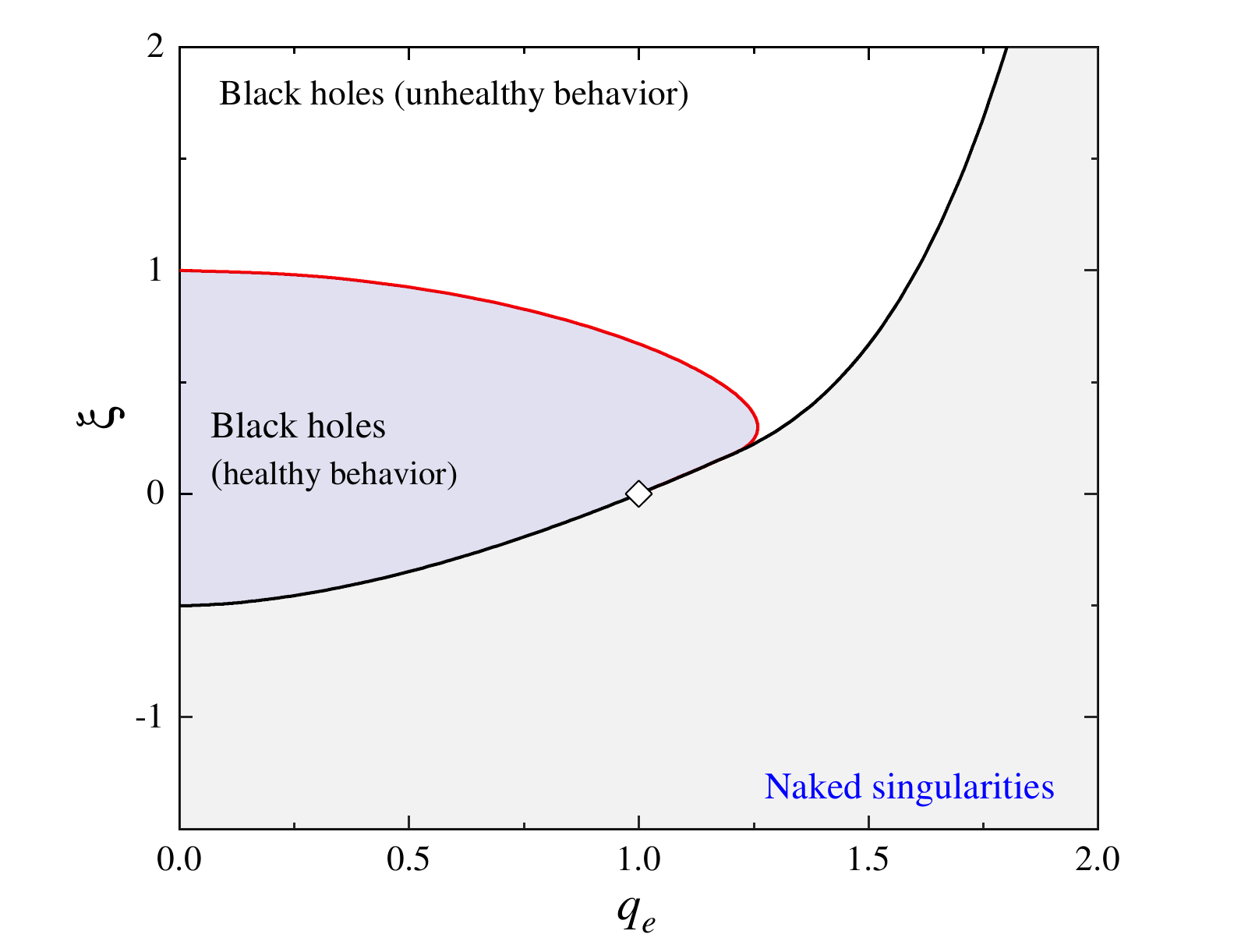}
    \includegraphics[width=0.49\linewidth]{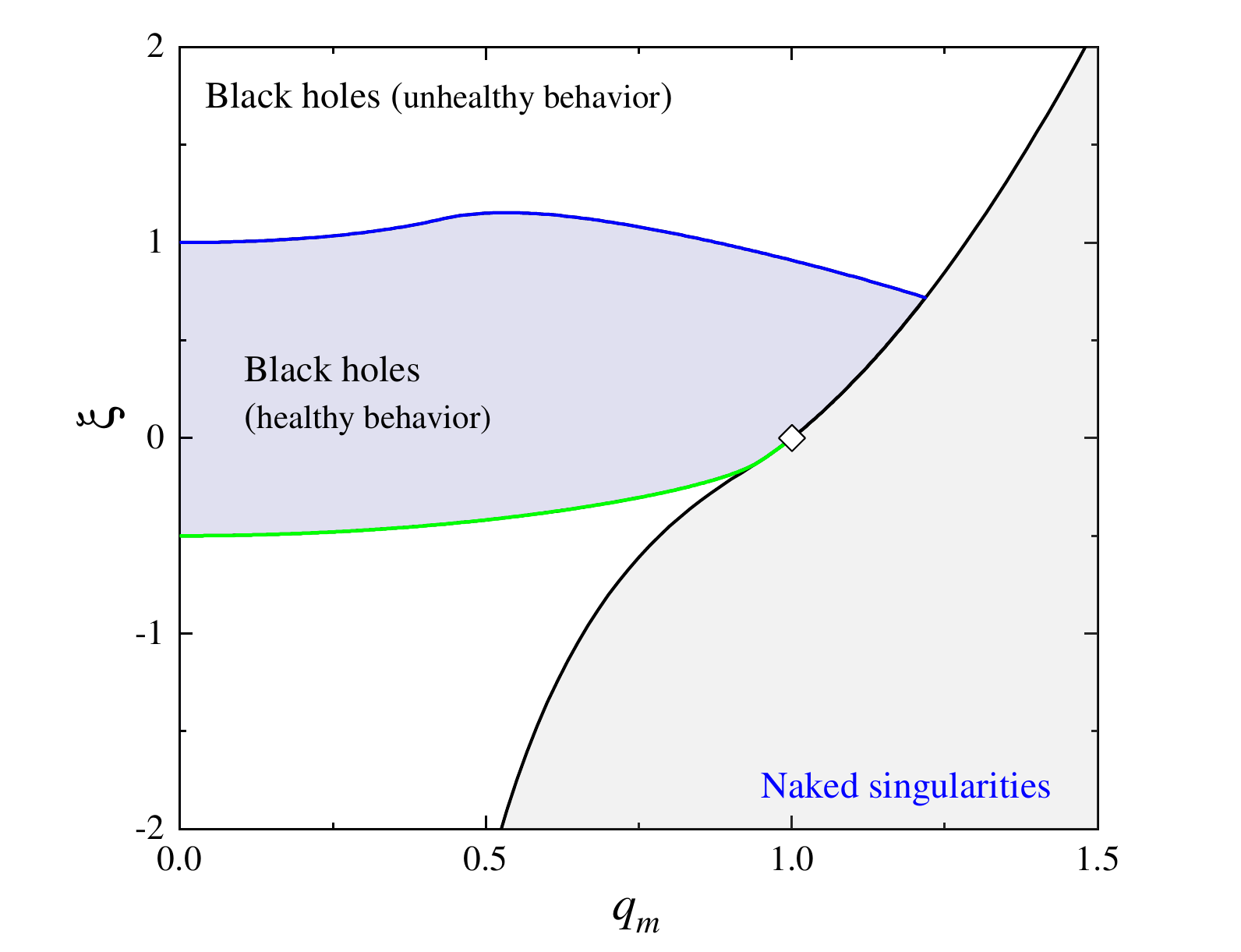}
    \caption{The parameter domains corresponding to BHs and NSs for purely electric (left panel) and purely magnetic (right panel) configurations, respectively. The rhombus marks the case of the extremal Reissner--Nordström BH. The red line corresponds to the boundary defined by  condition (\ref{eq:st_cond_el}) for electric BHs, while the green and blue lines correspond to the boundaries defined by~(\ref{eq:st_cond_mg2}, \ref{eq:st_cond_mg4}) for the magnetic BHs, respectively. (The blue line differs from \cite{Chen:2024hkm}, as satisfying conditions~(\ref{eq:st_cond_mg2}, \ref{eq:st_cond_mg4}) only at the outer horizon is not sufficient. The additional roots can appear for $r>r_h$ as demonstrated in \ref{sec:StbCond}).
    Note that the black line is the extremal BH boundary.}
    \label{fig:sols_domains}
\end{figure*}

One can clearly observe that the electric-magnetic charge duality is absent beyond the RN limit. Using (\ref{eq:AssInf1}, \ref{eq:AssInf2}) as initial conditions at some large $r$, we can numerically solve equations (\ref{eq:bkgEE1}--\ref{eq:bkgEE2}) by integrating backwards. Typical examples of the electric and magnetic metric functions are shown in Fig. \ref{fig:sols_example}. One can see that in the outer domain, the BH solutions closely resemble the RN BH, except in a small region near the horizon. In contrast to the RN black hole, these black holes can exist for charges $q_{e,m}/M>1$, depending on $\xi$. In Fig. \ref{fig:sols_domains}, the solid black line separates the parameter ranges corresponding to black holes and naked singularities. This line is also the extremal black hole case. (We note that the weak gravity conjecture predicts $q_{e,m}/M\geq1$ at the extremal value \cite{Arkani-Hamed:2006emk,Kats:2006xp}).
For the electric BH, this line has a finite limit as $q_e\to0$. Indeed, we should have $f(r_h)=f'(r_h)=0$, and from (\ref{eq:bkgEE2}, \ref{eq:denom_f}) we obtain $E(r_h)=r_h^2-8|\xi|=q_e^2$. In the limit $q_e\to0$, the leading-order expansion gives $r_h\to2M+O(q_e^2\ln(q_e))$, which yields $\xi\to-M^2/2$.

The detailed analysis of different aspects of the corresponding solutions was carried out in Refs.~\cite{Mueller-Hoissen:1988cpx,Balakin:2007am, BeltranJimenez2013btb,Brihaye_2020,Verbin_2022}.

Unless stated otherwise, all numerical results below are expressed in units $M=1$.
Thus, the quantities displayed in the figures and tables are the dimensionless combinations $\xi/M^2$, $q_e/M$, $q_m/M$, and $M\omega$. The mass $M$ is determined from the coefficient of the $1/r$ term in the asymptotic expansion~\eqref{eq:AssInf1}.


\section{Perturbations}\label{Perturbations}
To study the QNMs of black holes, we consider linear perturbations of a static and spherically symmetric spacetime. We expand the metric and the electromagnetic potential around their background values as
\eq{
g_{\mu\nu}=g^{(0)}_{\mu\nu}+h_{\mu\nu},\quad A_{\mu}=A^{(0)}_{\mu}+\delta A_{\mu},
}
where $g^{(0)}_{\mu\nu}$, $A^{(0)}_{\mu}$ correspond to the static spherically symmetric solution of (\ref{eq:bkgEE1}--\ref{eq:bkgA0}), while  $h_{\mu\nu}$, $\delta A_{\mu}$ represent  perturbations. Based on their transformation properties under parity on the two dimensional sphere, perturbations can be decomposed into axial (odd parity) and polar (even parity) sectors.
The background metric and the electric field are even under parity transformations, while the background magnetic field is odd. Consequently, in the presence of magnetic charge the gravitational and electromagnetic perturbations exhibit mixed axial–polar couplings, with polar gravitational modes coupling to axial electromagnetic modes and vice versa, while in the pure electric case the axial and polar sectors remain decoupled.

In the following, we mainly focus on the case with $\ell\geq 2$.
For the metric perturbations $h_{\mu\nu}$   one can choose the following gauge \cite{Chen:2024hkm,Zhang:2024cbw}

\eq{
h_{\mu \nu}= \left[
 \begin{array}{cccc}
 e^{-2\alpha}f\,H_0 & H_1 &0 & h_0\sin\theta{\partial_\vartheta}
\\ H_1 & H_2/f &K\partial_{\theta} & h_1\sin\theta{\partial_\vartheta}
\\ 0 & K\partial_{\theta} &0 & 0
\\ h_0 \sin\theta{\partial_\vartheta} & h_1 \sin\theta{\partial_\vartheta} &0 &0
\end{array}\right]
Y_{\ell m}, 
\label{eq:pert_polar}
}

For the electromagnetic potential perturbation $\delta A_{\mu}$, we use the ansatz\footnote{We can set $\delta A_{3}=0$ by using the $U(1)$ gauge symmetry.}
\eq{
\delta A_{\mu}
=[a_{1},
a_{2},
0,a_{4}\sin \theta \partial_{\theta}]^T
Y_{\ell m},
}
where $H_0$, $H_1$, $H_2$, $K$, $a_1$, and $a_2$ correspond to polar perturbations, while $h_0$, $h_1$, and $a_4$ correspond to axial perturbations. All these quantities are unknown functions of $t$ and $r$. Here $Y_{\ell m}=Y_{\ell m}(\theta,\phi)$ are spherical harmonics. Due to spherical symmetry, one can set $m=0$ without loss of generality.

After substituting this ansatz into the equations of motion \eqref{eq:EoM_gen} and linearizing them, we obtain a system of equations for the perturbation components. The explicit form of the coefficients is given in the supplementary Mathematica notebook. 

For both classes of perturbations, after introducing suitable master functions $\Psi=(\delta \psi_{grav},\delta \psi_{em})^T$, which are combinations of the perturbation components, the system can be reduced to two coupled master equations describing coupled gravitational and electromagnetic degrees of freedom.

The perturbation equations can be written schematically as a system of two coupled equations.
\eq{\label{eq:masterEq}
\frac{d^2}{dr^2}\bold{\Psi}+\hat{B}_1(r)\frac{d}{dr}\bold{\Psi}+\hat{B}_2(r,\omega^2)\bold{\Psi}=0,
}
where $\hat{B}_1, \hat{B}_2$ are 2 dimensional matrices of coefficients, and we separated the time-dependence as $\Psi=e^{-i\omega t} \Psi(r)$. 

\begin{figure}
    \includegraphics[width=0.495\textwidth]{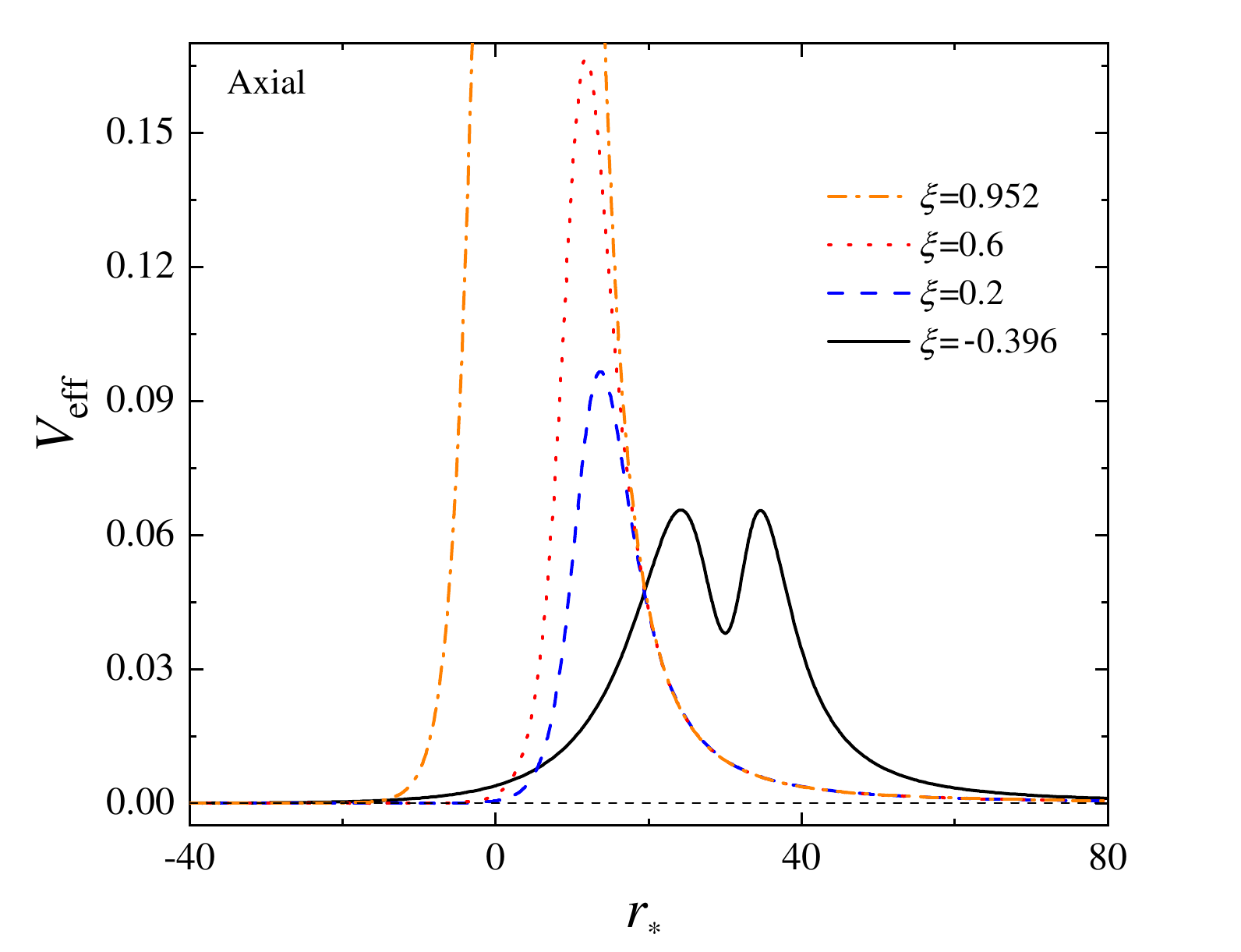}\\
    \includegraphics[width=0.495\textwidth]{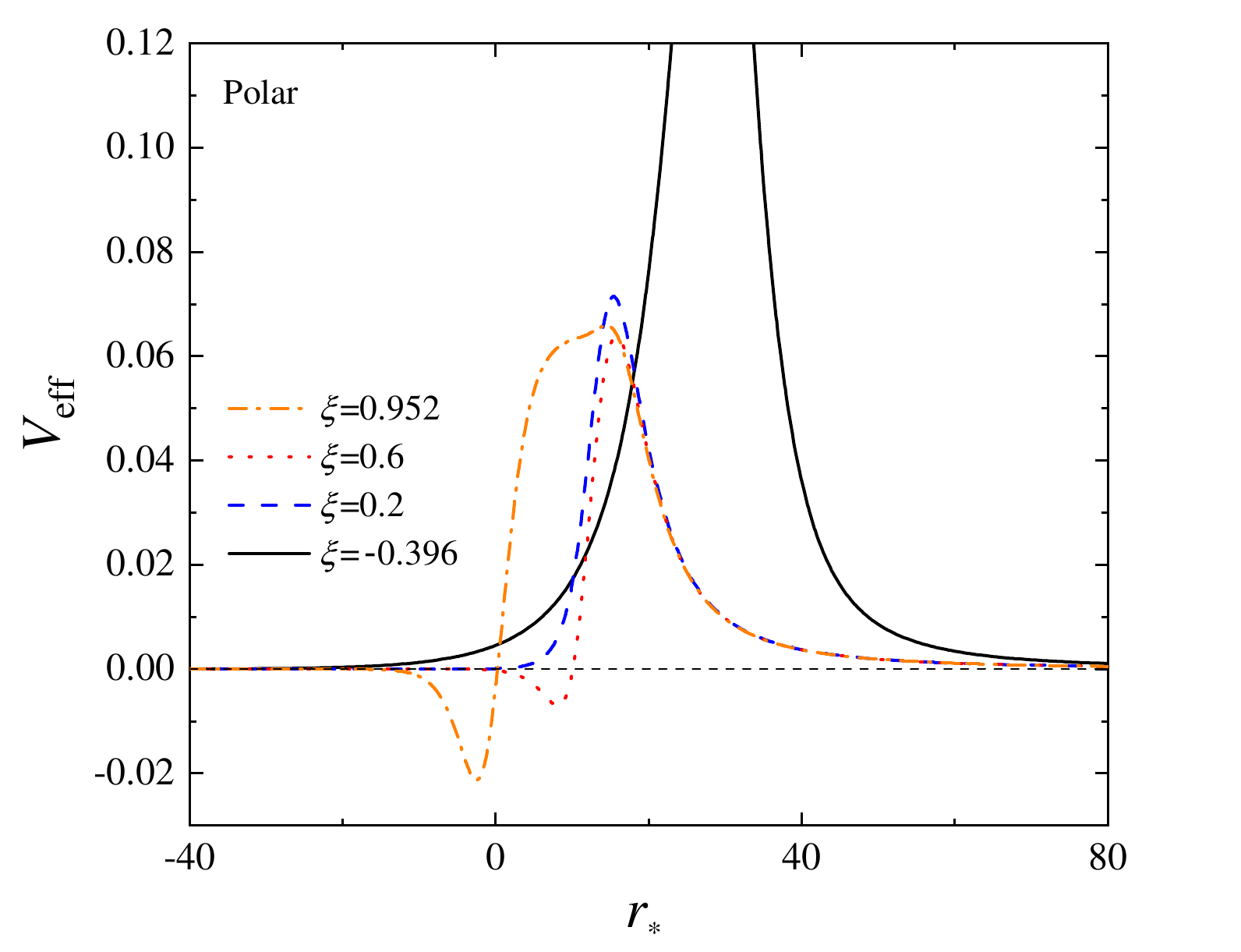}%
        
    \caption{The typical form of the effective potentials for the dipole perturbations for the electric black holes, $q_e=0.4$.}
\label{fig:effPotd1}
\end{figure}

\begin{figure}
    \includegraphics[width=0.495\textwidth]{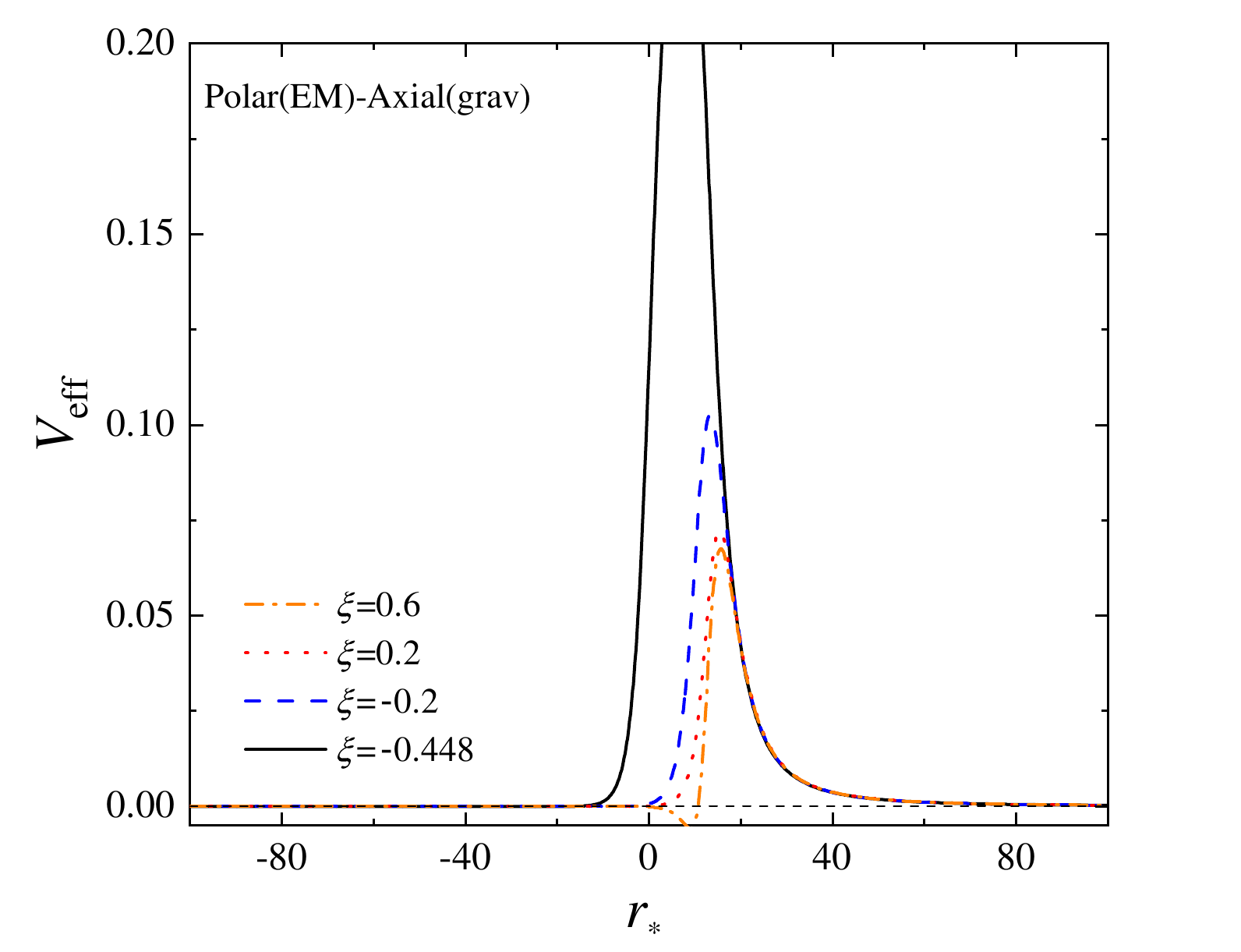}\\
    \includegraphics[width=0.495\textwidth]{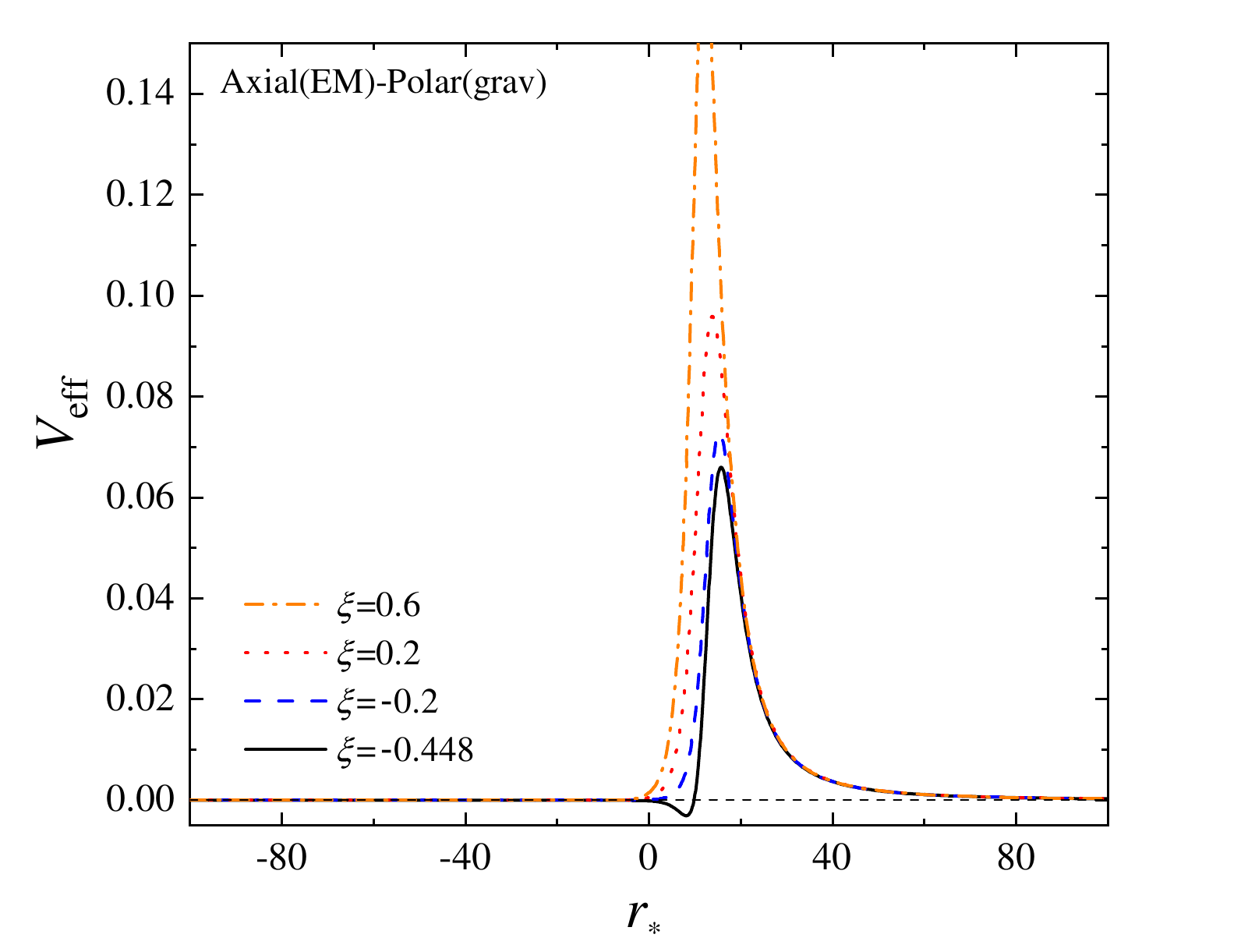}%
        
    \caption{The typical form of the effective potentials of the dipole perturbations for the magnetic black holes, $q_m=0.4$.}
\label{fig:effPotd2}
\end{figure}

The case with $\ell=1$ requires a separate treatment with another gauge, following \cite{Chen:2024hkm}, one can set $K=0$ in (\ref{eq:pert_polar}). The  equations of pertubations can then be reduced to a single master equation,
\eq{
\frac{d^2\Psi}{dr_*^2}+\left[\omega^2-V_{\rm eff}\right]\Psi=0,
}
where $dr_*/dr= e^{-\alpha}/(c_r^{(e,m)} f)$ defines the  tortoise coordinate, and \eq{
(c^{(e)}_{r})^2=1,~
(c^{(m)}_{r})^2=1+\frac{768\xi^2 q_m^2 f(r)}{M_2(r)},
} are the  radial propagation speed for electric and magnetic BHs, respectively.  $V_{\rm eff}$ is the effective potential for the specific type of perturbations. The typical behavior of the effective potential is shown in Figs. \ref{fig:effPotd1}--\ref{fig:effPotd2}. In most of the parameter space, the effective potentials represent a single peak potential barrier with (except for the axial perturbations of the electrically charged BHs), a shallow negative region near the horizon form single-peaked potential barriers with a shallow negative region near the horizon, except for axial perturbations of electrically charged BHs. As $\xi$ approaches the boundaries of its allowed range, the shape of the potential becomes more complicated and may develop several peaks. The exact form of $V_{\rm eff}$ can be found in  the supplementary Mathematica file.

\subsection{Stability conditions}
\label{sec:StbCond}
In \cite{Chen:2024hkm}, it was shown that the absence of ghost and gradient instabilities requires the following conditions for electrically charged black holes:
\eq{
E_1(r)&=r-4\xi f'>0,\label{eq:st_cond_el}\\
E_2(r)&=E(r)=r^2+8\xi(1-f)>0,\label{eq:st_cond_el_2}\\
E_3(r)&=r+8\xi f'=\frac{E_2^2-8\xi q_e^2}{rE_2}>0,\label{eq:st_cond_el_3}\\
E_4(r)&=E_3^3+\frac{576\xi^2rfq_e^2}{E_2^2}>0,\label{eq:st_cond_el_4}\\
E_5(r)&=rE_1E_3^2E_2+
192\xi^2q_e^2\left[\frac{E_2-q_e^2-2rfE_1}{E_2}\right]>0.
\label{eq:st_cond_el_5}
}
Here (\ref{eq:st_cond_el}--\ref{eq:st_cond_el_3}) are general no-ghost stability conditions for electric BHs, while (\ref{eq:st_cond_el_4}, \ref{eq:st_cond_el_5}) are angular Laplacian   stability conditions for axial and polar perturbations, correspondingly.

The condition $E_1>0$ is sufficient to satisfy the remaining $E_{2,3,4,5}>0$ stability conditions. Regularity of $F_{\mu\nu}F^{\mu\nu}$ together with  (\ref{eq:bkgA0}) requires  that $E_2>0$ for $r\geq r_h$. Using $E_2'(r)=2E_1(r)>0$, one can see that $E_2$ is a monotonically increasing function if (\ref{eq:st_cond_el}) is satisfied. At the outer horizon, $f'(r_h)\geq0$ and   (\ref{eq:bkgEE2}) yields  
\eq{\label{eq:estim}
E_2(r_h)\geq q_e^2,~r(1-f)=r_h+\int_{r_h}^{r}\frac{q_e^2\,dx}{E_2(x)}
\geq\frac{q_e^2r}{E_2(r)},
}
where we used that $E_2$ is monotonically increasing. From (\ref{eq:estim}) and (\ref{eq:bkgEE2}), we obtain  $f'(r)>0$ for $r>r_h$. Hence, $f(r)$ is a monotonically increasing function. Conditions (\ref{eq:st_cond_el_3}) and (\ref{eq:st_cond_el_4}) are then automatically satisfied.

For condition (\ref{eq:st_cond_el_5}), the second term can be negative.  Let us define
\eq{\label{eq:Ddef}
D=E_2-q_e^2-2rfE_1.
}
Then for $D\geq0$, the positivity of $E_5$ is trivial. Consider therefore $D<0$. 

For $\xi>0$, integrating (\ref{eq:st_cond_el}), 
we obtain 
\eq{\label{eq:fel_in}
8\xi f<r^2-r_h^2<r^2.}

From (\ref{eq:bkgEE2}, \ref{eq:fel_in}) and $D<0$, we have
\eq{
0<f'<\frac{f\left[r^2-8\xi(1-f)\right]}{r(r^2+8\xi)}<\frac{f}{r}<\frac{r}{8\xi},
}
This yields two additional inequalities $f'<r/(8\xi)$ and $1-f<r^2/(8\xi)$. 
Then the first term in (\ref{eq:st_cond_el_5}) can be estimated as 
\eq{
rE_1E_3^2E_2=r E_2\left[r^3+4\xi f'(3r^2-64\xi^2f'^2)\right]>r^6.
}
To estimate the second term, we  can drop  some positive terms in (\ref{eq:Ddef}) and (\ref{eq:bkgEE2}) and strengthen the inequalities   $D>-f\left[r^2-8\xi(1-f)\right]$, $q_e^2<E_2(1-f)$.\\ Therefore,
\eq{
\left|\frac{192\xi^2q_e^2D}{E_2}\right|=192\xi^2f(1-f)\left[r^2-8\xi(1-f)\right]<\frac{3}{4}r^6.
}
Hence $E_5>r^6/4>0$.

For $\xi<0$, from (\ref{eq:estim}) at the outer  horizon we have $q_e^2\leq r_h^2-8|\xi|$ and $r(1-f)\geq r_h$. Therefore, $q_e^2\leq r^2(1-f)^2-8|\xi|$. Then together with (\ref{eq:bkgEE2}), these yield $D>-r^2f^2$. Furthermore, $E_1>r$ and 
\[
rE_3=E_2+8|\xi|q_e^2/E_2\geq2\sqrt{8|\xi|}|q_e|.
\]
Consequently, $E_2^2>6|\xi|r^2f^2$. Hence, 
\eq{rE_1E_3^2E_2>32|\xi|q_e^2E_2>192|\xi|^2q_e^2r^2f^2/E_2.}
Using $D(r)>-r^2f^2$, we obtain $E_5>0$ for $\xi<0$. 

The case $\xi=0$ is trivial, since $E_5=r^6>0$. Thus, condition (\ref{eq:st_cond_el_5}) is automatically satisfied for both signs of $\xi$.

Differentiating $E_2'=2E_1$ and using (\ref{eq:bkgEE1}--\ref{eq:bkgA0}), we obtain
\eq{
E_1'+\left(\frac{2}{r}+\frac{8\xi q_e^2}{rE_2^2}\right)E_1=3,
}
This equation has the formal solution
\eq{
E_1(r)=e^{-N(r)}
\left[E_1(r_h)+3\int_{r_h}^{r}e^{N(x)}\,dx\right],
}
where $N(r)=\int_{r_h}^{r}\left(2/x+8\xi q_e^2/[xE_2^2(x)]\right)dx$. Hence, if  $E_1(r_h)>0$, then  $E_1(r)>0$ for all $r>r_h$. Therefore,  (\ref{eq:st_cond_el}) needs to be checked only at the event horizon.

For the magnetic black holes, we have another set of conditions \cite{Chen:2024hkm}
\eq{
M_1&=r^4+8\xi q_m^2>0,\label{eq:st_cond_mg1} \\
M_2&=r^6+4\xi r^4(f-1)+12\xi q_m^2 r^2-768\xi^2 q_m^2 f>0,\label{eq:st_cond_mg2} \\
M_3&=[\ell(\ell+1)-2]r^4 M_1+4q_m^2M_2>0,\label{eq:st_cond_mg3}\\
M_4&=r^2+8\xi(1-f)>0,\label{eq:st_cond_mg4} \\
M_5&=r^4 M_4+192\xi^2q_m^2f>0,\label{eq:st_cond_mg5} 
} 
Here (\ref{eq:st_cond_mg1}--\ref{eq:st_cond_mg3}) are general radial stability conditions for magnetic BHs, while (\ref{eq:st_cond_mg4}, \ref{eq:st_cond_mg5}) are conditions for the absence of Laplacian instabilities for axial-polar and polar-axial perturbations, respectively.

One can see that only (\ref{eq:st_cond_mg2}, \ref{eq:st_cond_mg4}) are necessary, since all other conditions are automatically satisfied. Using the background equations (\ref{eq:bkgEE1}, \ref{eq:bkgEE2}), one can also see that estimating the signs of (\ref{eq:st_cond_mg2}, \ref{eq:st_cond_mg4}) only at the outer horizon, as was done in Ref.~\cite{Chen:2024hkm}, is not enough. Let us expand (\ref{eq:st_cond_mg2}) near the horizon. From 
\eq{
M_2(r_h,q_0)=r_h^6-4\xi r_h^2(r_h^2-3q_0^2)=0,
}
we obtain 
\eq{
q_0^2=r_h^2(4\xi-r_h^2)/(12\xi),
}
for $\xi>r_h^2/4$. Then, we have 
\eq{M_2'(r_h,q_0)=3r_h^3(7r_h^2-24\xi)<0,
}
for $\xi>7r_h^2/24$. Hence, $M_2<0$ in some vicinity outside the outer horizon. Since $M_2(r)\simeq r^6>0$ as $r\to\infty$, there is at least one root outside this region. Now let $q_1^2=q_0^2+\delta$, where $\delta>0$ is sufficiently small. Then $M_2(r_h,q_1)=12\xi r_h^2\delta>0$, while by continuity $M_2$ remains negative at some $r>r_h$. Therefore, $M_2$ has two roots outside the horizon and is negative between them. Since this is just a local analysis, we verify the presence of both roots numerically.

In the domain with $M_2(r)>0$, we have 
\eq{
M_4'=\frac{2M_2+1152\xi^2q_m^2f}
{r(r^4+8\xi q_m^2)}>0,~M_4(r_h)=r_h^2+8\xi.
}
Hence, $M_4$ is monotonically increasing, and its enough to check the sign  at the horizon if $M_2(r)>0$.

\section{Quasinormal Mode spectra}
\label{sec:QNMs}
Quasinormal modes  are the eigenvalues $\omega$ of the master equation (\ref{eq:masterEq}), that  correspond to ingoing waves at the black hole horizon and outgoing waves at spatial infinity, respectively. They satisfy the following boundary conditions
\eq{
\Psi\sim  e^{-i\omega r_*}
\begin{pmatrix}
C^-_1\\
C^-_2
\end{pmatrix},\,\,\,\Psi\sim e^{i\omega r_*}\begin{pmatrix}
C^+_1\\
C^+_2
\end{pmatrix},
}
as $r_*\to\mp \infty$, respectively.

We numerically solve the corresponding boundary value problem using a pseudospectral method \cite{boyd2013chebyshev}. Introducing the ingoing Eddington--Finkelstein coordinate $v=t+r_*$, we remove the ingoing horizon behavior by defining a new radial master field $\boldsymbol{\Phi}(r)$ according to
\eq{\boldsymbol{\Psi}(r)=e^{-i\omega r_*}\boldsymbol{\Phi}(r).}
Then we factor out the singular contributions at spatial infinity as
\eq{
\boldsymbol{\Phi}\sim r^{4i\omega} e^{2i\omega r}\widetilde{\boldsymbol{\Psi}}(r),
}
where $\widetilde{\boldsymbol{\Psi}}(r)$ is regular at spatial infinity, and introduce the compactified coordinate $u$   by
\eq{
r=r_h+\frac{L u}{1-u},~u\in(0,1),
}
where $L$ is some constant that controls distribution of nodes. We choose values in  the interval $L\in(10^{-3},5)$.

We discretize Eq. (\ref{eq:masterEq}) on the Chebyshev-Lobatto grid with nodes
\eq{u_j=\frac{1}{2}\left(1-\cos\left[\frac{\pi j}{N}\right]\right),~~j=0,1...N.}
This yields  the matrix equation
\eq{\label{eq:discr_eq}
M(\omega)\tilde{\psi}=0,
}
where $M(\omega)$ is a discretized operator that depends quadratically on $\omega$ and  $\tilde{\psi}=[\psi_1(u_0),...\psi_1(u_N),\psi_2(u_0),...\psi_2(u_N)]^T$ is the vector of function values at the nodes $u_i$.  One can linearize (\ref{eq:discr_eq}) and reformulate it as a generalized eigenvalue problem. Alternatively, the corresponding eigenvalue $\omega$ and eigenvector $\tilde{\psi}$ are then obtained iteratively using Newton's method \cite{10.5555/1403886}.

To avoid spurious eigenvalues, we perform calculations on multiple grids with typical sizes
$N=50-500$ points and with increased numerical precision. We mostly use a single-domain realization; however in the vicinity of the  limiting values of $\xi$ we  use a multi-domain realization or significantly increase the number of nodes. 

\subsection{Results}
\label{sec:results}

\begin{figure}
    \includegraphics[width=0.49\textwidth]{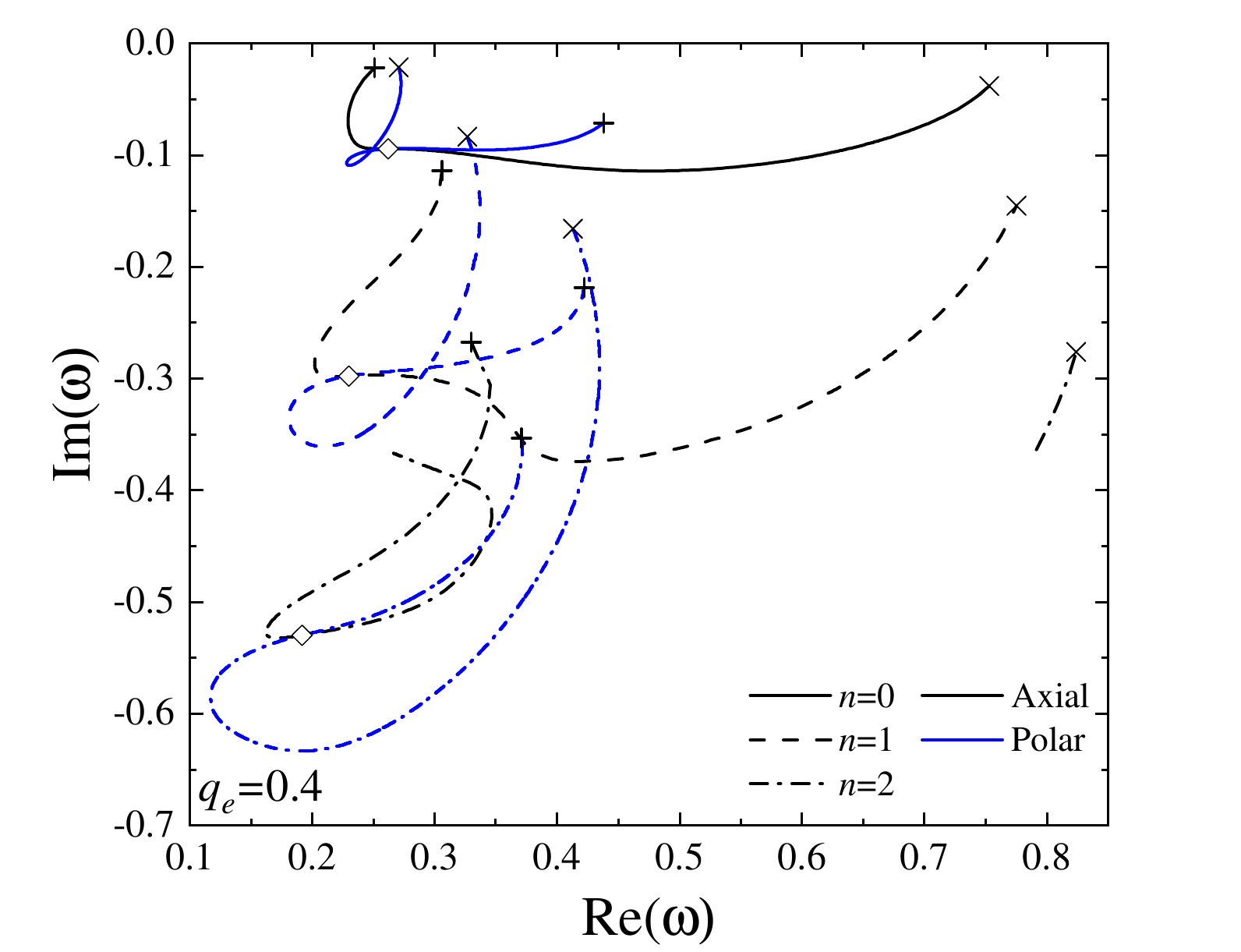}
   \\ \includegraphics[width=0.49\textwidth]{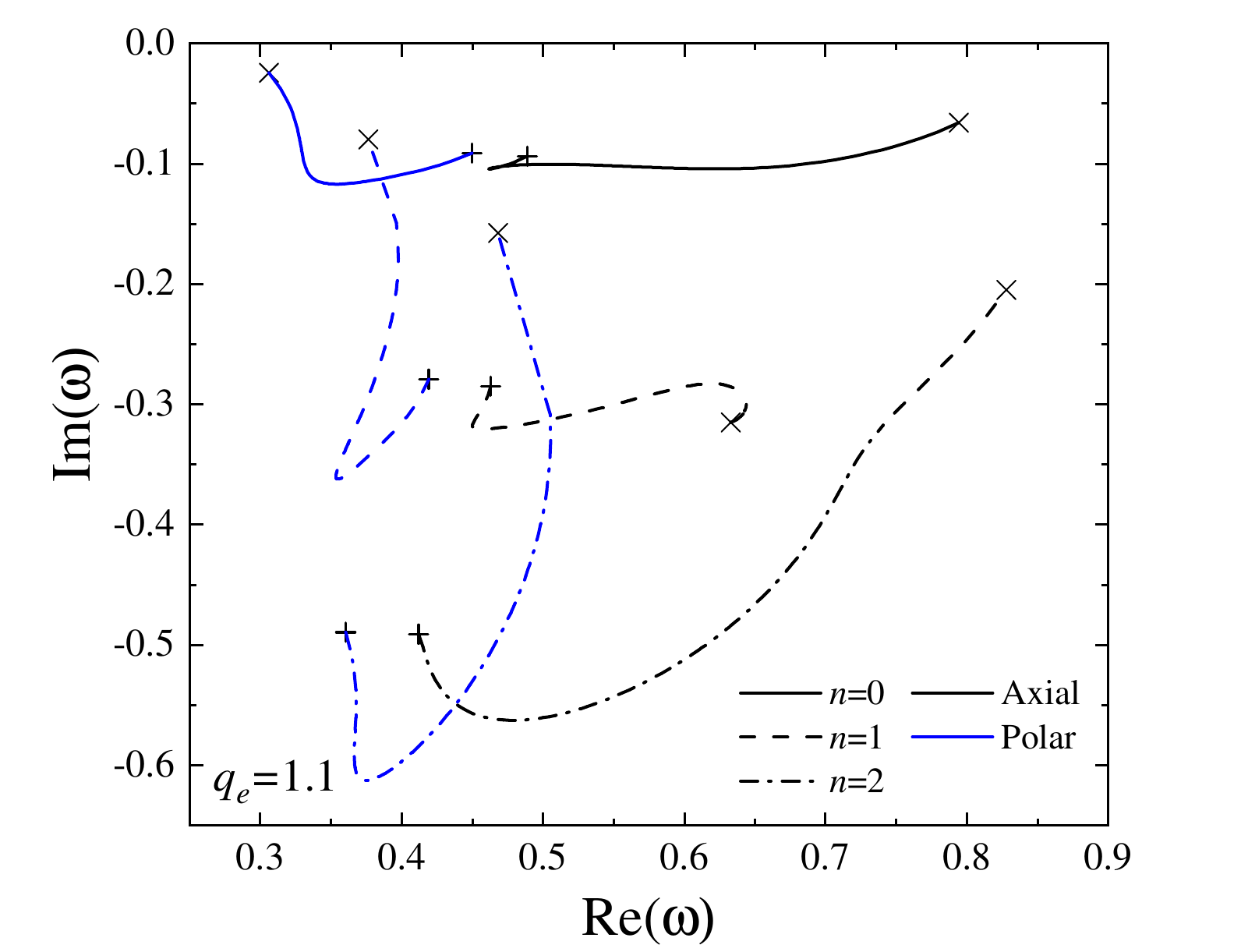}
    \caption{The QNM trajectories in the complex  $\omega$ plane for $q_e=0.4$ (top) and $q_e=1.1$ (bottom) for the dipole perturbations of the electric black holes ($\ell=1$). The diamond markers  indicate the Reissner–Nordstrom modes. The cross and plus markers indicate the endpoints of the continuation at the maximal and minimal values of $\xi$ respectively.}
    \label{fig:WIReldip}
\end{figure}

\begin{figure}
    \includegraphics[width=0.49\textwidth]{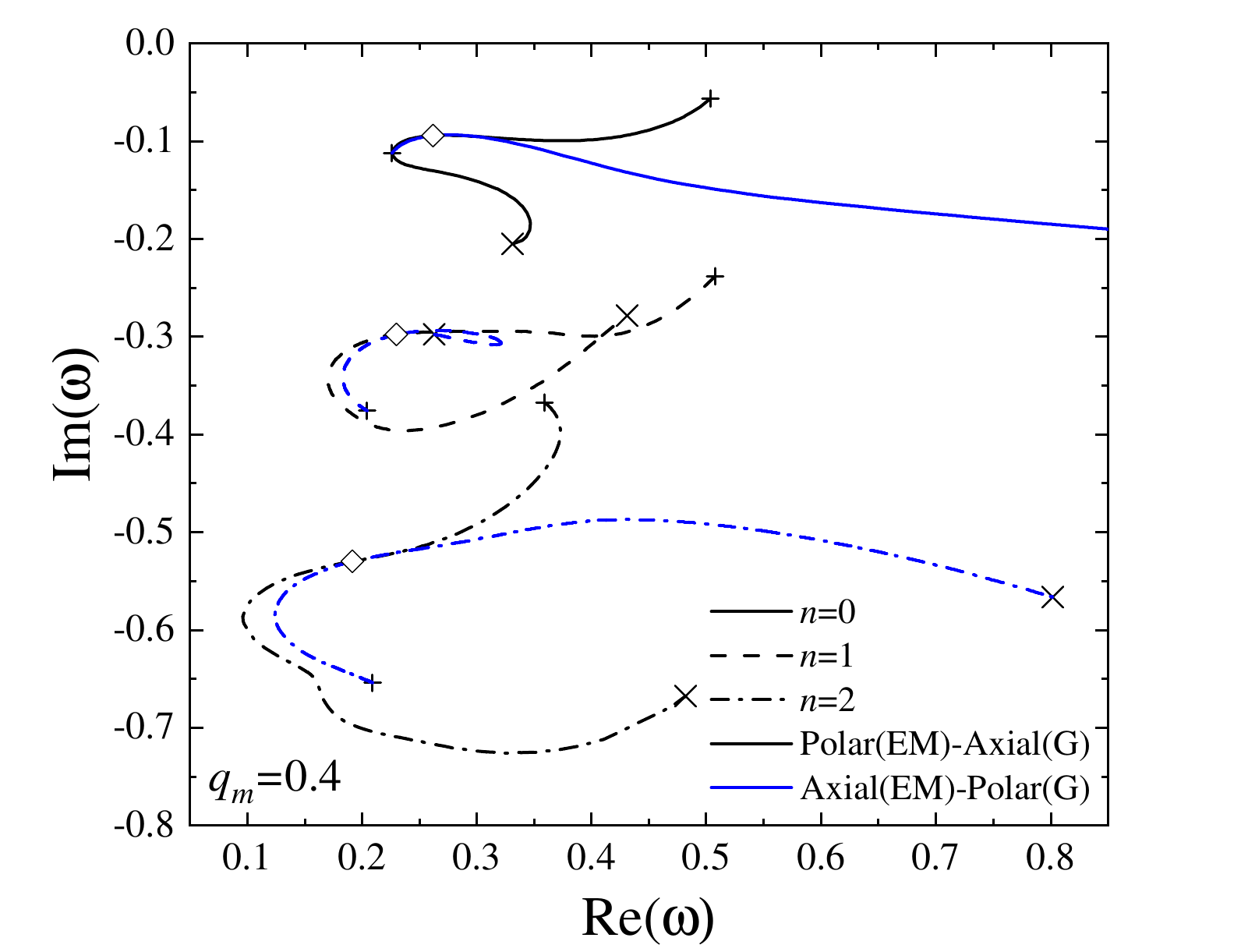}
   \\ \includegraphics[width=0.49\textwidth]{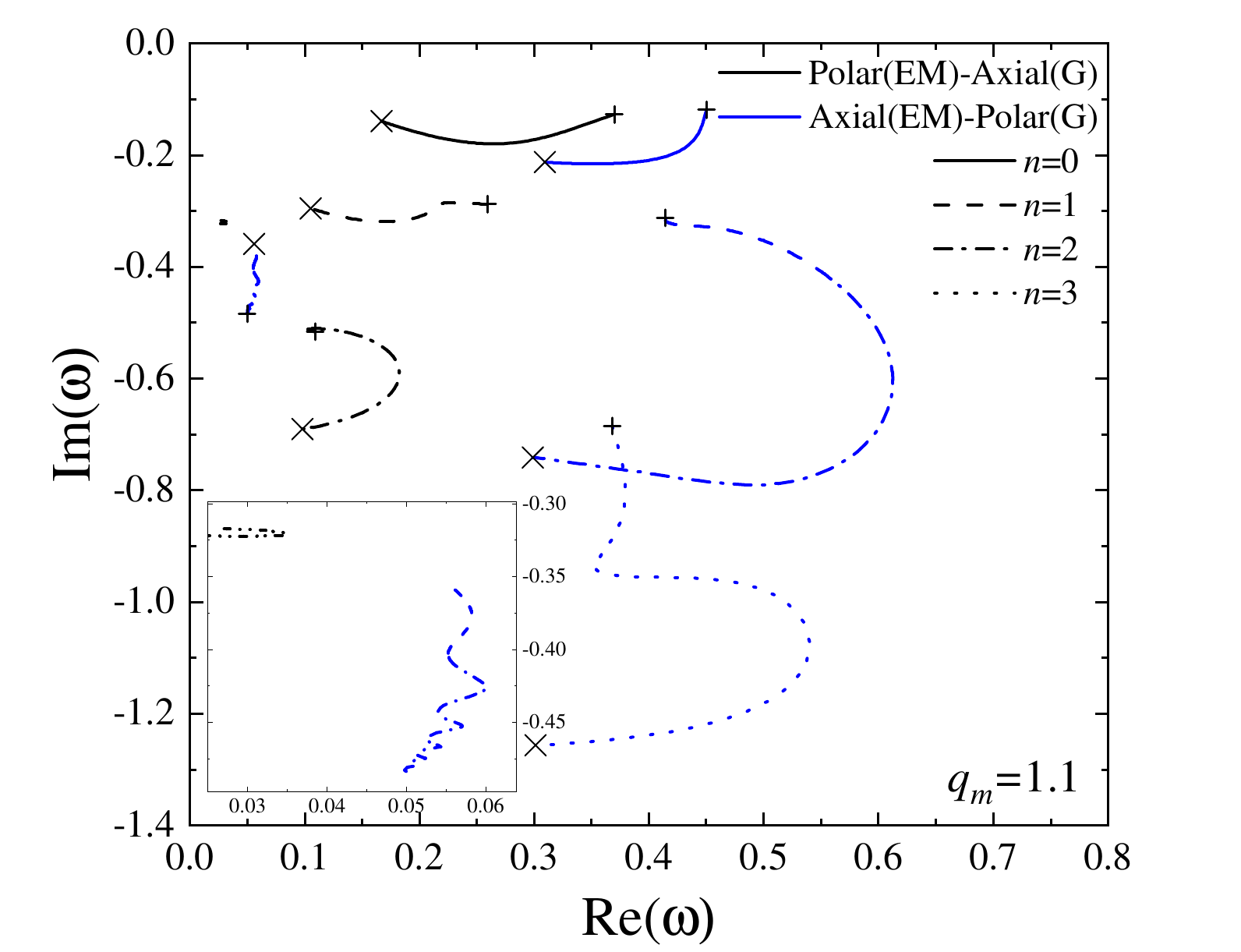}
    \caption{The QNMs trajectories in the complex  $\omega$ plane for $q_m=0.4$ (top) and $q_m=1.1$ (bottom) for the dipole perturbations of the magnetic black holes ( $\ell=1$). The diamond markers  indicate the Reissner–Nordstrom modes. The cross and plus markers indicate the endpoints of the continuation at the maximal and minimal values of $\xi$ respectively.}
    \label{fig:WIRmagdip}
\end{figure}

\begin{figure*}
    \includegraphics[width=0.495\textwidth]{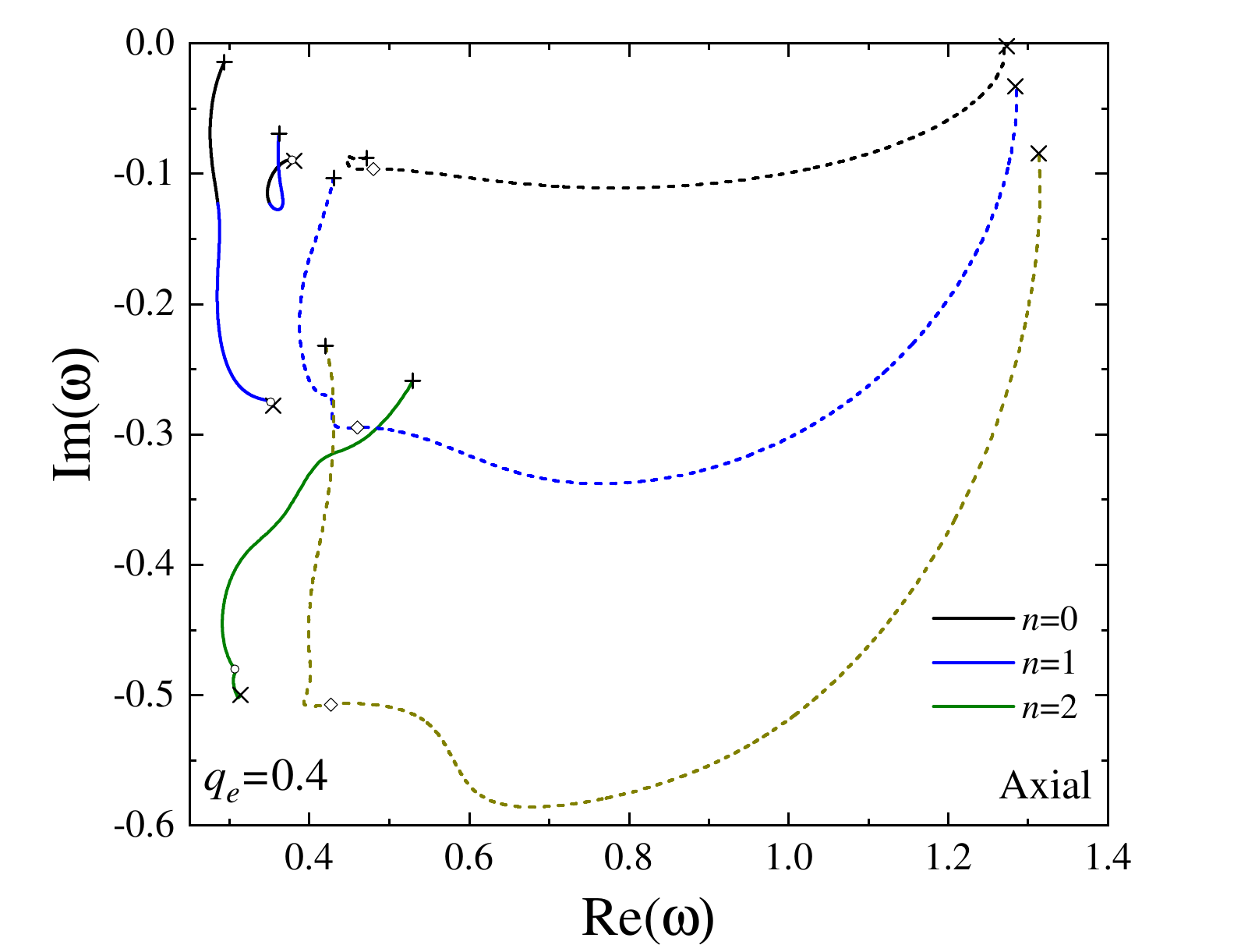}%
    \includegraphics[width=0.495\textwidth]{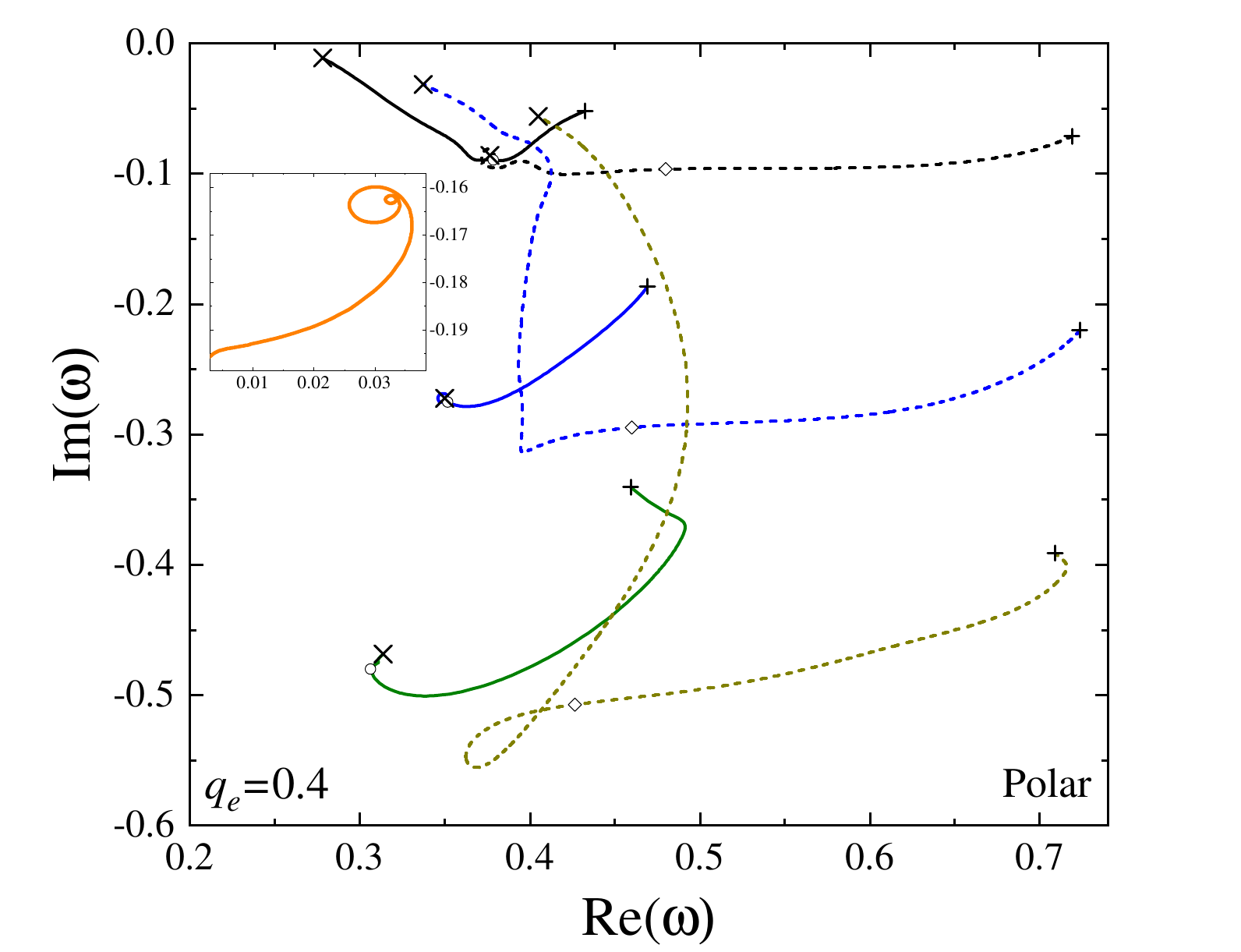}
    \includegraphics[width=0.495\textwidth]{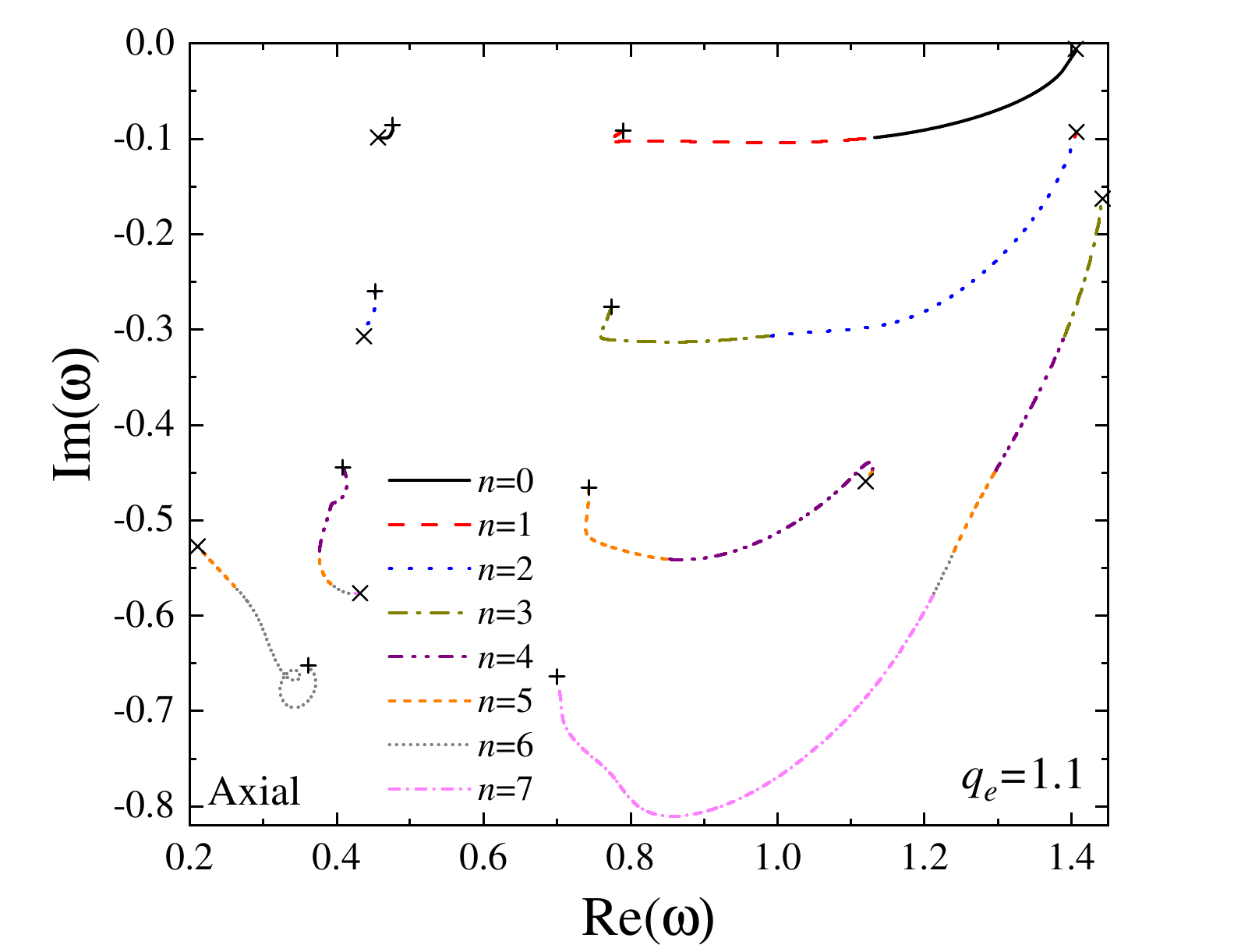}%
        \includegraphics[width=0.495\textwidth]{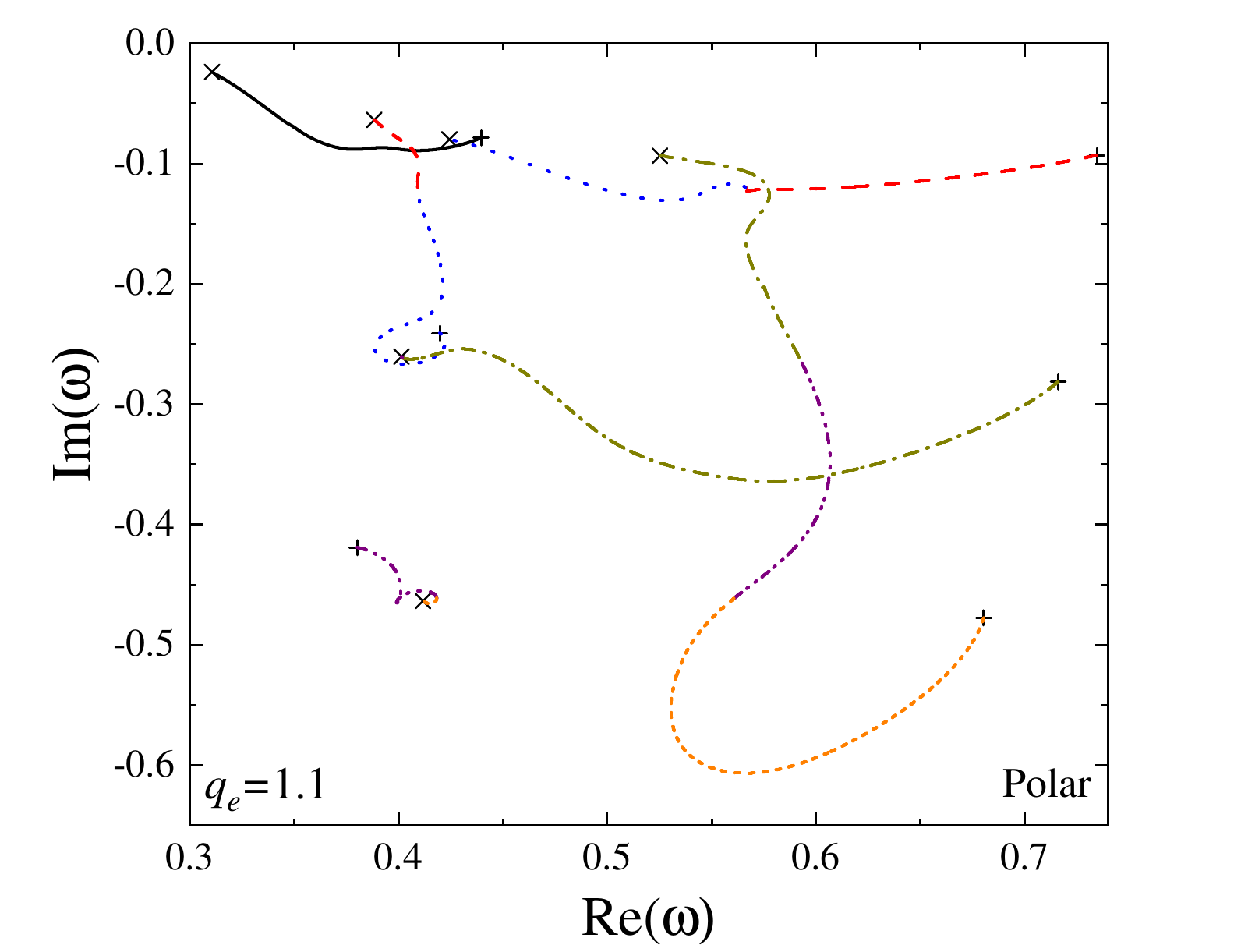}%
    \caption{The QNMs trajectories in the complex  $\omega$ plane for $q_e=0.4$ (top) and $q_e=1.1$ (bottom); $\ell=2$. The circle and diamond markers in the left panel indicate the Reissner–Nordstrom modes.  The cross and plus markers indicate the endpoints of the continuation at the maximal and minimal values of $\xi$ respectively.}
\label{fig:ImRe_electric}
\end{figure*}

\begin{figure*}
    \includegraphics[width=0.495\textwidth]{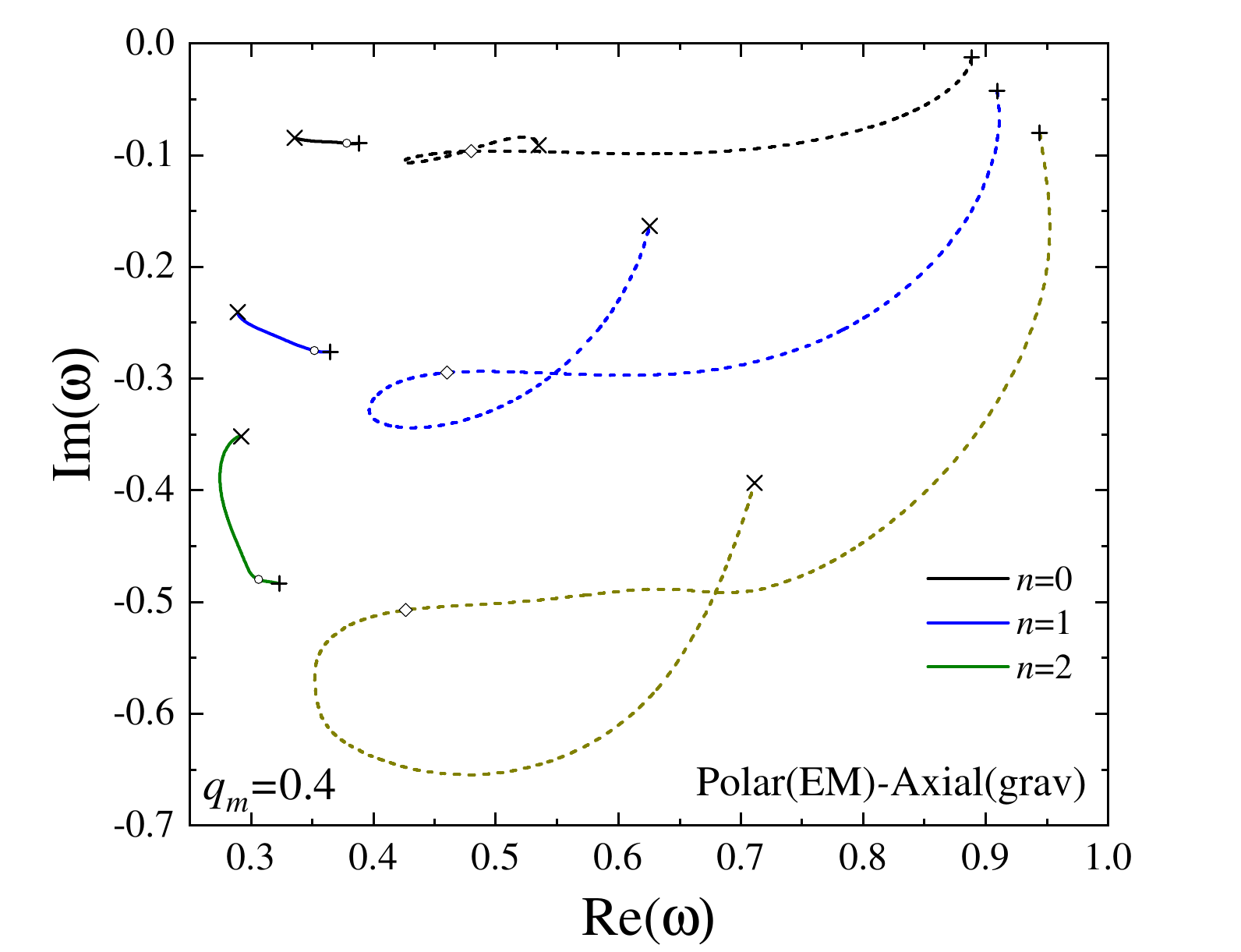}%
    \includegraphics[width=0.495\textwidth]{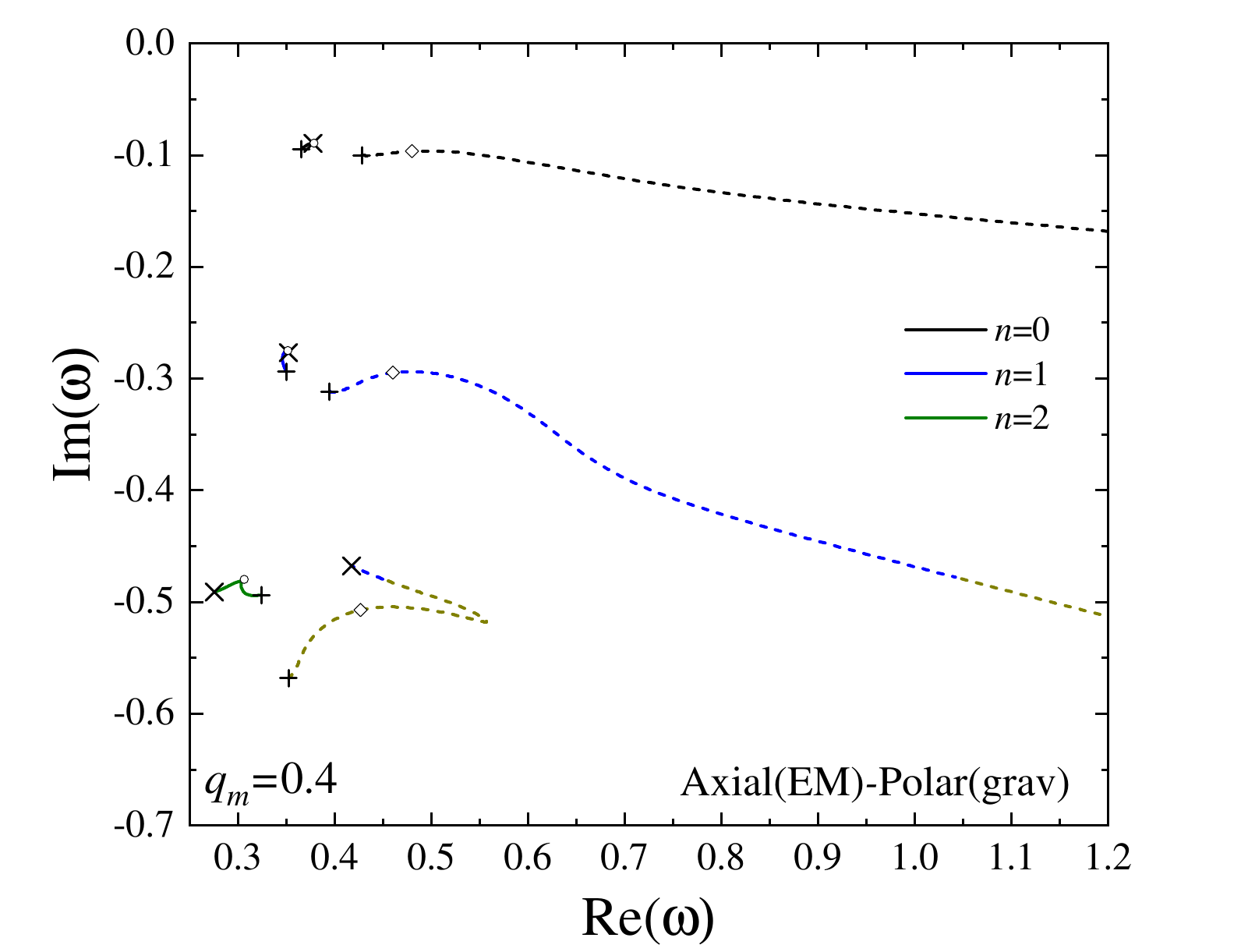}
    \includegraphics[width=0.495\textwidth]{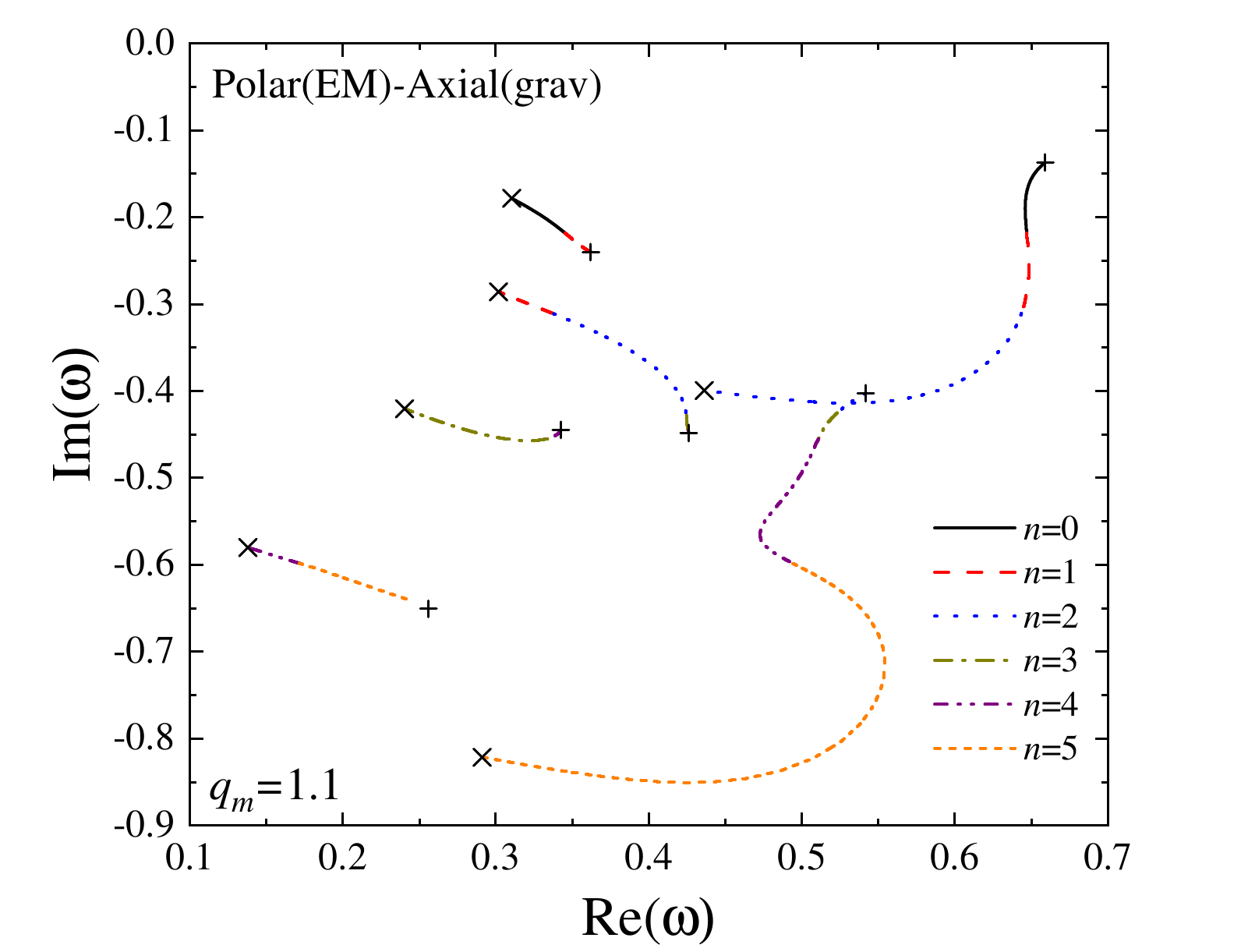}%
    \includegraphics[width=0.495\textwidth]{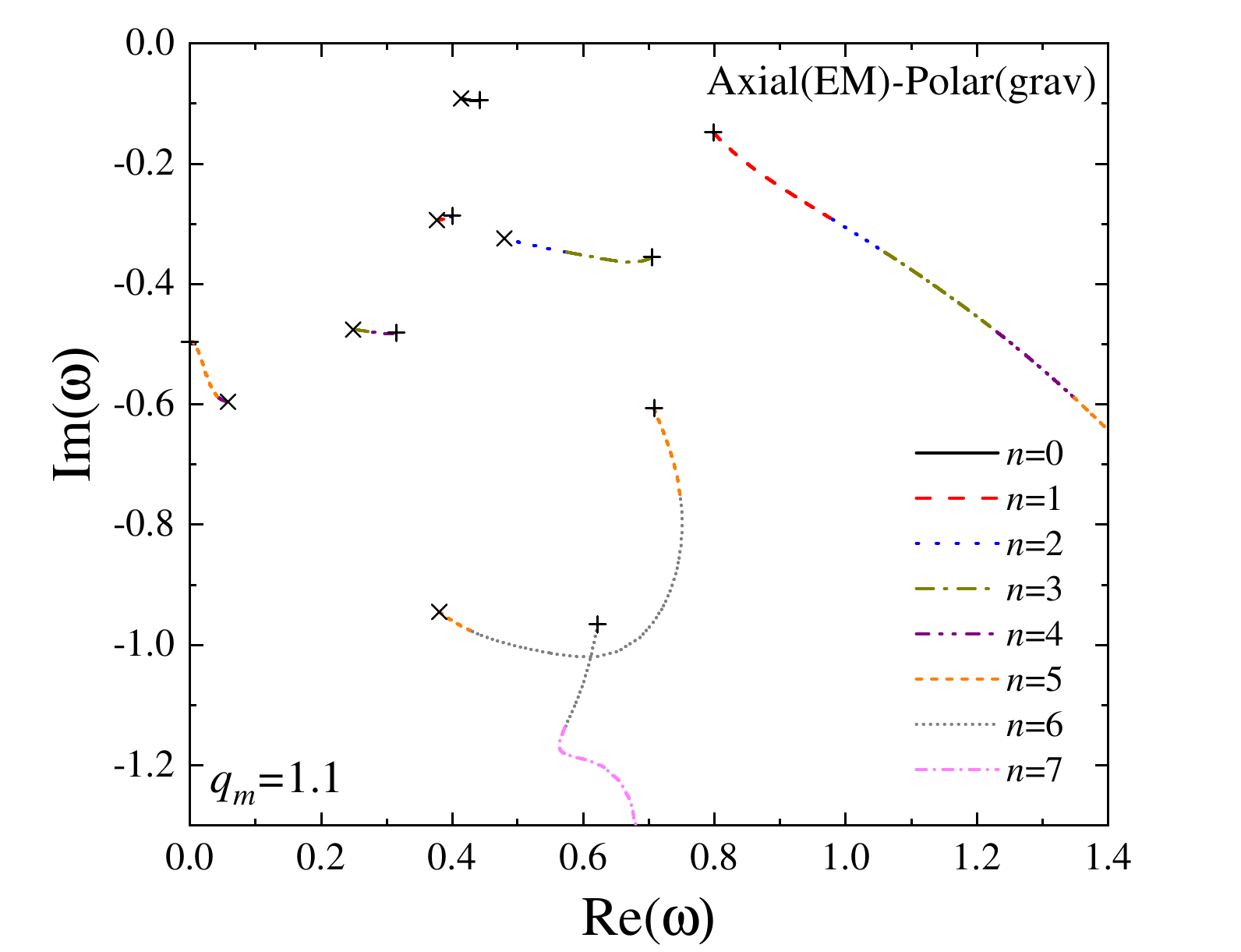}%
    \caption{The QNM trajectories in the complex  $\omega$ plane for $q_m=0.4$ (top) and $q_m=1.1$ (bottom); $\ell=2$. The circle and diamond markers in the left panel indicate the Reissner–Nordstrom modes.  The cross and plus markers indicate the endpoints of the continuation at the maximal and minimal values of $\xi$ respectively.}
    \label{fig:ImRe_magnetic}
\end{figure*}

The QNM trajectories in the complex-frequency plane for dipole perturbations ($\ell=1$) are shown in Figs.~\ref{fig:WIReldip} and \ref{fig:WIRmagdip}, while those for higher multipoles ($\ell\geq2$) are shown in Figs.~\ref{fig:ImRe_electric} and \ref{fig:ImRe_magnetic} for electrically and magnetically charged BHs, respectively. The dependence of the real and imaginary parts of the quadrupole ($\ell=2$) frequencies as functions of $\xi$ is shown in Figs.~\ref{fig:Re_electric} and \ref{fig:Im_electric} for electric BHs and in Figs.~\ref{fig:Re_magnetic} and \ref{fig:Im_magnetic} for magnetic BHs.

Examples of QNM frequencies are given in Tables~\ref{tab:TabEld}--\ref{tab:TabMagd} and Tables~\ref{tab:TabEl} and \ref{tab:TabMg}.

For $\ell\geq2$ and $\xi=0$, the master equations~(\ref{eq:masterEq}) can be decoupled into gravitational and electromagnetic sectors. The corresponding RN frequencies are indicated by circles and diamonds, respectively.

For $\xi\neq0$, the gravitational and electromagnetic perturbations are coupled, and such a separation is unavailable. However, we denote the branches that start from the corresponding RN frequencies at $\xi=0$ as gravitational and electromagnetic branches.\footnote{This classification is not always unambiguous, since the gravitational and electromagnetic branches can cross.}

For fixed $q_{e,m}$, the allowed interval $\xi\in(\xi_1,\xi_2)$ is determined by the general stability conditions~(\ref{eq:st_cond_el}, \ref{eq:st_cond_mg2}, \ref{eq:st_cond_mg4}) and by the boundary separating BHs from naked singularities. The endpoints $\xi_1$ and $\xi_2$ are marked by plus and cross symbols, respectively. For $q_{e,m}>M$, the RN solution contains a naked singularity, and therefore there are no reference RN frequencies for the BH case.

All perturbation sectors of electric and magnetic BHs share several common features. For both electric and magnetic BHs, the isospectrality between the two parity sectors is broken for $\xi\neq0$. In addition, the electric and magnetic QNM spectra are no longer related by the electric--magnetic duality of the RN case~\cite{De_Felice_2024}. 

Additional spectral branches without RN counterparts appear (see Fig. \ref{fig:WIRmagdip}). In the dipole case ($\ell=1$), their appearance can be understood from the significant deformation of the effective potential $V_{\rm eff}$. In addition, multiple QNM branch reconnections may occur near both boundaries of the allowed interval $\xi\in(\xi_1,\xi_2)$.

For $\ell\geq2$ perturbations, the fundamental gravitational modes change only weakly over most of the allowed parameter range. However  the overtones which are expected to probe near horizon geometry deviate more strongly, and one can observe an overtone outburst \cite{Konoplya:2022pbc,Konoplya:2022hll}.

\subsection{Electric Black Holes}
For the particular example $q_e=0.4$, the allowed interval is $\xi_1\simeq-0.397<\xi<\xi_2\simeq0.953$, where $\xi_1$ and $\xi_2$ correspond to the BH--NS boundary and the boundary defined by condition (\ref{eq:st_cond_el}), respectively.

In the axial sector, the fundamental QNMs and the first overtones deviate only weakly from their RN values for $\xi>0$, apart from the electromagnetic branches. For these branches, $\re(\omega)$ grows rapidly, whereas $\im(\omega)$ remains close to its RN value over most of this interval before tending to zero as $\xi\to\xi_2$ (see Fig.~\ref{fig:Im_electric}).

For $\xi<0$, both the gravitational and electromagnetic branches change more strongly, with several branch reconnections and $\im(\omega)\to0$ as $\xi\to\xi_1$. This behavior is expected and is similar to that of RN BHs, where near-zero-damped modes are present~\cite{Zimmerman:2015trm,Yang:2012pj}.
In the polar sector, the gravitational branches mainly follow those in the axial sector. However, the behavior of $\re(\omega)$ for the electromagnetic branches is opposite to that in the axial sector: $\re(\omega)$ grows rapidly as $\xi\to\xi_1$. In addition, new QNM branches that do not originate from RN modes appear near $\xi_1$.

The coefficients of the master equations (\ref{eq:masterEq}) have poles at the points where the conditions  \eqref{eq:st_cond_el} changes sign at the outer horizon. While the solutions remain analytic for $\xi=\xi_2$ an effective barrier similar to that arising for massive fields appears. This may explain why $|\im(\omega)|$ tends to zero.

\subsection{Magnetic Black Holes}

For magnetic BHs, gravitational and electromagnetic perturbations of opposite parity are coupled. Polar electromagnetic perturbations are coupled to axial gravitational perturbations, while axial electromagnetic perturbations are coupled to polar gravitational perturbations. We denote these two systems as Polar(EM)-Axial(G) and Axial(EM)-Polar(G), respectively.

For the particular example $q_m=0.4$, the allowed interval is $\xi_1\simeq-0.45<\xi<\xi_2\simeq1.1$, where $\xi_1$ and $\xi_2$ correspond to the lower and upper boundaries defined by conditions (\ref{eq:st_cond_mg2}) and (\ref{eq:st_cond_mg4}), respectively\footnote{Note that near the boundaries of the allowed interval, the numerical convergence deteriorates.}.

For the Polar(EM)-Axial(G) perturbations, the gravitational branches remain close to their RN values over most of the interval. Their $\re(\omega)$ generally decreases as $\xi$ increases, although the overtone branches turn near $\xi_2$. Their $\im(\omega)$ does not tend to  zero at either boundary. The electromagnetic branches change more strongly. For $\xi<0$, their $\re(\omega)$ increases rapidly as $\xi\to\xi_1$. For $\xi>0$, their $\re(\omega)$ first decreases and then increases near $\xi_2$. 

For the Axial(EM)-Polar(G) perturbations, both $\re(\omega)$ and $\im(\omega)$ of the gravitational branches change weakly over most of the interval. The electromagnetic branches deviate moderately from their RN values for $\xi<0$. For $\xi>0$, the $\re(\omega)$ of the fundamental mode and the first overtone increase rapidly, while the next overtone reaches a maximum and then decreases. At the same time, their $\im(\omega)$ remains finite and becomes more negative, especially for the overtones.

Typical time evolution of the  Gaussian wave packets governed by ~(\ref{eq:masterEq}) with the  time-dependence restored are shown in Figs.~\ref{fig:wave_el_axial}--\ref{fig:wave_mag_el_ax} for several cases of electric and magnetic BHs, respectively. One can observe the ringdown behaviour followed by standard Price tails. Since we deal with a coupled system, the fundamental modes and first overtones associated with different branches compete with one another and may be simultaneously visible in the ringdown profiles (see Fig. \ref{fig:wave_mag_el_ax}). We verify our calculations by fitting the ringdown profiles and comparing the extracted frequencies with those obtained using the pseudospectral method. The frequencies obtained using the two methods are in perfect agreement.
\FloatBarrier

\section{Conclusions}\label{Conclusions}

We have studied linear gravitational perturbations and quasinormal modes spectra of electrically and magnetically charged black holes in a higher-order action in which the electromagnetic field is coupled non-minimally to the double dual Riemann tensor (in the vector Horndeski form with two-derivative equations of motion). We mapped the black-hole solutions in the charge--coupling parameter space and refined the regions satisfying the no-ghost and gradient-stability conditions \cite{Chen:2024hkm}. For both types of black holes, we strictly demonstrated  that the conditions (\ref{eq:st_cond_el}) and (\ref{eq:st_cond_mg2}, \ref{eq:st_cond_mg4}) are   sufficient to guarantee the remaining conditions, as well as for  magnetic black holes  checking these conditions only at the outer horizon is insufficient.

For nonzero $\xi$ (the new interaction's coupling), the gravitational and electromagnetic perturbations are coupled. The interaction breaks both the axial--polar isospectrality and the electric--magnetic duality as can be observed in the Reissner--Nordstr\"om spectrum \cite{De_Felice_2024}. The fundamental gravitational modes remain close to their Reissner--Nordstr\"om values over much of the allowed parameter space, whereas the electromagnetic branches and the higher overtones are considerably more sensitive to the coupling. In particular, we find overtone outbursts, multiple reconnections of spectral branches, and additional branches without Reissner--Nordstr\"om counterparts. 

Near the boundaries of the allowed parameter domains, the spectra show qualitatively different behavior in the electric and magnetic cases. Some electric branches become weakly damped, with $|\operatorname{Im}\omega|$ tending to zero, whereas the imaginary parts for the  the magnetic BHs  do not show a similar tendency toward zero.
Additionally, we have scanned the allowed parameter domains $(\xi,q_{e,m})$ and found no exponentially growing modes with ${\rm Im}(\omega)>0$. Although our numerical analysis was restricted to the lowest relevant multipoles, $\ell=1$ and $\ell=2$, instabilities at higher $\ell$ are expected to be less likely, since the increasing angular-momentum contribution raises the effective centrifugal barrier, while the absence of angular gradient instabilities rules out instabilities in the eikonal regime. This numerical evidence for mode stability is distinct from the analytic no-ghost and gradient-stability conditions; taken together, these results support the linear viability of the configurations examined here.

Although electrically charged black holes constitute well-defined solutions, astrophysical black holes are generally expected to be uncharged or to carry only a negligibly small net electric charge. This is because different physical mechanisms, including electron-positron production and neutralization from the surrounding plasma, tends to efficiently neutralize any substantial charge \cite{10.1093/mnras/stz1904}. Nevertheless, electrically charged black hole solutions remain of considerable theoretical interest \cite{Arkani-Hamed:2006emk,PhysRevLett.123.051601}.

Magnetically charged black holes are particularly interesting from an astrophysical and cosmological perspective. In contrast to electric charge, there is no efficient way for magnetic charge to be neutralized. Their discharge requires magnetically charged states, such as magnetic monopoles, and can therefore be strongly suppressed if the lightest monopoles are sufficiently massive. Primordial black holes formed after monopole producing phase transitions in the early universe \cite{PhysRevLett.43.1365,Guth:1979bh}  may acquire a net magnetic charge. They may then retain their charge as they cannot easily neutralize. This is a particularly interesting direction for future observations and constraints.

A natural direction for further work is to extend the analysis of this paper to rotating charged black holes, where the interplay between spin, charge, and the non-minimal coupling may lead to qualitatively new features.

\section*{Acknowledgments}
M.~P.~H. is supported in part by National Science Foundation grants PHY-2310572 and PHY-2609905.
O.~S.~S. is supported in part
by the Tufts Scholar at Risk Program and the Harvard Scholar at Risk Program.
We thank Grant Remmen for discussion.

\appendix

\section{Additional Figures}\label{AppExtra}

Below, we present additional figures that provide further numerical details

\nopagebreak

\begin{figure*}
    \includegraphics[width=0.495\textwidth]{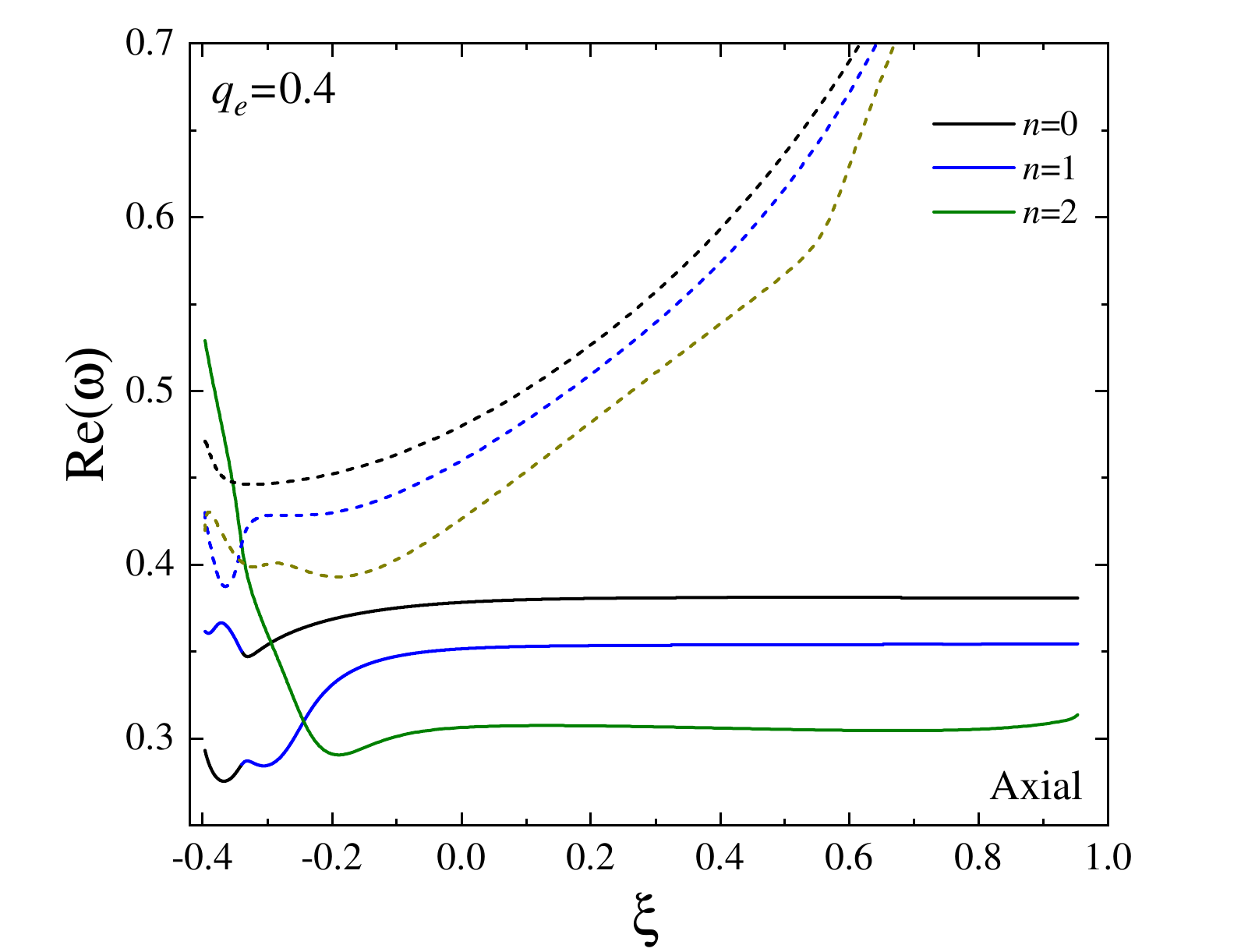}%
    \includegraphics[width=0.495\textwidth]{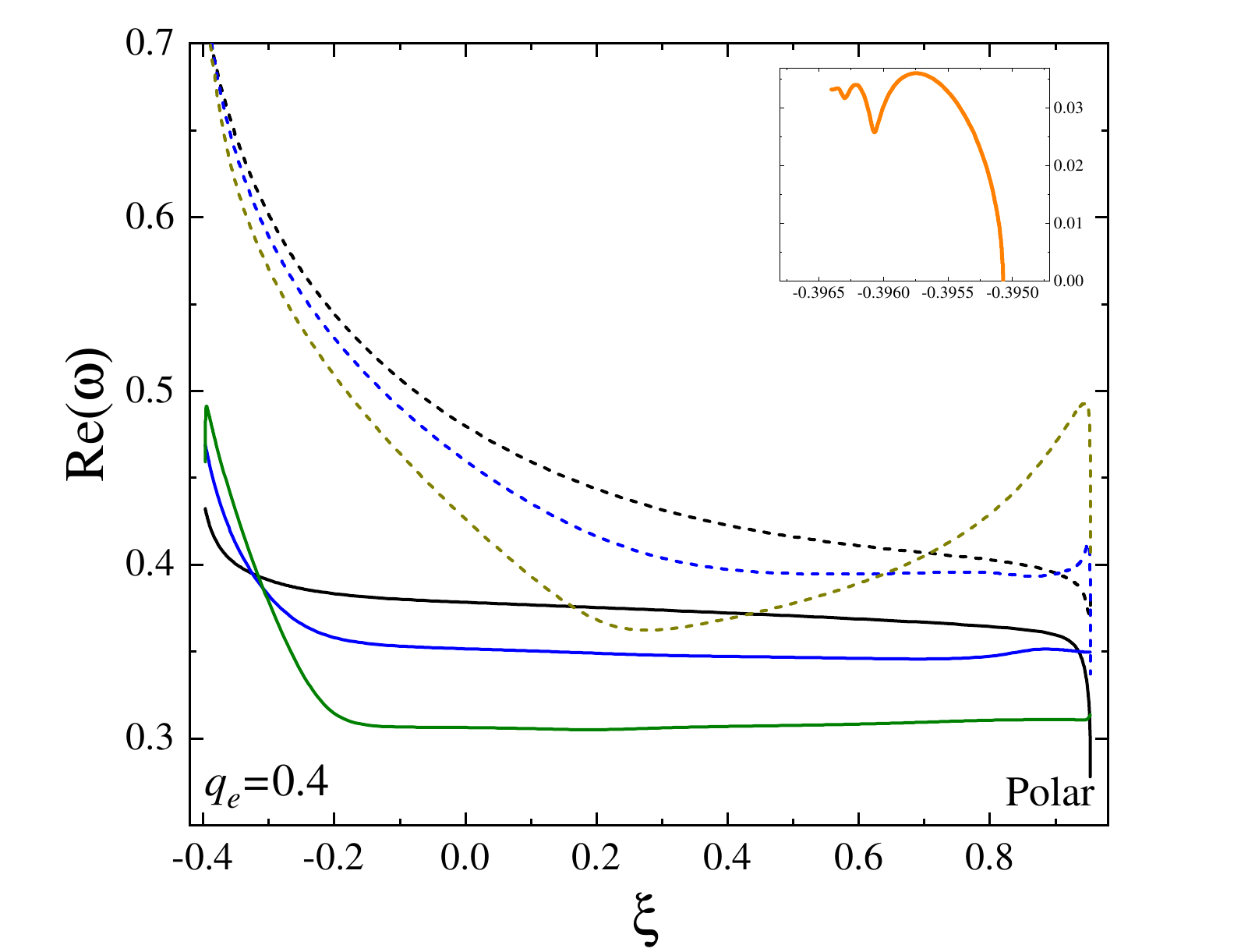}
    \includegraphics[width=0.495\textwidth]{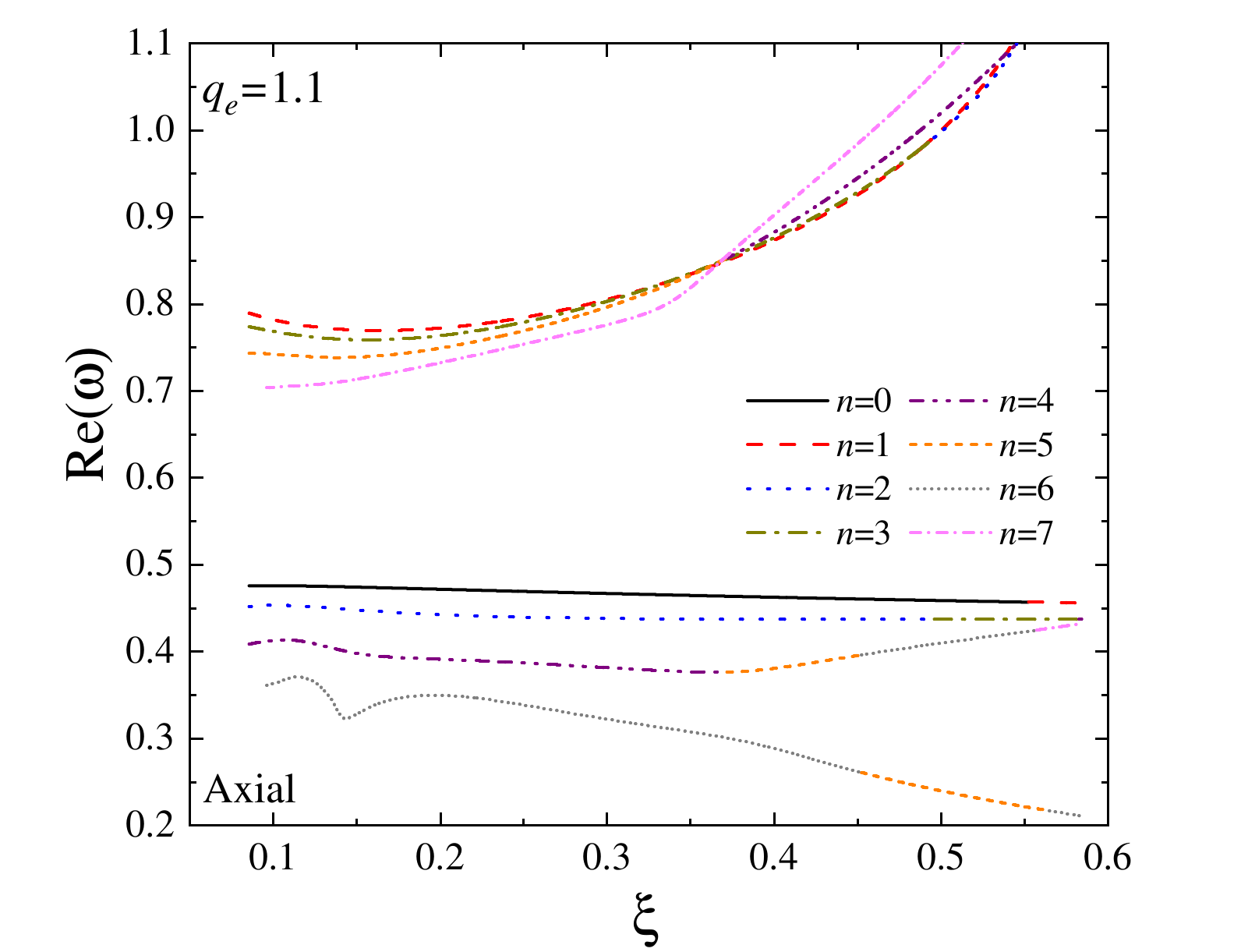}%
        \includegraphics[width=0.495\textwidth]{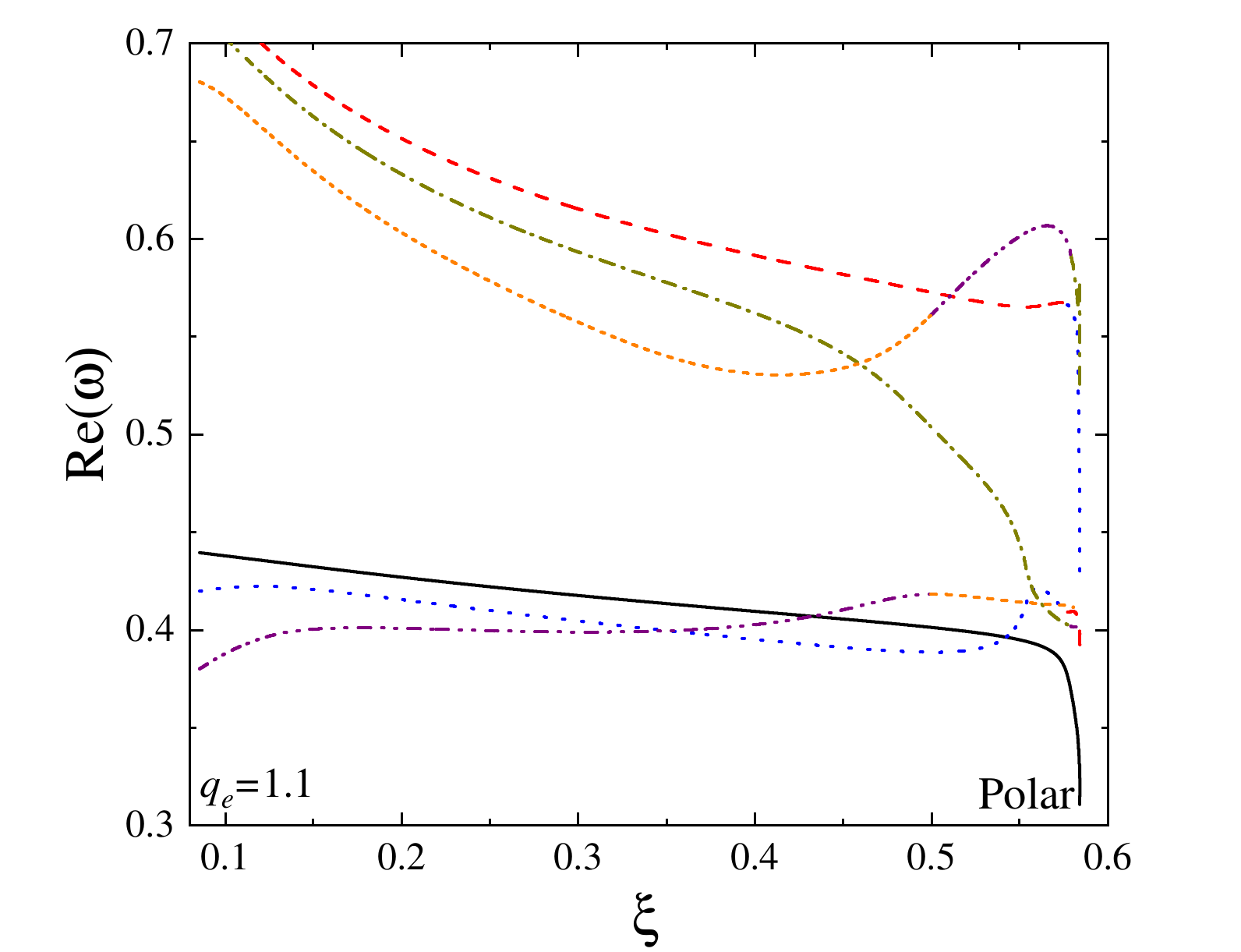}%
    \caption{Typical examples of the behavior of the real parts of the fundamental quasinormal modes and  first overtones for $q_e=0.4$ (top) and $q_e=1.1$ (bottom); $\ell=2$.
Solid curves correspond to the gravitational channel branch, while dashed curves correspond to the electromagnetic channel branch.}
\label{fig:Re_electric}
\end{figure*}
\begin{figure*}
    \includegraphics[width=0.495\textwidth]{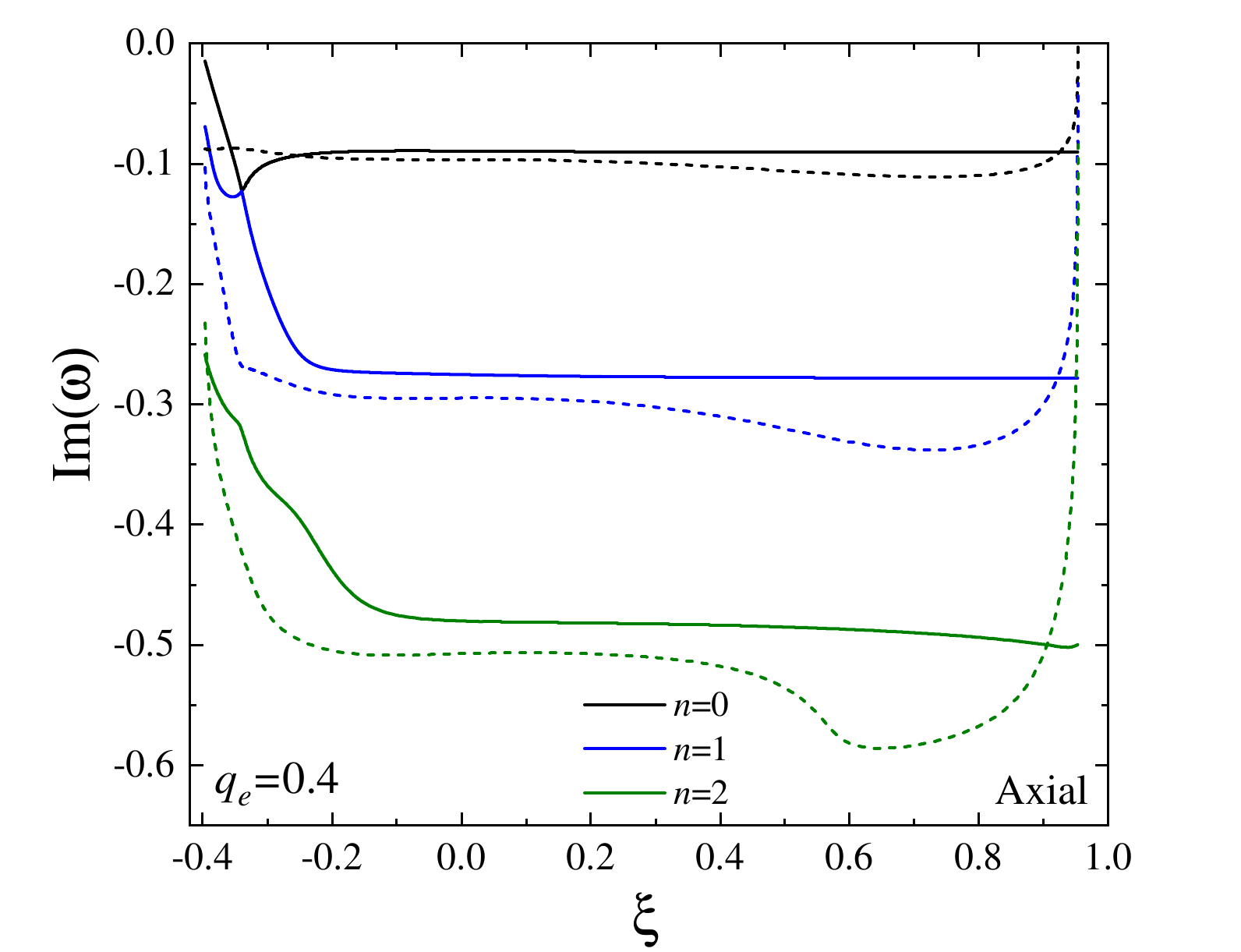}%
    \includegraphics[width=0.495\textwidth]{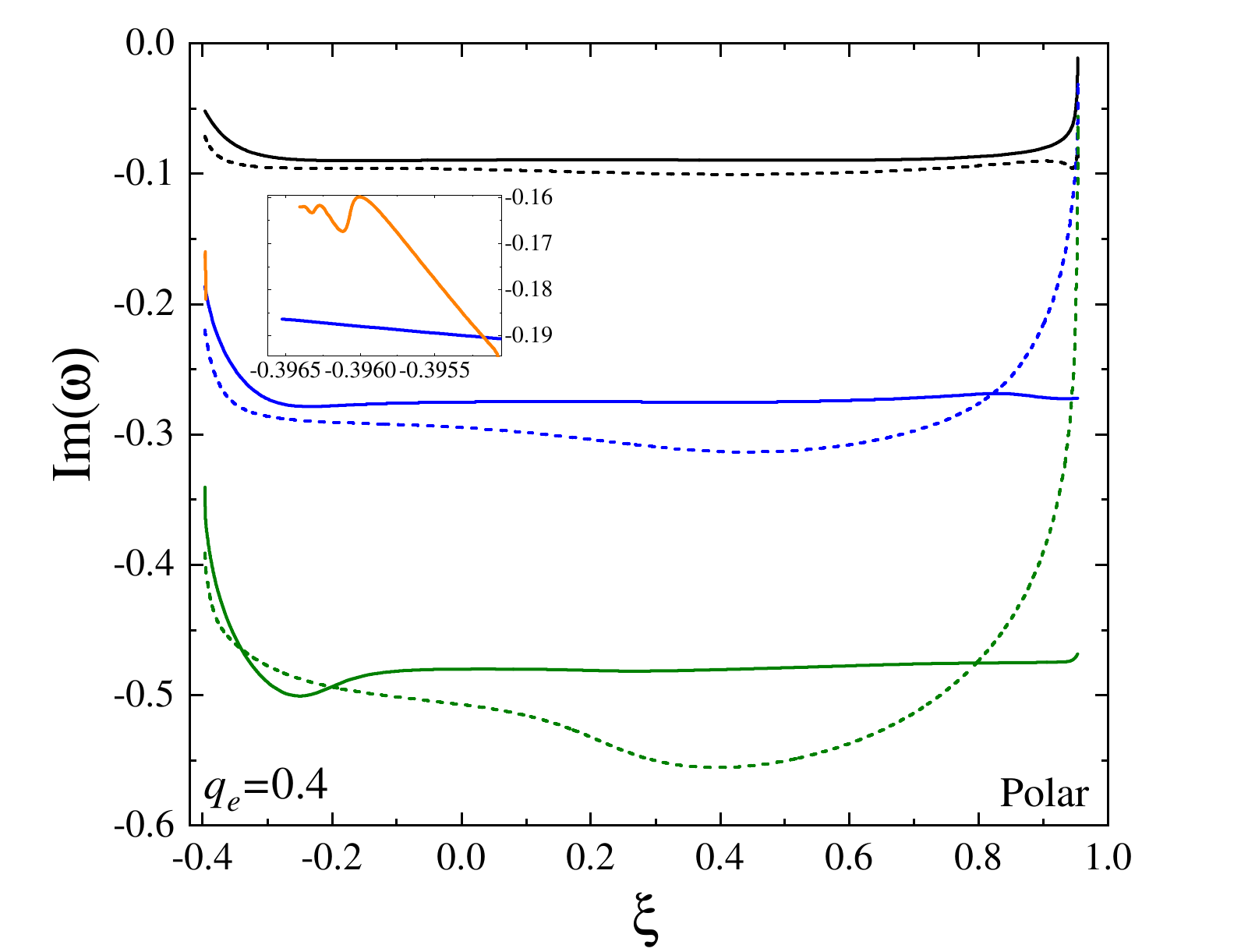}
    \includegraphics[width=0.495\textwidth]{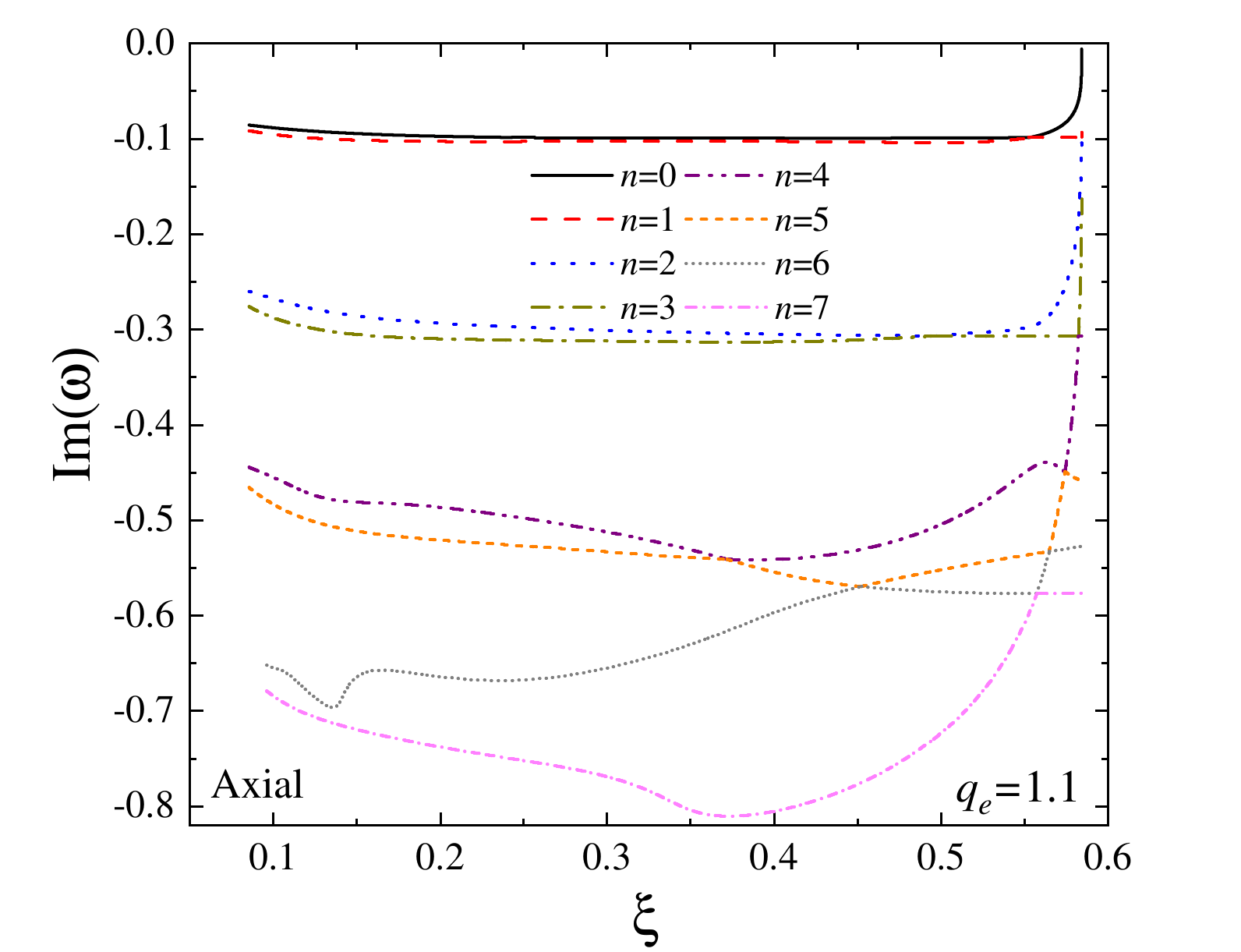}%
        \includegraphics[width=0.495\textwidth]{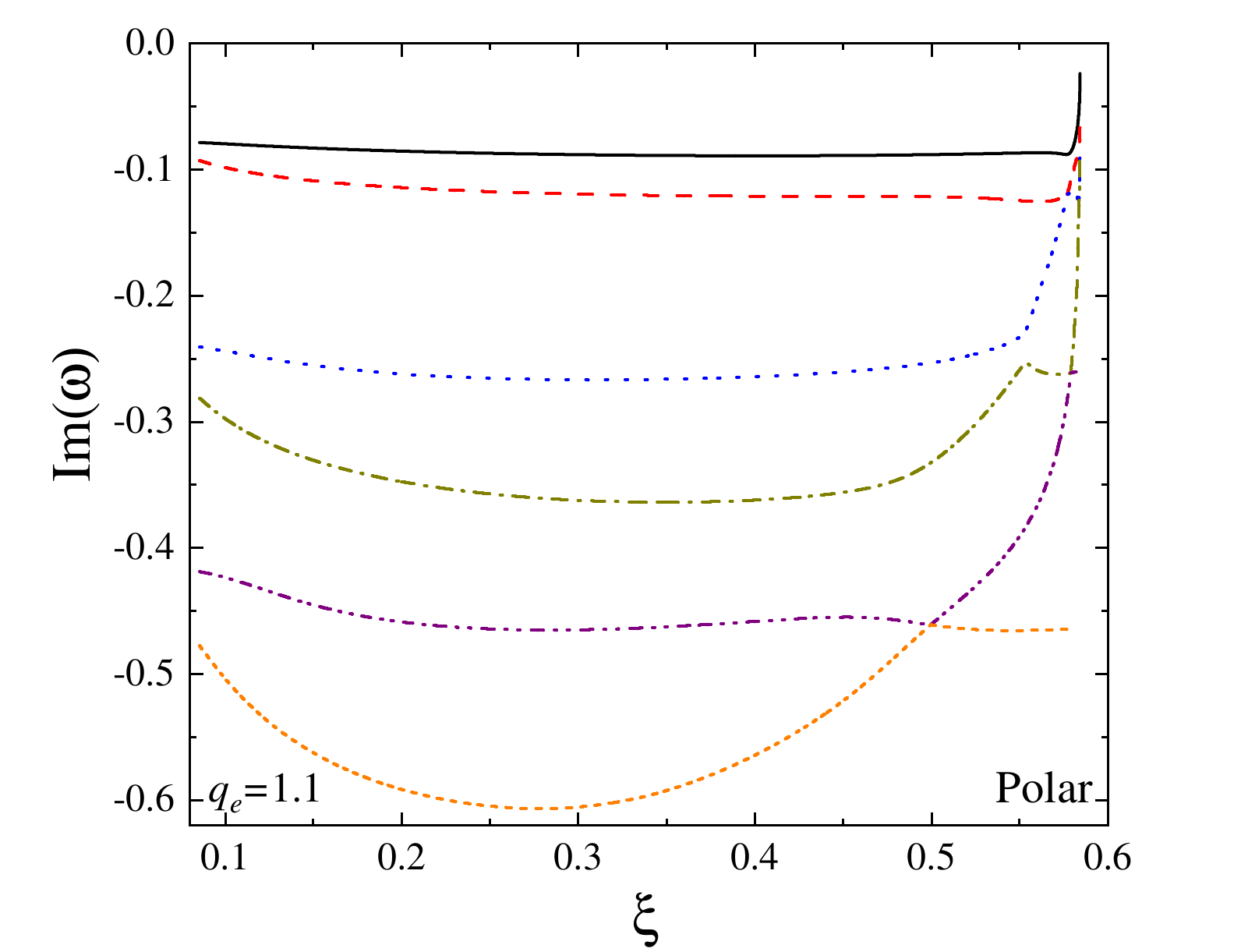}%
    \caption{Typical examples of the behavior of the imaginary parts of the fundamental quasinormal mode and the first overtones for $q_e=0.4$ (top) and $q_e=1.1$ (bottom); $\ell=2$.
Solid curves correspond to the gravitational channel branch, while dashed curves correspond to the electromagnetic channel branch.}
\label{fig:Im_electric}
\end{figure*}

\begin{figure*}
    \includegraphics[width=0.495\textwidth]{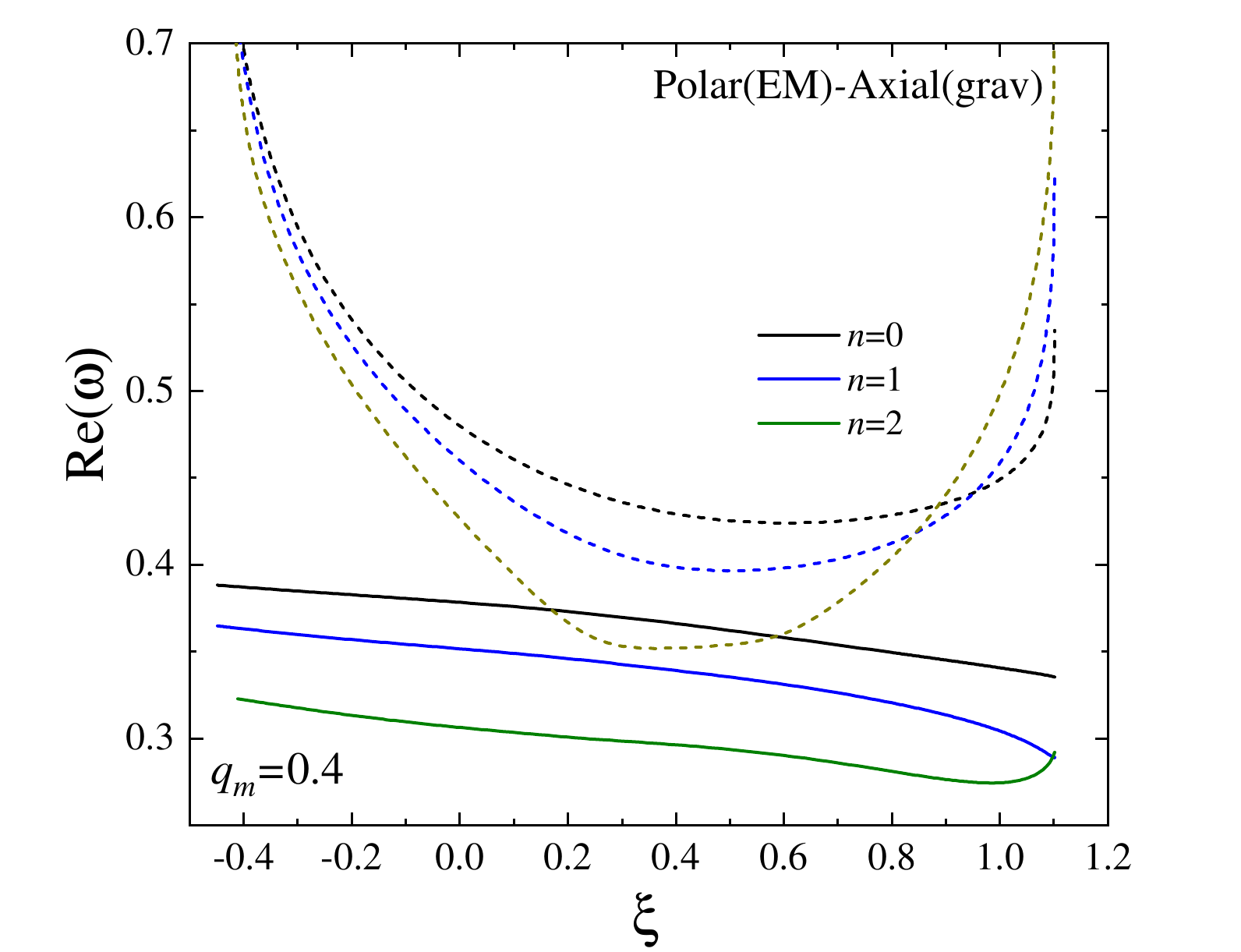}%
    \includegraphics[width=0.495\textwidth]{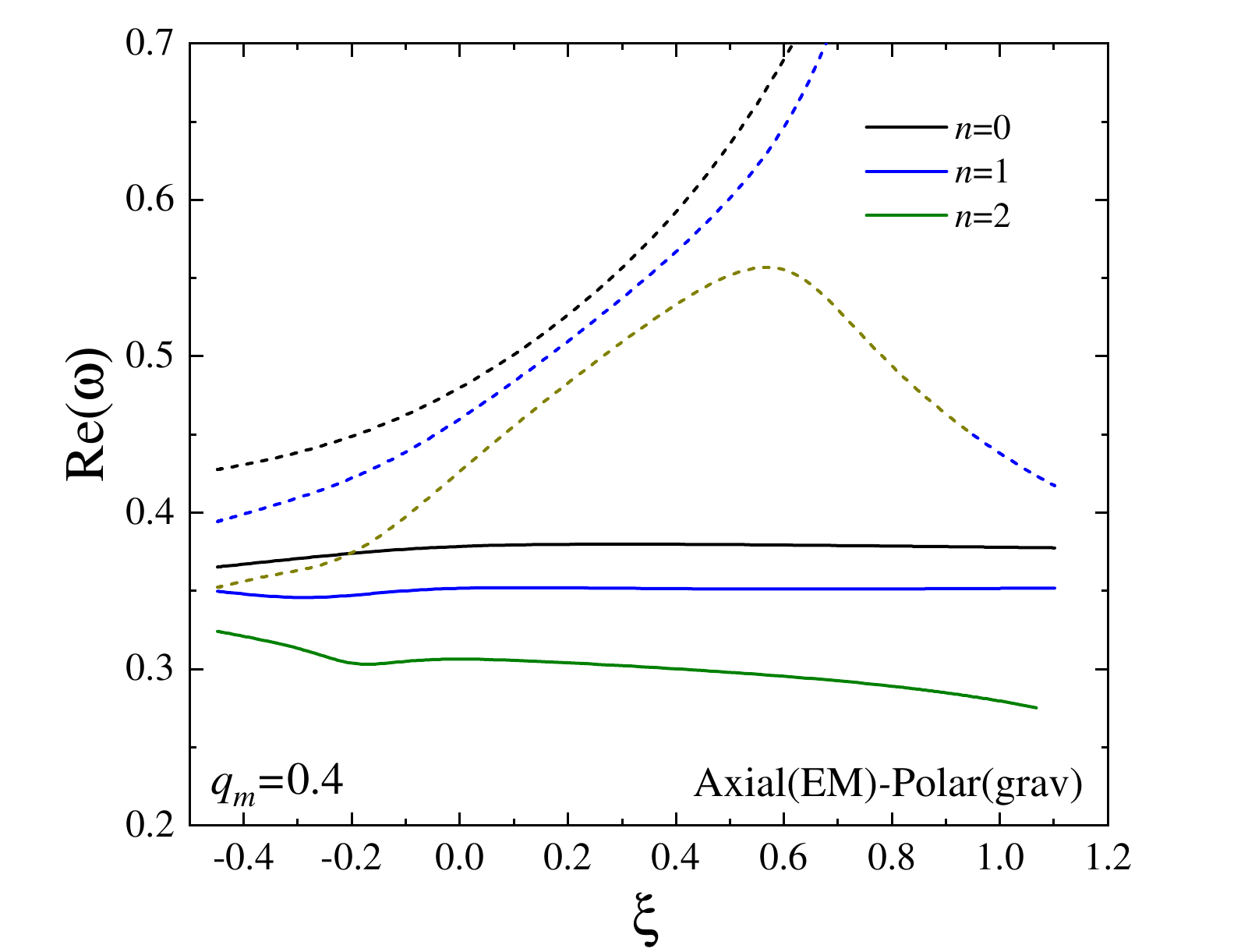}
    \includegraphics[width=0.495\textwidth]{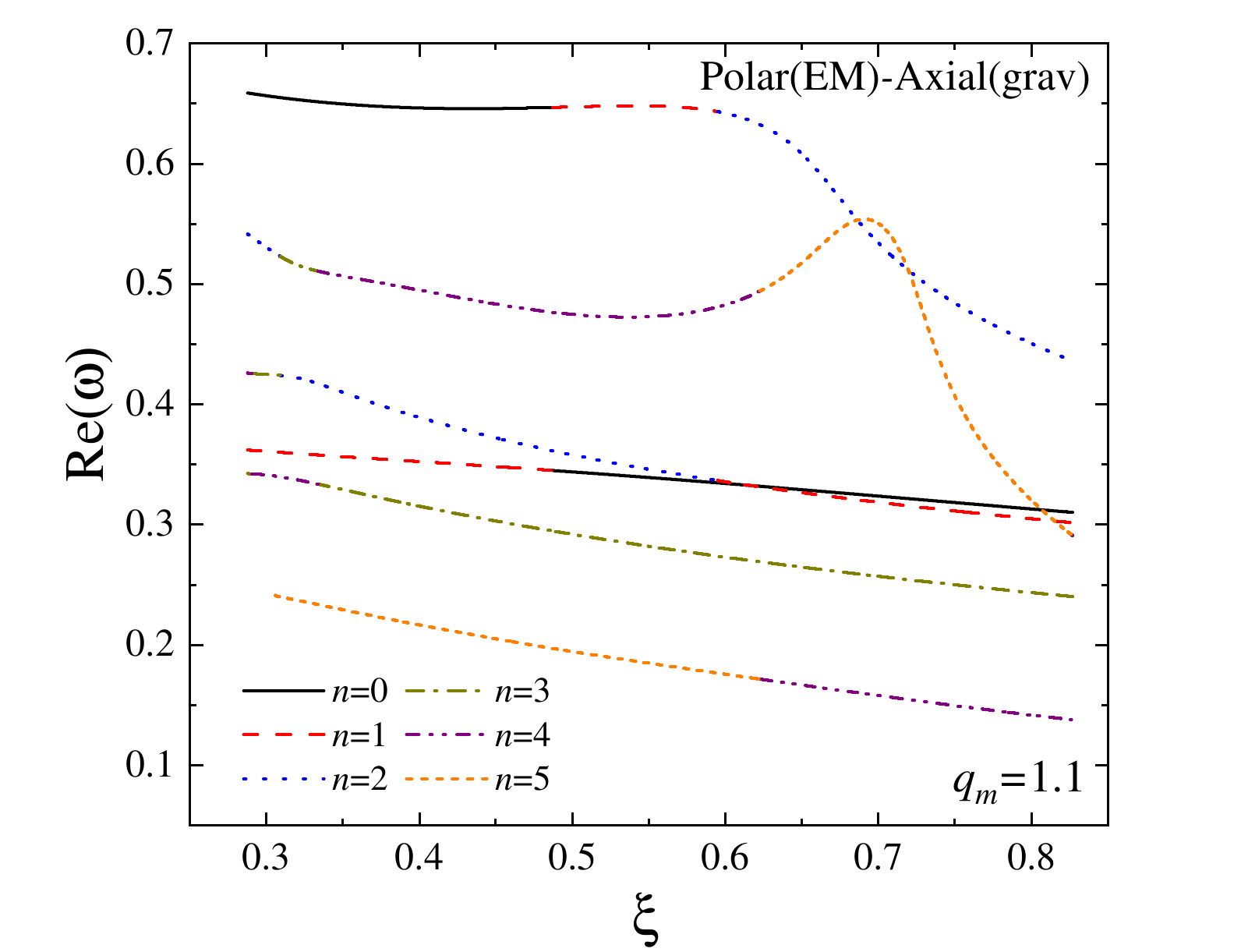}%
    \includegraphics[width=0.495\textwidth]{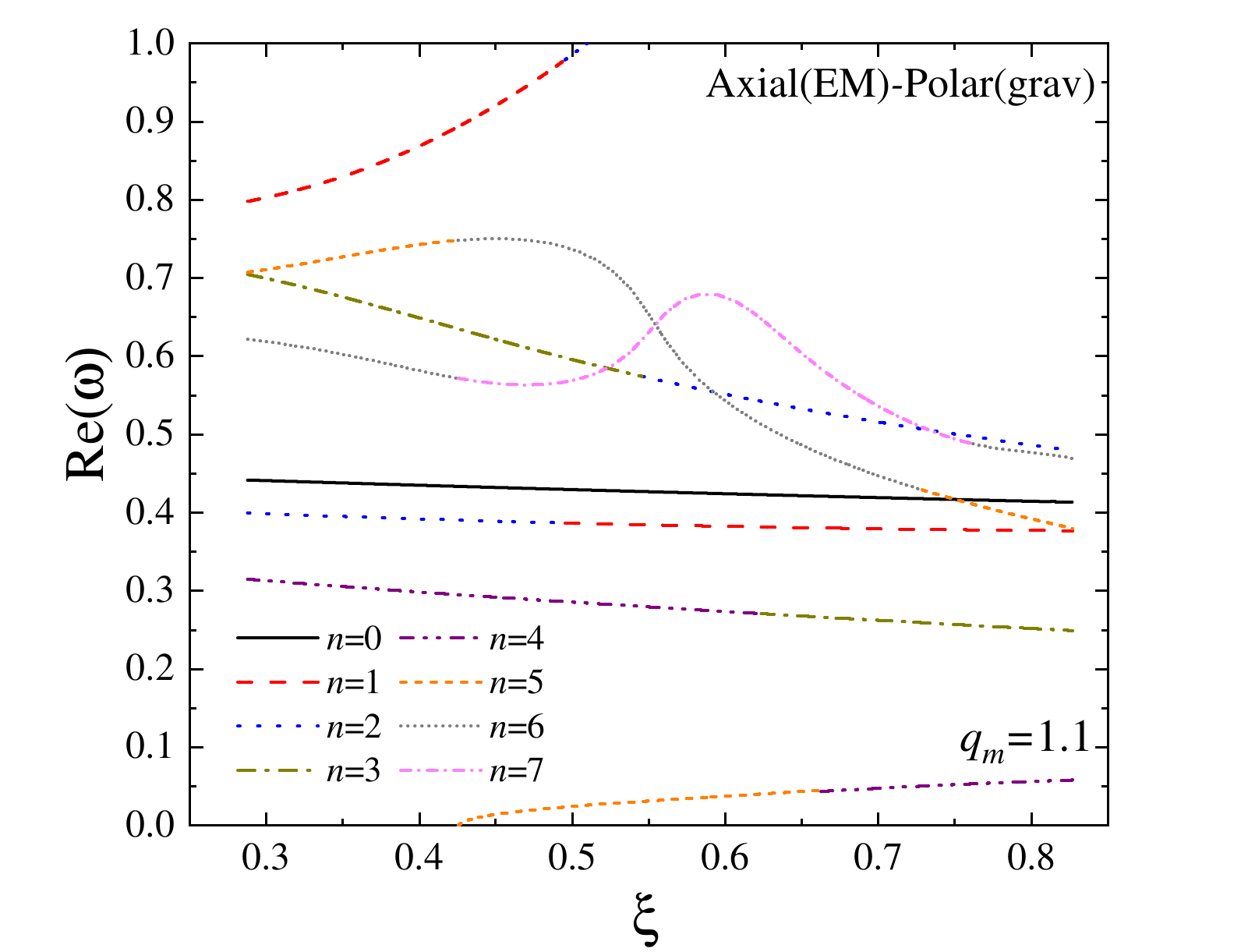}%
    \caption{Typical examples of the behavior of the real parts of the fundamental quasinormal modes and  first overtones for $q_m=0.4$ (top) and $q_m=1.1$ (bottom); $\ell=2$.
Solid curves correspond to the gravitational channel branch, while dashed curves correspond to the electromagnetic channel branch.}
\label{fig:Re_magnetic}
\end{figure*}
\begin{figure*}
    \includegraphics[width=0.495\textwidth]{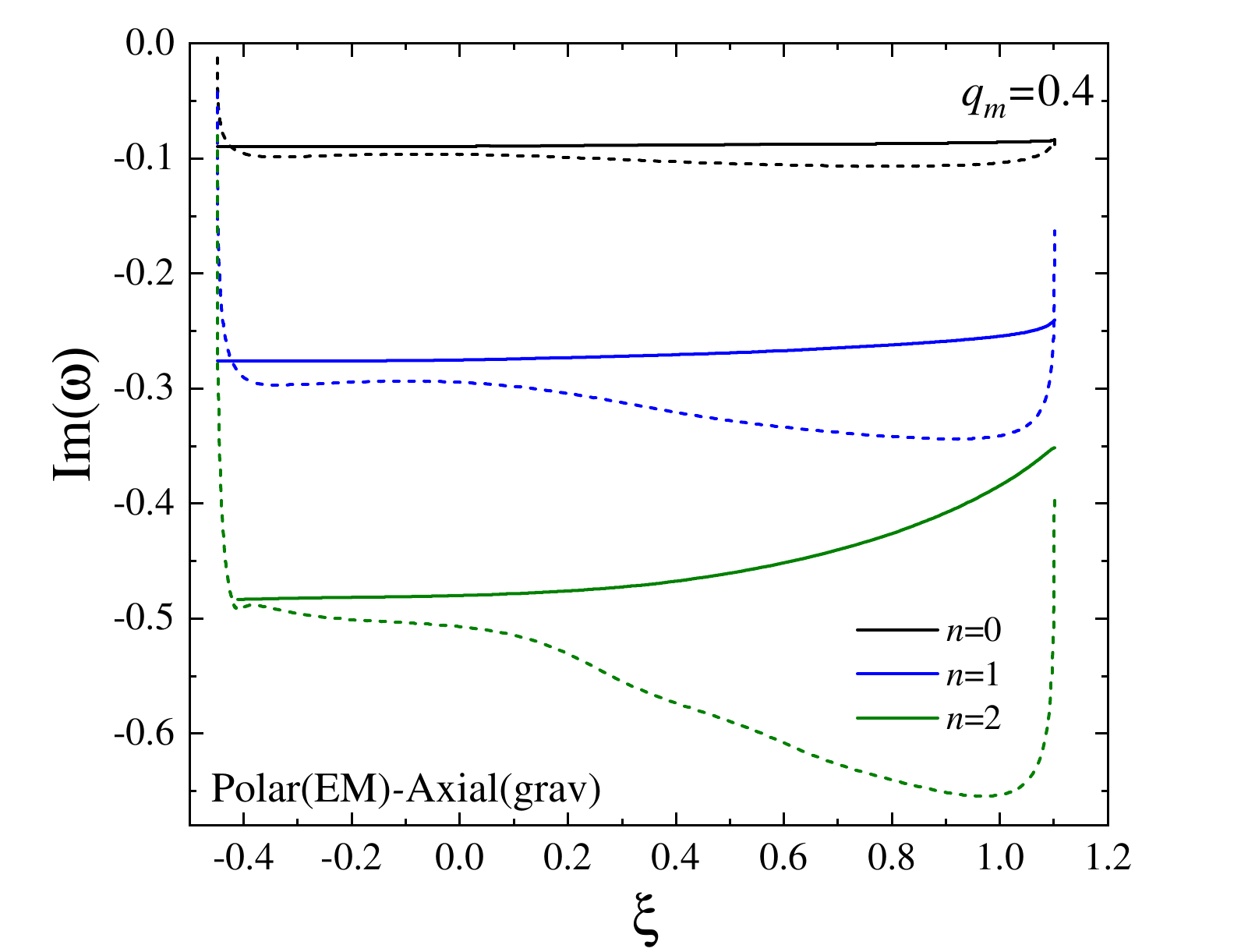}%
    \includegraphics[width=0.495\textwidth]{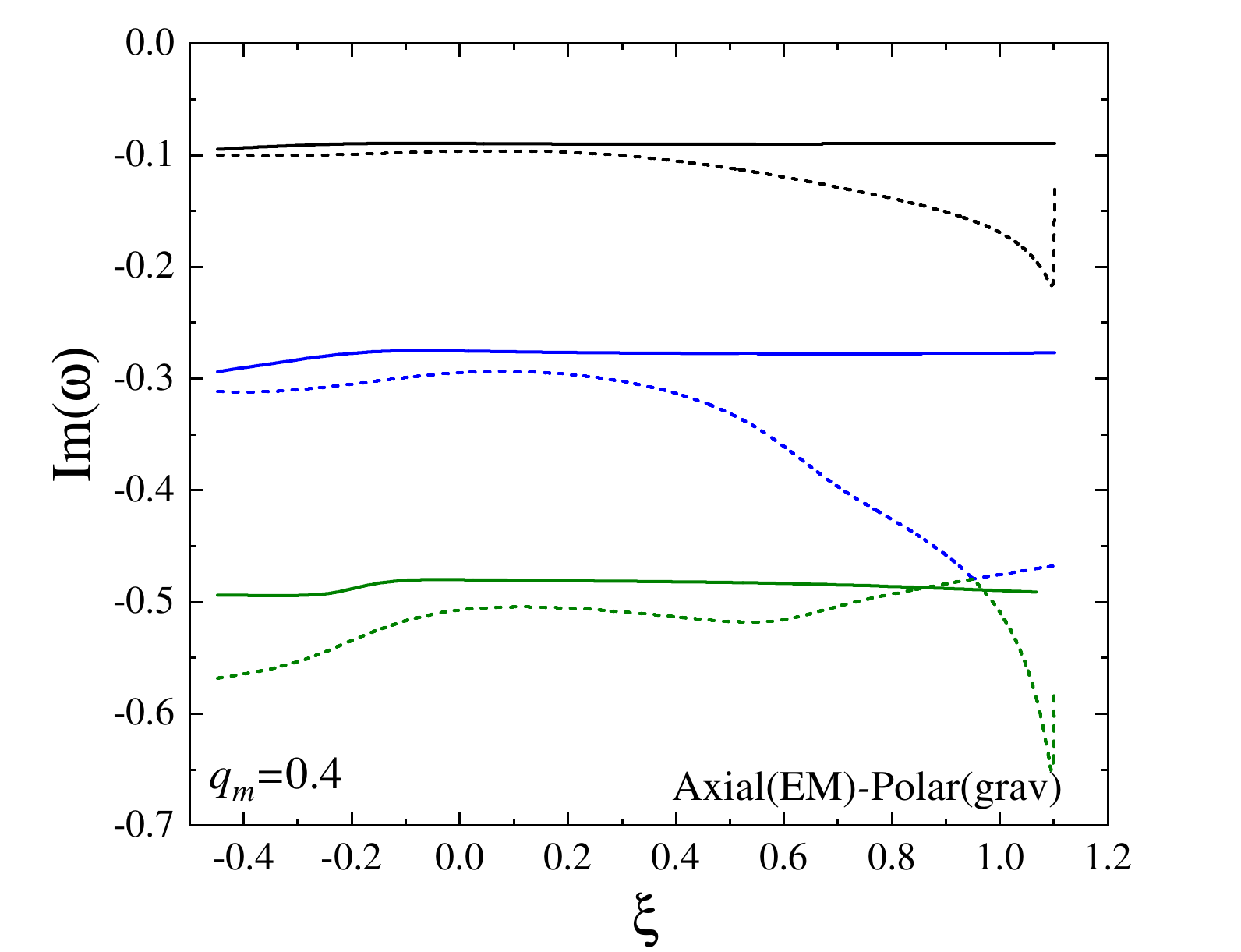}
    \includegraphics[width=0.495\textwidth]{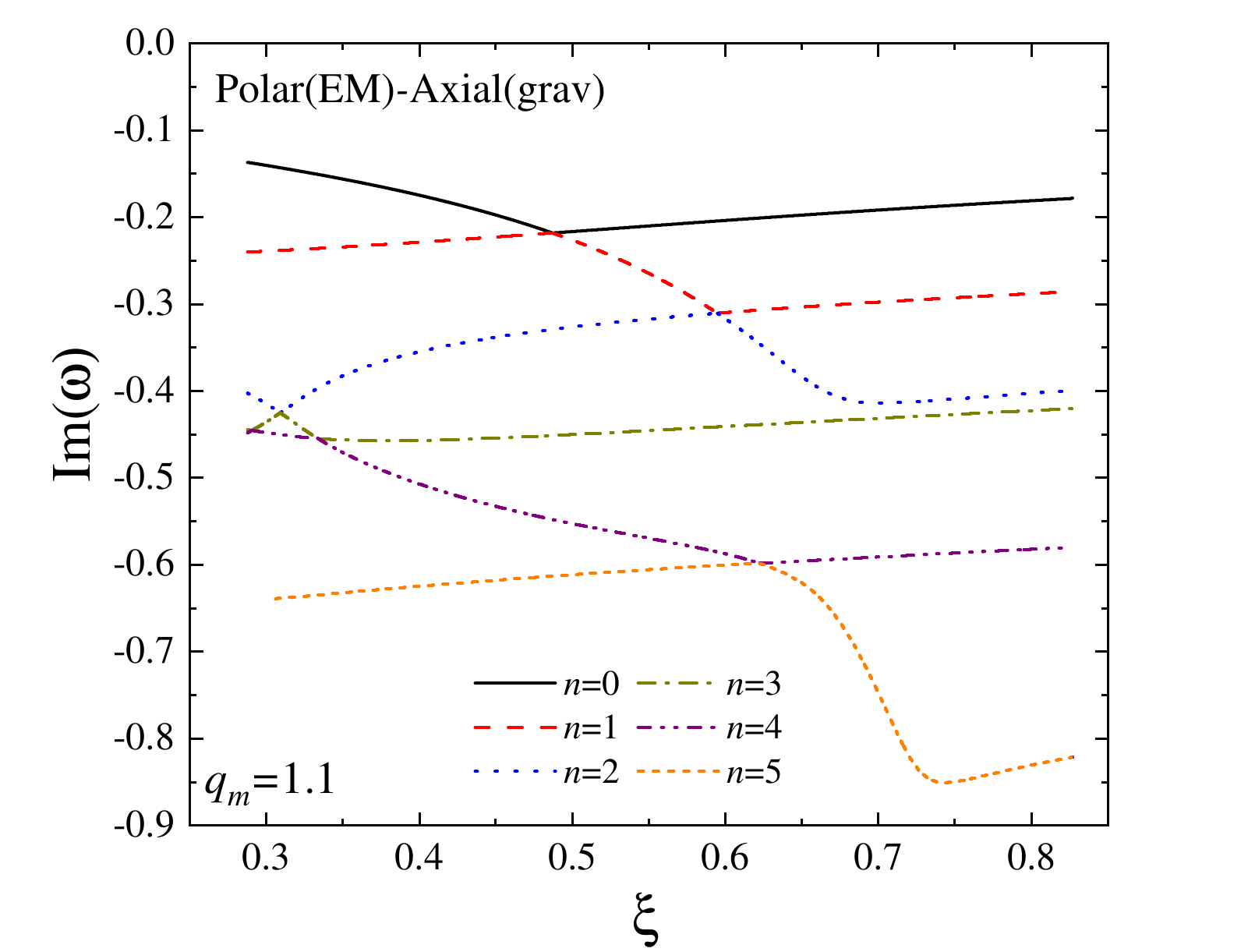}%
    \includegraphics[width=0.495\textwidth]{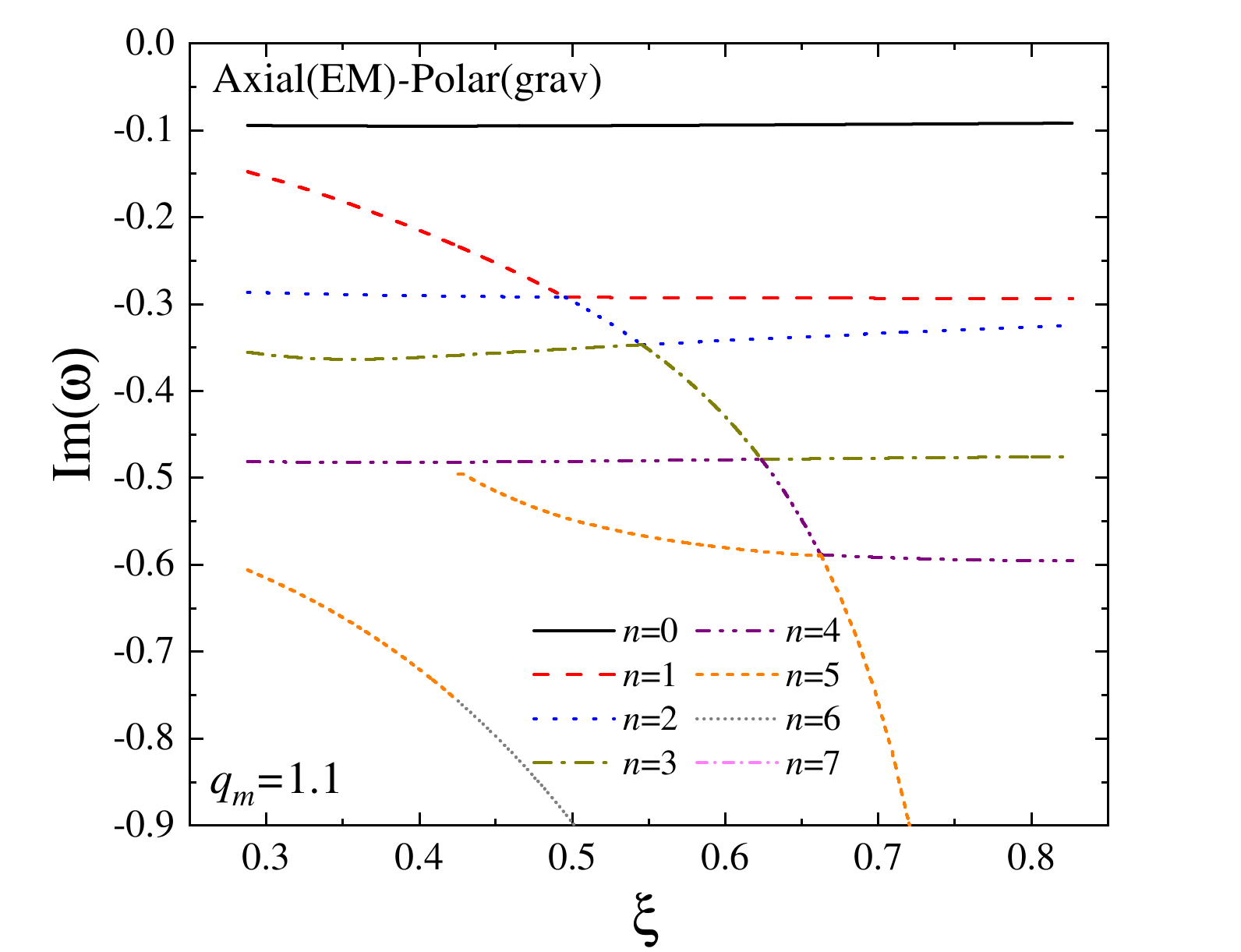}%
    \caption{Typical examples of the behavior of the imaginary parts of the fundamental quasinormal modes and  first overtones for $q_m=0.4$ (top) and $q_m=1.1$ (bottom); $\ell=2$.
Solid curves correspond to the gravitational channel branch, while dashed curves correspond to the electromagnetic channel branch.}
\label{fig:Im_magnetic}
\end{figure*}

\FloatBarrier

\clearpage
\onecolumngrid

\section{Precise Values of QNMs}\label{AppPrecise}
\centering
\begin{table}[h]
\caption{The accurate values of QNMs for electric BHs for the dipole perturbations ($\ell=1$) for different $\xi$ and  $q_e=0.7$.}
\begin{tabular}{|c|c|c|}
\hline
\hline
$\xi$ & Axial & Polar \\
\hline
\multirow{3}{*}{$-0.22$} & $0.312280 - 0.055865i$ & $0.412405 - 0.080883i$ \\
 & $0.329305 - 0.189625i$ & $0.394023 - 0.242267i$ \\
 & $0.334527 - 0.360275i$ & $0.366289 - 0.402745i$ \\
\hline
\multirow{3}{*}{$-0.1$} & $0.294570 - 0.092095i$ & $0.330905 - 0.095216i$ \\
 & $0.274517 - 0.288448i$ & $0.311792 - 0.294672i$ \\
 & $0.254502 - 0.509551i$ & $0.289799 - 0.511618i$ \\
\hline
\multirow{3}{*}{RN} & $0.299483 - 0.097382i$ & $0.299483 - 0.097382i$ \\
 & $0.272088 - 0.304516i$ & $0.272088 - 0.304516i$ \\
 & $0.237671 - 0.535370i$ & $0.237671 - 0.535370i$ \\
\hline
\multirow{3}{*}{$0.3$} & $0.342492 - 0.100899i$ & $0.256279 - 0.107345i$ \\
 & $0.319837 - 0.312389i$ & $0.227235 - 0.362133i$ \\
 & $0.288854 - 0.541196i$ & $0.239247 - 0.638442i$ \\
\hline
\multirow{3}{*}{$0.6$} & $0.436229 - 0.110135i$ & $0.255891 - 0.108911i$ \\
 & $0.404985 - 0.330164i$ & $0.273912 - 0.345313i$ \\
 & $0.372234 - 0.527027i$ & $0.322216 - 0.591654i$ \\
\hline
\multirow{3}{*}{$0.85$} & $0.768515 - 0.045420i$ & $0.281849 - 0.033147i$ \\
 & $0.800330 - 0.141103i$ & $0.346692 - 0.111999i$ \\
 & $0.855904 - 0.244799i$ & $0.440575 - 0.206139i$ \\
\hline
\multirow{3}{*}{$0.8505$} & $0.781973 - 0.030174i$ & $0.279543 - 0.022874i$ \\
 & $0.811642 - 0.095879i$ & $0.337015 - 0.079118i$ \\
 & $0.861510 - 0.170275i$ & $0.420415 -  0.148752i$ \\
\hline
\hline
\end{tabular}
\label{tab:TabEld}
\end{table}
\begin{table}[h]
\caption{The accurate values of QNMs for magnetic BHs for the dipole perturbations ($\ell=1$) for different $\xi$, $q_m=0.7$.}
\centering
\begin{tabular}{|c|c|c|}
\hline
\hline
$\xi$ & Polar(EM)-Axial(grav) & Axial(EM)-Polar(grav) \\
\hline
\multirow{3}{*}{$-0.333$} & $0.542830 - 0.034351i$ & $0.293317 - 0.116419i$ \\
 & $0.586516 - 0.113656i$ & $0.293733 - 0.372377i$ \\
 & $0.651524 - 0.218903i$ & $0.315652 - 0.649411i$ \\
\hline
\multirow{3}{*}{$-0.15$} & $0.339796 - 0.097348i$ & $0.290353 - 0.105460i$ \\
 & $0.322317 - 0.300584i$ & $0.256949 - 0.344316i$ \\
 & $0.307222 - 0.518570i$ & $0.252891 - 0.623926i$ \\
\hline
\multirow{3}{*}{RN} & $0.299483 - 0.097382i$ & $0.299483 - 0.097382i$ \\
 & $0.272088 - 0.304516i$ & $0.272088 - 0.304516i$ \\
 & $0.237671 - 0.535370i$ & $0.237671 - 0.535370i$ \\
\hline
\multirow{3}{*}{$0.3$} & $0.258635 - 0.110321i$ & $0.338391 - 0.100125i$ \\
 & $0.160408 - 0.342283i$ & $0.316222 - 0.300703i$ \\
 & $0.074166 - 0.465292i$ & $0.300338 - 0.538478i$ \\
\hline
\multirow{3}{*}{$0.6$} & $0.242979 - 0.144851i$ & $0.409684 - 0.149317i$ \\
 & $0.163561 - 0.483297i$ & $0.341068 - 0.296040i$ \\
 & $0.198517 - 0.803669i$ & $0.379454 - 0.611825i$ \\
\hline
\multirow{3}{*}{$0.9$} & $0.236370 - 0.224580i$ & $0.297809 - 0.255202i$ \\
 & $0.221178 - 0.556687i$ & $0.672165 - 0.326057i$ \\
 & $0.257730 - 0.922166i$ & $0.440224 - 0.808377i$ \\
\hline
\hline
\end{tabular}
\label{tab:TabMagd}
\end{table}

\begin{table}[!ht]
\caption{The accurate values of QNMs for electric BHs for different $\xi$, $q_e=0.7$.}
\centering
\begin{tabular}{|c|c|c|c|c|}
\hline
\hline
 & \multicolumn{2}{c|}{Axial} & \multicolumn{2}{c|}{Polar} \\
\hline

$\xi$ & $\ell=2$ & $\ell=3$ & $\ell=2$ & $\ell=3$ \\
\hline
\multirow{4}{*}{$-0.22$} & $0.541785-0.074137 i$ & $0.767783-0.080889 i$ & $0.424332-0.079666 i$ & $0.678054-0.077178 i$ \\
 & $0.323698-0.079500 i$ & $0.574587-0.094523 i$ & $0.678414-0.081456 i$ & $0.937350-0.083493 i$ \\
 & $0.359645-0.137078 i$ & $0.531618-0.153967 i$ & $0.669728-0.246075 i$ & $0.674844-0.235002 i$ \\
 & $0.527678-0.215292 i$ & $0.751908-0.246212 i$ & $0.415757-0.249298 i$ & $0.930696-0.251251 i$ \\
\hline
\multirow{4}{*}{$-0.1$} & $0.382841-0.088239 i$ & $0.617093-0.091952 i$ & $0.399905-0.090293 i$ & $0.645758-0.092693 i$ \\
 & $0.529231-0.095358 i$ & $0.751657-0.095946 i$ & $0.578645-0.095781 i$ & $0.813512-0.096145 i$ \\
 & $0.353042-0.264579 i$ & $0.598783-0.277000 i$ & $0.374575-0.277841 i$ & $0.631607-0.281036 i$ \\
 & $0.514035-0.290153 i$ & $0.740321-0.289958 i$ & $0.566795-0.290792 i$ & $0.804548-0.290253 i$ \\
\hline
\multirow{4}{*}{$0$} & $0.392498-0.089904 i$ & $0.631873-0.093455 i$ & $0.392498-0.089904 i$ & $0.631873-0.093455 i$ \\
 & $0.536508-0.098771 i$ & $0.760502-0.098981 i$ & $0.536508-0.098771 i$ & $0.760502-0.098981 i$ \\
 & $0.367133-0.276182 i$ & $0.616412-0.283258 i$ & $0.367133-0.276182 i$ & $0.616412-0.283258 i$ \\
 & $0.519564-0.300767 i$ & $0.748179-0.299230 i$ & $0.519564-0.300767 i$ & $0.748179-0.299230 i$ \\
\hline
\multirow{4}{*}{$0.3$} & $0.392498-0.089904 i$ & $0.642205-0.096277 i$ & $0.380827-0.089583 i$ & $0.597077-0.093278 i$ \\
 & $0.536508-0.098771 i$ & $0.866218-0.103266 i$ & $0.474428-0.105029 i$ & $0.690216-0.104154 i$ \\
 & $0.367133-0.276182 i$ & $0.625608-0.292232 i$ & $0.357386-0.276379 i$ & $0.582839-0.283725 i$ \\
 & $0.519564-0.300767 i$ & $0.854195-0.311188 i$ & $0.448614-0.324323 i$ & $0.670247-0.315806 i$ \\
\hline
\multirow{4}{*}{$0.6$} & $0.399433-0.092719 i$ & $0.639942-0.095833 i$ & $0.371884-0.088950 i$ & $0.562577-0.093072 i$ \\
 & $0.609852-0.102606 i$ & $1.094585-0.110717 i$ & $0.450171-0.103903 i$ & $0.670210-0.104670 i$ \\
 & $0.371337-0.284416 i$ & $0.625668-0.291862 i$ & $0.352584-0.271266 i$ & $0.547722-0.279268 i$ \\
 & $0.594364-0.310801 i$ & $1.087357-0.334033 i$ & $0.433961-0.315391 i$ & $0.646838-0.316191 i$ \\
\hline
\multirow{4}{*}{$0.85$} & $0.399046-0.092784 i$ & $1.78330-0.04154 i$ & $0.315411-0.038908 i$ & $0.325495-0.051486 i$ \\
 & $0.770603-0.110495 i$ & $0.637516-0.095273 i$ & $0.431769-0.083141 i$ & $0.545161-0.092555 i$ \\
 & $0.372417-0.286902 i$ & $1.79858-0.12647 i$ & $0.375000-0.083144 i$ & $0.662961-0.104181 i$ \\
 & $0.758646-0.335306 i$ & $1.82686-0.21570 i$ & $0.492809-0.149676 i$ & $0.465163-0.104436 i$ \\
\hline
\multirow{4}{*}{$0.8505$} & $1.275331-0.042473 i$ & $1.81188-0.02702 i$ & $0.301131-0.024965 i$ & $0.307565-0.031333 i$ \\
 & $0.398009-0.092382 i$ & $1.82702-0.08344 i$ & $0.376237-0.068614 i$ & $0.414761-0.074078 i$ \\
 & $1.295774-0.130322 i$ & $0.637511-0.095272 i$ & $0.398582-0.078402 i$ & $0.526399-0.082753 i$ \\
 & $1.332611-0.224180 i$ & $1.85405-0.14480 i$ & $0.471543-0.097999 i$ & $0.664625-0.103098 i$ \\
\hline
\hline
\end{tabular}
\label{tab:TabEl}
\end{table}

\begin{table}
\caption{The accurate values of QNMs for magnetic BHs for different $\xi$, $q_m=0.7$.}
\begin{tabular}{|c|c|c|c|c|}
\hline
\hline
 & \multicolumn{2}{c|}{EM(polar)--Grav(axial)} & \multicolumn{2}{c|}{EM(axial)--Grav(polar)} \\
 \hline
$\xi$ & $\ell=2$ & $\ell=3$ & $\ell=2$ & $\ell=3$ \\
\hline
\multirow{4}{*}{$-0.333$} & $0.914645-0.029531 i$ & $1.282605-0.028432 i$ & $0.378103-0.091797 i$ & $0.595773-0.097089 i$ \\
 & $0.427148-0.087352 i$ & $0.688838-0.090685 i$ & $0.515618-0.103940 i$ & $0.737503-0.099097 i$ \\
 & $0.945669-0.096531 i$ & $1.306202-0.090692 i$ & $0.372041-0.287790 i$ & $0.590256-0.298662 i$ \\
 & $0.995180-0.179833 i$ & $1.34607-0.16373 i$ & $0.496330-0.322345 i$ & $0.723300-0.299439 i$ \\
\hline
\multirow{4}{*}{$-0.15$} & $0.405265-0.090000 i$ & $0.654614-0.093241 i$ & $0.386545-0.089338 i$ & $0.617804-0.093718 i$ \\
 & $0.596513-0.098367 i$ & $0.840404-0.098726 i$ & $0.521456-0.101695 i$ & $0.741935-0.099879 i$ \\
 & $0.383369-0.276955 i$ & $0.641933-0.282831 i$ & $0.363784-0.276662 i$ & $0.603675-0.285229 i$ \\
 & $0.585203-0.298491 i$ & $0.831844-0.297960 i$ & $0.502298-0.311650 i$ & $0.728875-0.302458 i$ \\
\hline
\multirow{4}{*}{$0$} & $0.392498-0.089904 i$ & $0.631873-0.093455 i$ & $0.392498-0.089904 i$ & $0.631873-0.093455 i$ \\
 & $0.536508-0.098771 i$ & $0.760502-0.098981 i$ & $0.536508-0.098771 i$ & $0.760502-0.098981 i$ \\
 & $0.367133-0.276182 i$ & $0.616412-0.283258 i$ & $0.367133-0.276182 i$ & $0.616412-0.283258 i$ \\
 & $0.519564-0.300767 i$ & $0.748179-0.299230 i$ & $0.519564-0.300767 i$ & $0.748179-0.299230 i$ \\
\hline
\multirow{4}{*}{$0.3$} & $0.370324-0.087481 i$ & $0.585578-0.091642 i$ & $0.394882-0.092178 i$ & $0.636537-0.095722 i$ \\
 & $0.485925-0.108829 i$ & $0.703166-0.106772 i$ & $0.608105-0.106535 i$ & $0.866245-0.109053 i$ \\
 & $0.341701-0.266651 i$ & $0.567192-0.277668 i$ & $0.363890-0.282474 i$ & $0.617953-0.290306 i$ \\
 & $0.448338-0.334542 i$ & $0.680553-0.324547 i$ & $0.583253-0.314538 i$ & $0.843721-0.323399 i$ \\
\hline
\multirow{4}{*}{$0.6$} & $0.350118-0.085842 i$ & $0.544672-0.091365 i$ & $0.392085-0.091749 i$ & $0.629827-0.094433 i$ \\
 & $0.482628-0.129712 i$ & $0.708098-0.121103 i$ & $0.772255-0.158864 i$ & $1.110882-0.159806 i$ \\
 & $0.315990-0.259243 i$ & $0.525218-0.277329 i$ & $0.361008-0.284782 i$ & $0.613907-0.288180 i$ \\
 & $0.246995-0.404722 i$ & $0.677334-0.370363 i$ & $0.57402-0.40164 i$ & $1.05390-0.49136 i$ \\
\hline
\multirow{4}{*}{$0.9$} & $0.333204-0.085655 i$ & $0.513384-0.091499 i$ & $0.388224-0.090433 i$ & $0.621822-0.092659 i$ \\
 & $0.519018-0.195987 i$ & $0.775138-0.185768 i$ & $0.361377-0.284219 i$ & $0.608224-0.281771 i$ \\
 & $0.296615-0.258043 i$ & $0.495582-0.277247 i$ & $1.234764-0.316501 i$ & $1.77264-0.31477 i$ \\
 & $0.177689-0.350349 i$ & $0.106713-0.431478 i$ & $0.47332-0.39186 i$ & $0.6316-0.4867 i$ \\
\hline
\hline
\end{tabular}
\label{tab:TabMg}
\end{table}

\begin{figure*}
    \includegraphics[width=0.32\linewidth]{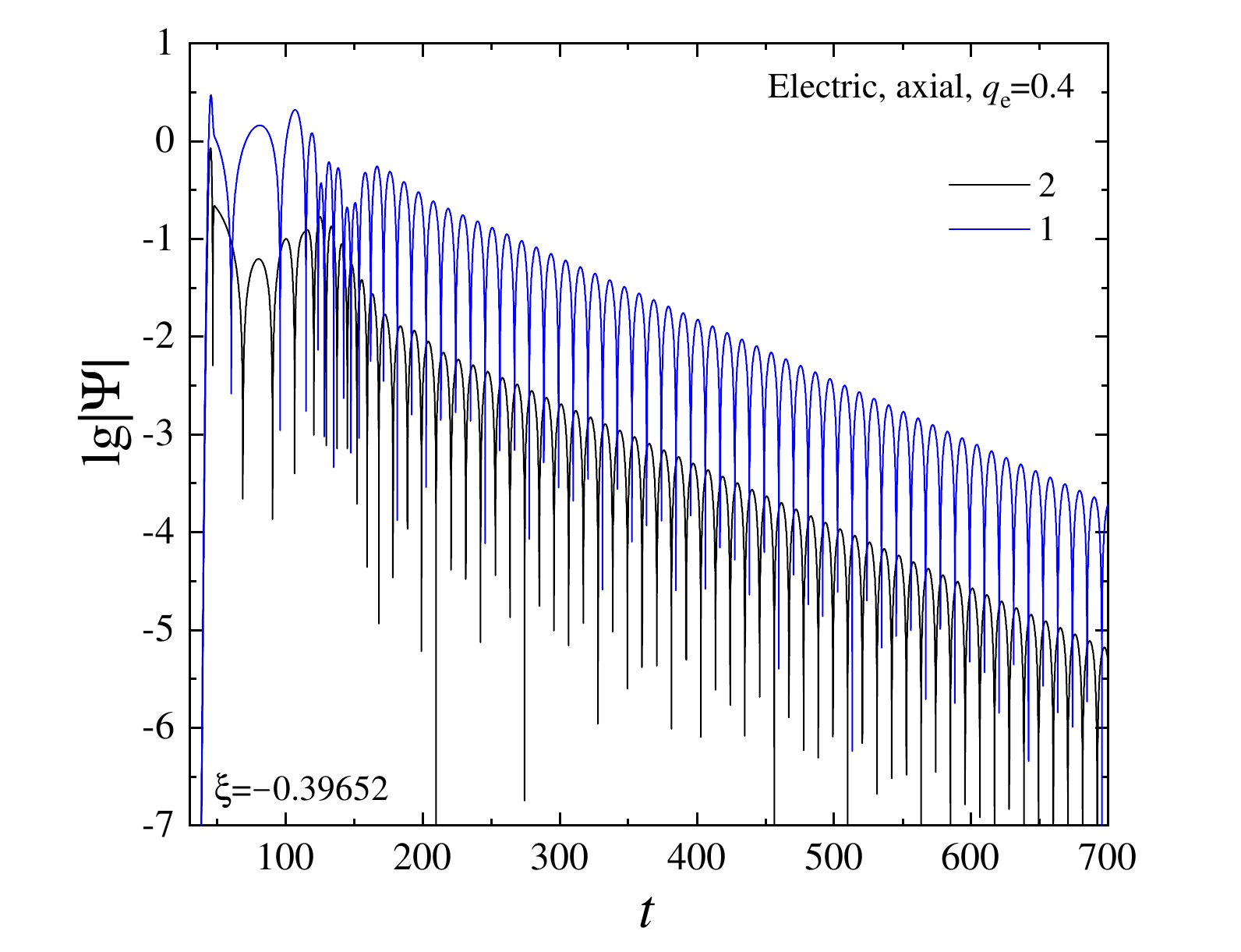}
    \includegraphics[width=0.32\linewidth]{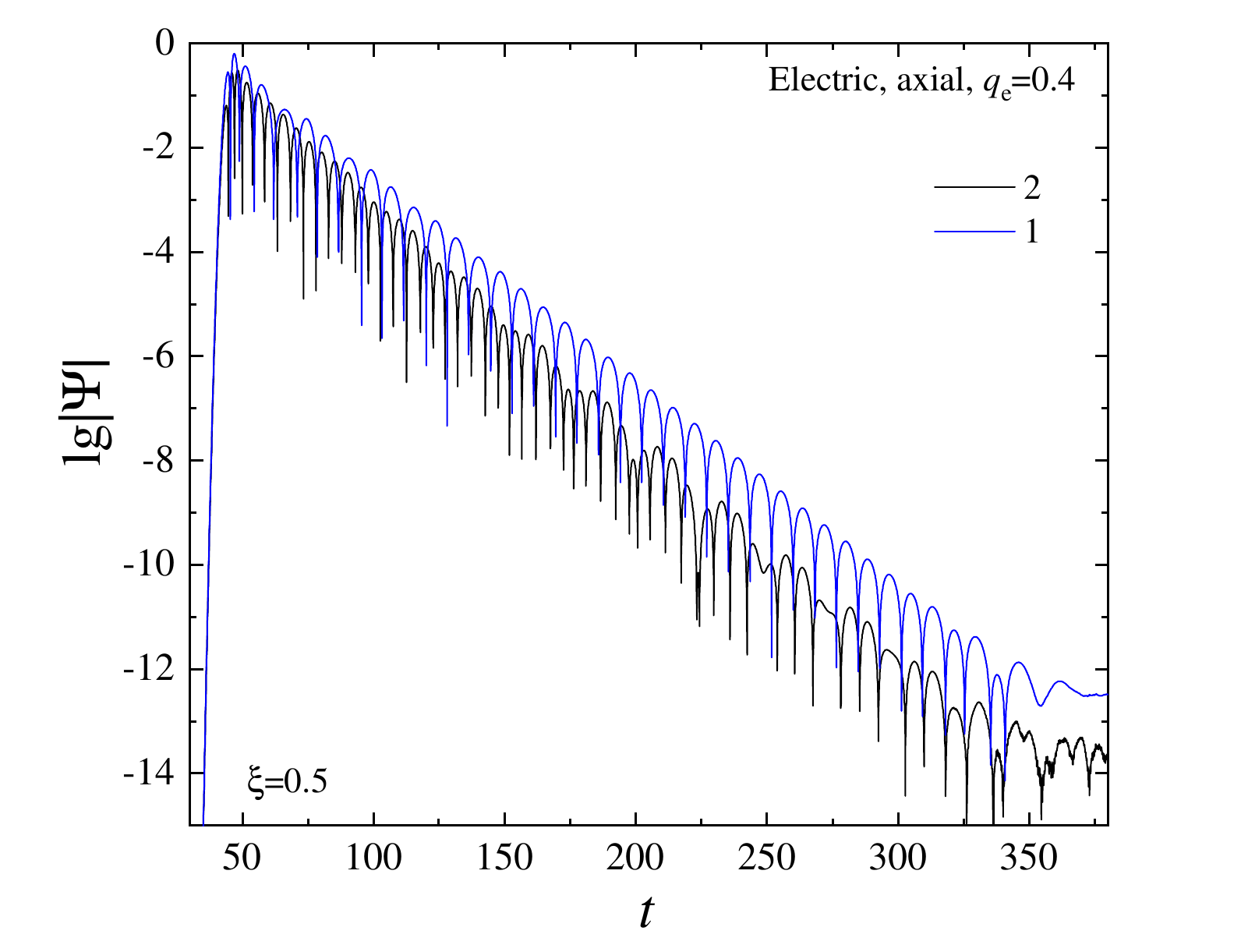}
    \includegraphics[width=0.32\linewidth]{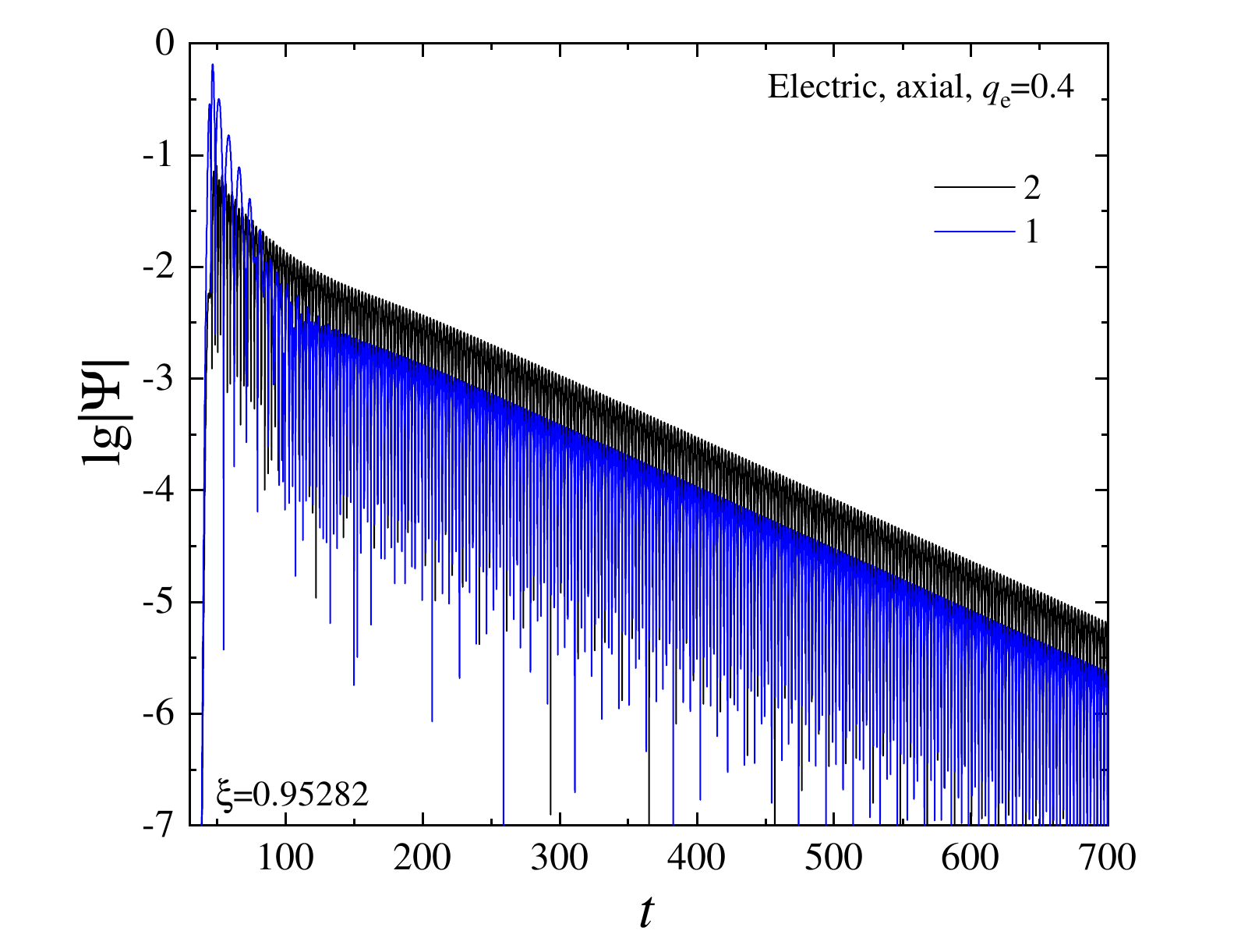}
   \caption{Typical time evolution of the components $\Psi_i$ for the coupled system of master equations  describing axial perturbations of electric BHs for different $\xi$, with  the initial Gaussian packet $\Psi_i = 2 e^{-(r_* - 5)^2/2}$ and $\ell=2$.}
    \label{fig:wave_el_axial}
\end{figure*}

\begin{figure*}
    \includegraphics[width=0.32\linewidth]{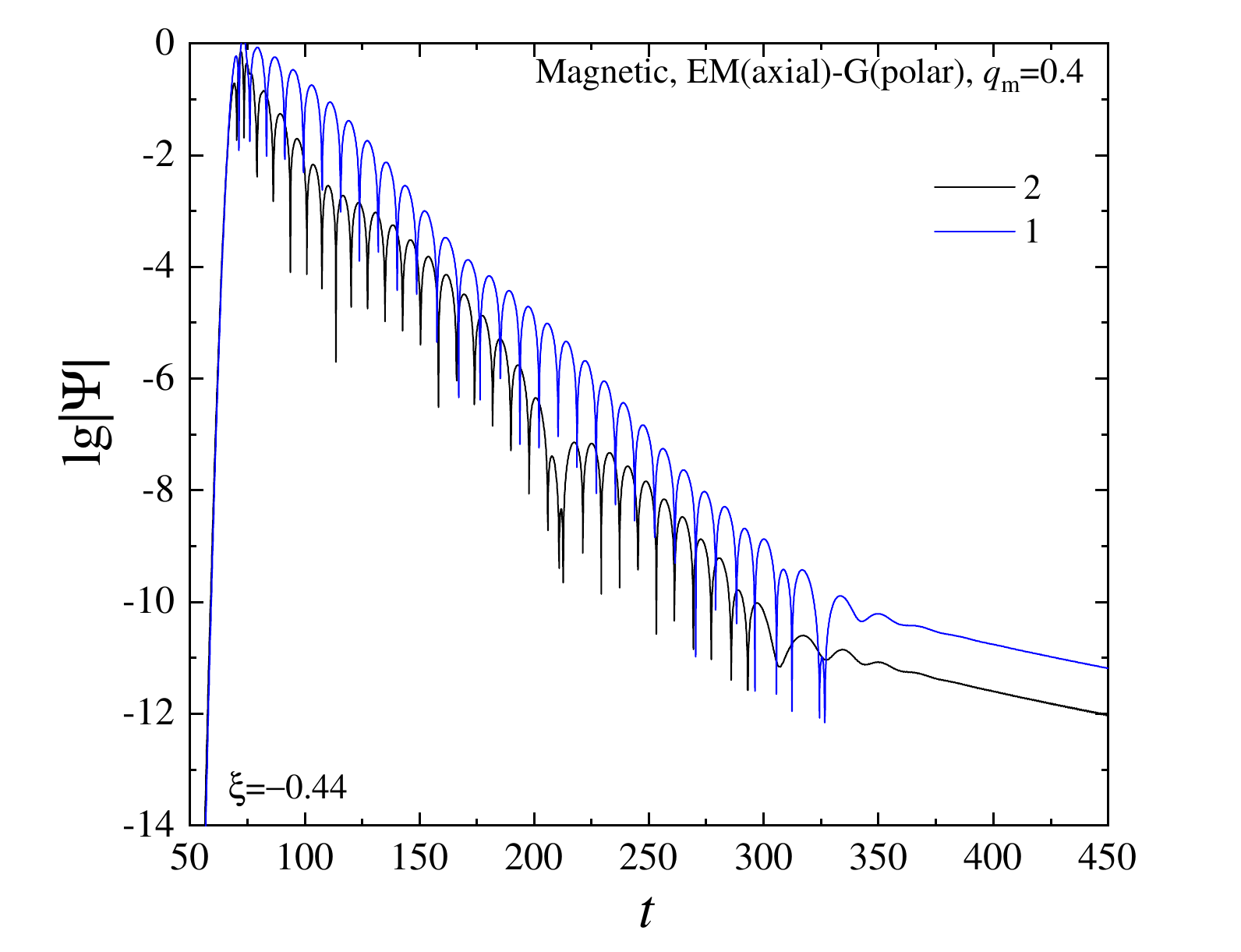}
    \includegraphics[width=0.32\linewidth]{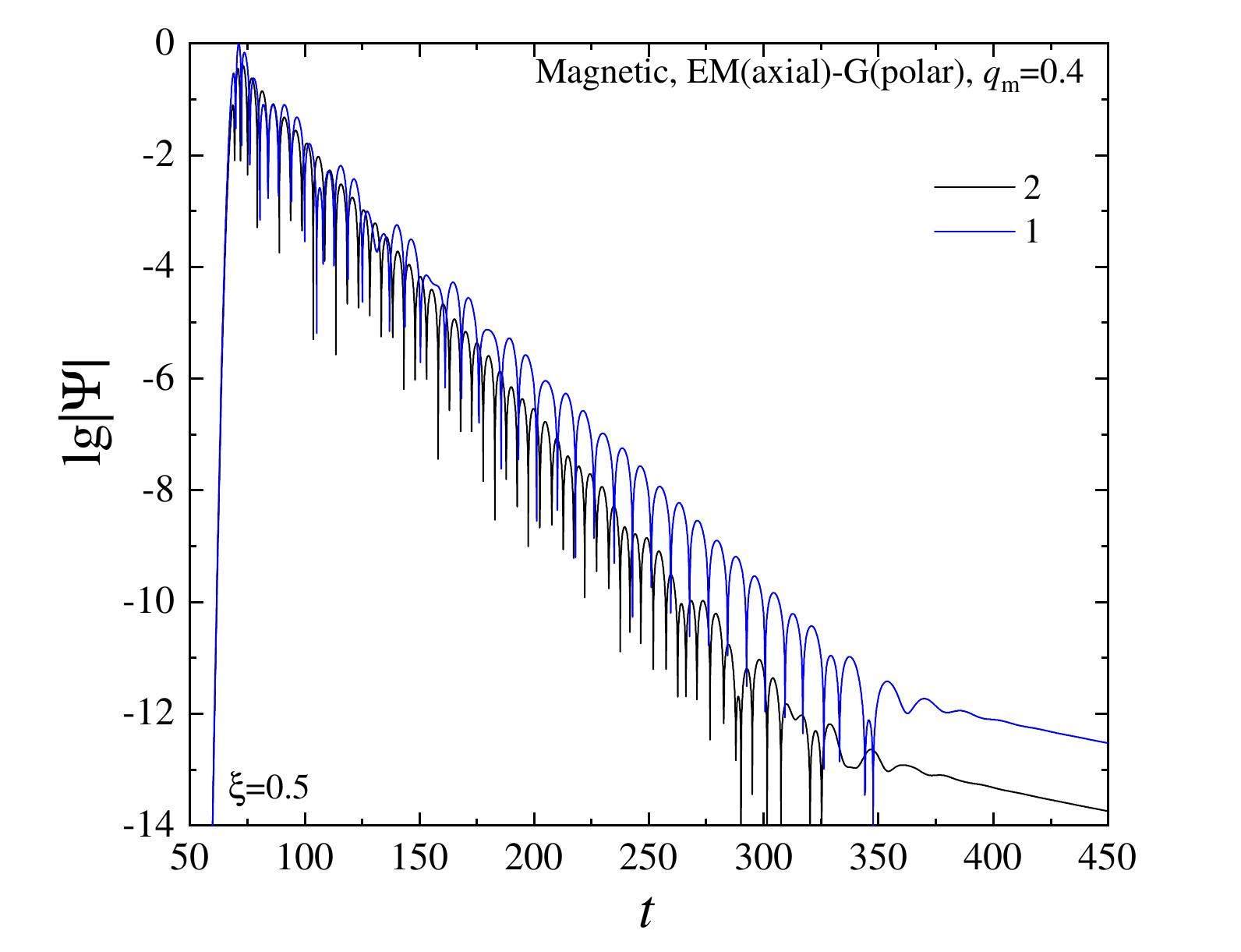}
    \includegraphics[width=0.32\linewidth]{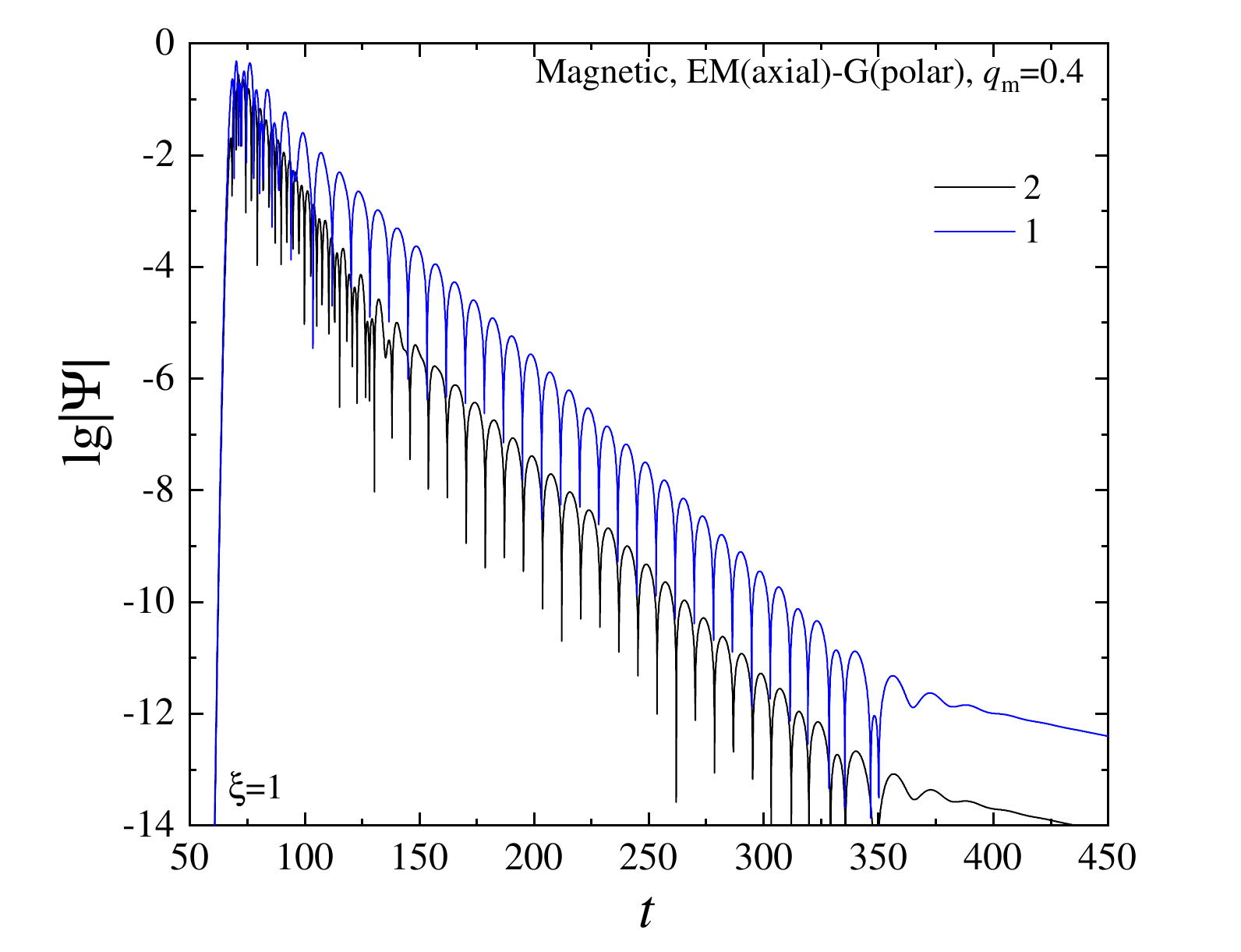}
    \caption{Typical time evolution of the components $\Psi_i$ for the coupled system of master equations describing axial electromagnetic and polar gravitational perturbations of magnetic BHs for different $\xi$, with the initial Gaussian packet $\Psi_i = 2 e^{-(r_* - 10)^2/2}$ and $\ell=2$.}
    \label{fig:wave_mag_el_ax}
\end{figure*}
\FloatBarrier
\bibliography{references.bib}

@article{Carballo-Rubio2025zwz,
    author = {Carballo-Rubio, Ra{\'u}l and Delaporte, H{\'e}lo{\"\i}se and Eichhorn, Astrid and Fernandes, Pedro G. S.},
    title = "{Non-minimal light-curvature couplings and black-hole imaging}",
    eprint = "2505.21431",
    archivePrefix = "arXiv",
    primaryClass = "astro-ph.HE",
    journal = "",
    month = "5",
    year = "2025"
}

@article{BeltranJimenez2013btb,
    author = "Beltran Jimenez, Jose and Durrer, Ruth and Heisenberg, Lavinia and Thorsrud, Mikjel",
    title = "{Stability of Horndeski vector-tensor interactions}",
    eprint = "1308.1867",
    archivePrefix = "arXiv",
    primaryClass = "hep-th",
    doi = "10.1088/1475-7516/2013/10/064",
    journal = "JCAP",
    volume = "10",
    pages = "064",
    year = "2013"
}

@article{Chen:2024hkm,
    author = "Chen, Che-Yu and De Felice, Antonio and Tsujikawa, Shinji",
    title = "{Linear stability of vector Horndeski black holes}",
    eprint = "2404.09377",
    archivePrefix = "arXiv",
    primaryClass = "gr-qc",
    reportNumber = "RIKEN-iTHEMS-Report-24, YITP-24-45, WUCG-24-04",
    doi = "10.1088/1475-7516/2024/07/022",
    journal = "JCAP",
    volume = "07",
    pages = "022",
    year = "2024"
}

@article{Chiang:2025gpa,
    author = "Chiang, Hsu-Wen and Garcia-Saenz, Sebastian and Sang, Aofei",
    title = "{Black hole destabilization via trapped quasinormal modes}",
    eprint = "2504.04779",
    archivePrefix = "arXiv",
    primaryClass = "gr-qc",
    doi = "10.1103/ksyy-wdls",
    journal = "Phys. Rev. D",
    volume = "112",
    number = "2",
    pages = "024017",
    year = "2025"
}

@article{Garcia-Saenz:2022wsl,
    author = "Garcia-Saenz, Sebastian and Held, Aaron and Zhang, Jun",
    title = "{Schwarzschild quasi-normal modes of non-minimally coupled vector fields}",
    eprint = "2202.07131",
    archivePrefix = "arXiv",
    primaryClass = "gr-qc",
    doi = "10.1007/JHEP05(2022)139",
    journal = "JHEP",
    volume = "05",
    pages = "139",
    year = "2022"
}

@article{Garcia-Saenz:2021uyv,
    author = "Garcia-Saenz, Sebastian and Held, Aaron and Zhang, Jun",
    title = "{Destabilization of Black Holes and Stars by Generalized Proca Fields}",
    eprint = "2104.08049",
    archivePrefix = "arXiv",
    primaryClass = "gr-qc",
    reportNumber = "Imperial/TP/2021/SGS/01",
    doi = "10.1103/PhysRevLett.127.131104",
    journal = "Phys. Rev. Lett.",
    volume = "127",
    number = "13",
    pages = "131104",
    year = "2021"
}

@article{Verbin_2022,
   title={Magnetic and electric black holes in the vector-tensor Horndeski theory},
   volume={106},
   ISSN={2470-0029},
   url={http://dx.doi.org/10.1103/PhysRevD.106.024057},
   DOI={10.1103/physrevd.106.024057},
   number={2},
   journal={Physical Review D},
   publisher={American Physical Society (APS)},
   author={Verbin, Y.},
   year={2022},
   month=jul }

@article{Brihaye_2020,
   title={Scalarized compact objects in a vector-tensor Horndeski gravity},
   volume={102},
   ISSN={2470-0029},
   url={http://dx.doi.org/10.1103/PhysRevD.102.124021},
   DOI={10.1103/physrevd.102.124021},
   number={12},
   journal={Physical Review D},
   publisher={American Physical Society (APS)},
   author={Brihaye, Y. and Verbin, Y.},
   year={2020},
   month=dec }

@article{Boyce_2025,
   title={EFT corrections to charged black hole quasinormal modes},
   volume={2025},
   ISSN={1029-8479},
   url={http://dx.doi.org/10.1007/JHEP12(2025)017},
   DOI={10.1007/jhep12(2025)017},
   number={12},
   journal={Journal of High Energy Physics},
   publisher={Springer Science and Business Media LLC},
   author={Boyce, William L. and Santos, Jorge E.},
   year={2025},
   month=dec }

@article{Mueller-Hoissen:1988cpx,
    author = "Mueller-Hoissen, Folkert and Sippel, Reinhard",
    title = "{Spherically Symmetric Solutions of the Nonminimally Coupled Einstein-maxwell Equations}",
    reportNumber = "GOE-ITP-15/88",
    doi = "10.1088/0264-9381/5/11/010",
    journal = "Class. Quant. Grav.",
    volume = "5",
    pages = "1473",
    year = "1988"
}

@article{Balakin:2007am,
    author = "Balakin, Alexander B. and Bochkarev, Vladimir V. and Lemos, Jose P. S.",
    title = "{Non-minimal coupling for the gravitational and electromagnetic fields: Black hole solutions and solitons}",
    eprint = "0712.4066",
    archivePrefix = "arXiv",
    primaryClass = "gr-qc",
    doi = "10.1103/PhysRevD.77.084013",
    journal = "Phys. Rev. D",
    volume = "77",
    pages = "084013",
    year = "2008"
}

@article{LIGOScientific:2016aoc,
    author = "Abbott, B. P. and others",
    collaboration = "LIGO Scientific, Virgo",
    title = "{Observation of Gravitational Waves from a Binary Black Hole Merger}",
    eprint = "1602.03837",
    archivePrefix = "arXiv",
    primaryClass = "gr-qc",
    reportNumber = "LIGO-P150914",
    doi = "10.1103/PhysRevLett.116.061102",
    journal = "Phys. Rev. Lett.",
    volume = "116",
    number = "6",
    pages = "061102",
    year = "2016"
}

@article{LIGOScientific:2017vwq,
    author = "Abbott, B. P. and others",
    collaboration = "LIGO Scientific, Virgo",
    title = "{GW170817: Observation of Gravitational Waves from a Binary Neutron Star Inspiral}",
    eprint = "1710.05832",
    archivePrefix = "arXiv",
    primaryClass = "gr-qc",
    reportNumber = "LIGO-P170817",
    doi = "10.1103/PhysRevLett.119.161101",
    journal = "Phys. Rev. Lett.",
    volume = "119",
    number = "16",
    pages = "161101",
    year = "2017"
}

@article{LIGOScientific:2020zkf,
    author = "Abbott, R. and others",
    collaboration = "LIGO Scientific, Virgo",
    title = "{GW190814: Gravitational Waves from the Coalescence of a 23 Solar Mass Black Hole with a 2.6 Solar Mass Compact Object}",
    eprint = "2006.12611",
    archivePrefix = "arXiv",
    primaryClass = "astro-ph.HE",
    reportNumber = "LIGO-P190814",
    doi = "10.3847/2041-8213/ab960f",
    journal = "Astrophys. J. Lett.",
    volume = "896",
    number = "2",
    pages = "L44",
    year = "2020"
}

@article{KAGRA:2013rdx,
    author = "Abbott, B. P. and others",
    collaboration = "KAGRA, LIGO Scientific, Virgo",
    title = "{Prospects for observing and localizing gravitational-wave transients with Advanced LIGO, Advanced Virgo and KAGRA}",
    eprint = "1304.0670",
    archivePrefix = "arXiv",
    primaryClass = "gr-qc",
    reportNumber = "LIGO-P1200087, VIR-0288A-12, JGW-P1808427",
    doi = "10.1007/s41114-020-00026-9",
    journal = "Living Rev. Rel.",
    volume = "19",
    pages = "1",
    year = "2016"
}

@article{Kokkotas:1999bd,
    author = "Kokkotas, Kostas D. and Schmidt, Bernd G.",
    title = "{Quasinormal modes of stars and black holes}",
    eprint = "gr-qc/9909058",
    archivePrefix = "arXiv",
    doi = "10.12942/lrr-1999-2",
    journal = "Living Rev. Rel.",
    volume = "2",
    pages = "2",
    year = "1999"
}

@article{Berti:2009kk,
    author = "Berti, Emanuele and Cardoso, Vitor and Starinets, Andrei O.",
    title = "{Quasinormal modes of black holes and black branes}",
    eprint = "0905.2975",
    archivePrefix = "arXiv",
    primaryClass = "gr-qc",
    doi = "10.1088/0264-9381/26/16/163001",
    journal = "Class. Quant. Grav.",
    volume = "26",
    pages = "163001",
    year = "2009"
}

@article{Konoplya:2011qq,
    author = "Konoplya, R. A. and Zhidenko, A.",
    title = "{Quasinormal modes of black holes: From astrophysics to string theory}",
    eprint = "1102.4014",
    archivePrefix = "arXiv",
    primaryClass = "gr-qc",
    doi = "10.1103/RevModPhys.83.793",
    journal = "Rev. Mod. Phys.",
    volume = "83",
    pages = "793--836",
    year = "2011"
}

@article{Bolokhov:2025rng,
    author = "Bolokhov, Sergei V. and Skvortsova, Milena",
    title = "{Review of Analytic Results on Quasinormal Modes of Black Holes}",
    doi = "10.1134/S0202289325700306",
    journal = "Grav. Cosmol.",
    volume = "31",
    number = "4",
    pages = "423--446",
    year = "2025"
}

@article{Yang:2012pj,
    author = "Yang, Huan and Zhang, Fan and Zimmerman, Aaron and Nichols, David A. and Berti, Emanuele and Chen, Yanbei",
    title = "{Branching of quasinormal modes for nearly extremal Kerr black holes}",
    eprint = "1212.3271",
    archivePrefix = "arXiv",
    primaryClass = "gr-qc",
    doi = "10.1103/PhysRevD.87.041502",
    journal = "Phys. Rev. D",
    volume = "87",
    number = "4",
    pages = "041502",
    year = "2013"
}

@article{Zimmerman:2015trm,
    author = "Zimmerman, Aaron and Mark, Zachary",
    title = "{Damped and zero-damped quasinormal modes of charged, nearly extremal black holes}",
    eprint = "1512.02247",
    archivePrefix = "arXiv",
    primaryClass = "gr-qc",
    doi = "10.1103/PhysRevD.93.044033",
    journal = "Phys. Rev. D",
    volume = "93",
    number = "4",
    pages = "044033",
    year = "2016",
    note = "[Erratum: Phys.Rev.D 93, 089905 (2016)]"
}

@misc{barbosa2026runninglovenumberscharged,
      title={Running Love Numbers of Charged Black Holes}, 
      author={Sergio Barbosa and Sylvain Fichet and Lucas de Souza},
      year={2026},
      eprint={2602.00349},
      archivePrefix={arXiv},
      primaryClass={hep-th},
      url={https://arxiv.org/abs/2602.00349}, 
}

@book{10.5555/1403886,
author = {Press, William H. and Teukolsky, Saul A. and Vetterling, William T. and Flannery, Brian P.},
title = {Numerical Recipes 3rd Edition: The Art of Scientific Computing},
year = {2007},
isbn = {0521880688},
publisher = {Cambridge University Press},
address = {USA},
edition = {3}
}

@book{boyd2013chebyshev,
  title={Chebyshev and Fourier Spectral Methods: Second Revised Edition},
  author={Boyd, J.P.},
  isbn={9780486141923},
  series={Dover Books on Mathematics},
  url={https://books.google.com/books?id=b4TCAgAAQBAJ},
  year={2013},
  publisher={Dover Publications}
}

@article{Noumi:2026shc,
    author = "Noumi, Toshifumi and Wong, Sam S. C.",
    title = "{Extremal Love: tidal/electromagnetic deformability, logarithmic running and the weak gravity conjecture}",
    eprint = "2601.20962",
    archivePrefix = "arXiv",
    primaryClass = "hep-th",
    journal="",
    month = "1",
    year = "2026"
}

@article{Pereniguez:2025jxq,
    author = "Pere{\~n}iguez, David and Karnickis, Edgars",
    title = "{On the non-zero Love numbers of magnetic black holes}",
    eprint = "2509.12418",
    archivePrefix = "arXiv",
    primaryClass = "gr-qc",
    journal="",
    month = "9",
    year = "2025"
}

@article{Horndeski:1976gi,
    author = "Horndeski, G. W.",
    title = "{Conservation of Charge and the Einstein-Maxwell Field Equations}",
    doi = "10.1063/1.522837",
    journal = "J. Math. Phys.",
    volume = "17",
    pages = "1980--1987",
    year = "1976"
}

@article{Drummond:1980,
  title = {QED vacuum polarization in a background gravitational field and its effect on the velocity of photons},
  author = {Drummond, I. T. and Hathrell, S. J.},
  journal = {Phys. Rev. D},
  volume = {22},
  issue = {2},
  pages = {343--355},
  numpages = {0},
  year = {1980},
  month = {Jul},
  publisher = {American Physical Society},
  doi = {10.1103/PhysRevD.22.343},
  url = {https://link.aps.org/doi/10.1103/PhysRevD.22.343}
}

@article{PhysRevD.101.064055,
  title = {Lovelock black $p$-branes with fluxes},
  author = {Cisterna, Adolfo and Fuenzalida, Sebasti\'an and Oliva, Julio},
  journal = {Phys. Rev. D},
  volume = {101},
  issue = {6},
  pages = {064055},
  numpages = {9},
  year = {2020},
  month = {Mar},
  publisher = {American Physical Society},
  doi = {10.1103/PhysRevD.101.064055},
  url = {https://link.aps.org/doi/10.1103/PhysRevD.101.064055}
}

@article{Feng:2015sbw,
    author = "Feng, Xing-Hui and Lu, H.",
    title = "{Higher-Derivative Gravity with Non-minimally Coupled Maxwell Field}",
    eprint = "1512.09153",
    archivePrefix = "arXiv",
    primaryClass = "hep-th",
    doi = "10.1140/epjc/s10052-016-4007-y",
    journal = "Eur. Phys. J. C",
    volume = "76",
    number = "4",
    pages = "178",
    year = "2016"
}

@ARTICLE{Akiyama2019,
 author = {{ K. {Akiyama} and \textit{et al}}  } ,
collaboration = {Event Horizon Telescope Collaboration},
            title = "{First M87 Event Horizon Telescope Results. I. The Shadow of the Supermassive Black Hole}",
      journal = {The Astrophysical Journal Letters},
         year = 2019,
        month = apr,
       volume = {875},
       number = {1},
          eid = {L1},
        pages = {L1},
          doi = {10.3847/2041-8213/ab0ec7},
}

@ARTICLE{EHT2022,
 author = {{ K. {Akiyama} and \textit{et al}}  } ,
collaboration = {Event Horizon Telescope Collaboration},
            title = "{First Sagittarius A* Event Horizon Telescope Results. I. The Shadow of the Supermassive Black Hole in the Center of the Milky Way}",
      journal = {The Astrophysical Journal Letters},
         year = 2022,
        month = may,
       volume = {930},
       number = {2},
          eid = {L12},
        pages = {L12},
          doi = {10.3847/2041-8213/ac6674},
       url = {https://link.aps.org/doi/10.1103/PhysRevX.13.041039},
      }

@article{Zhang:2024cbw,
    author = "Zhang, Chao and Kase, Ryotaro",
    title = "{Even-parity stability of hairy black holes in U(1) gauge-invariant scalar-vector-tensor theories}",
    eprint = "2404.11910",
    archivePrefix = "arXiv",
    primaryClass = "gr-qc",
    doi = "10.1103/PhysRevD.110.044047",
    journal = "Phys. Rev. D",
    volume = "110",
    number = "4",
    pages = "044047",
    year = "2024"
}

@article{De_Felice_2024,
   title={Can we distinguish black holes with electric and magnetic charges from quasinormal modes?},
   volume={109},
   ISSN={2470-0029},
   url={http://dx.doi.org/10.1103/PhysRevD.109.084022},
   DOI={10.1103/physrevd.109.084022},
   number={8},
   journal={Physical Review D},
   publisher={American Physical Society (APS)},
   author={De Felice, Antonio and Tsujikawa, Shinji},
   year={2024},
   month=Apr }

@article{Konoplya:2022pbc,
    author = "Konoplya, R. A. and Zhidenko, A.",
    title = "{First few overtones probe the event horizon geometry}",
    eprint = "2209.00679",
    archivePrefix = "arXiv",
    primaryClass = "gr-qc",
    doi = "10.1016/j.jheap.2024.10.015",
    journal = "JHEAp",
    volume = "44",
    pages = "419--426",
    year = "2024"
}

@article{Konoplya:2022hll,
    author = "Konoplya, R. A. and Zinhailo, A. F. and Kunz, J. and Stuchlik, Z. and Zhidenko, A.",
    title = "{Quasinormal ringing of regular black holes in asymptotically safe gravity: the importance of overtones}",
    eprint = "2206.14714",
    archivePrefix = "arXiv",
    primaryClass = "gr-qc",
    doi = "10.1088/1475-7516/2022/10/091",
    journal = "JCAP",
    volume = "10",
    pages = "091",
    year = "2022"
}

@article{Hertzberg:2025heq,
    author = "Hertzberg, Mark P. and Nathan, Rachel and Semaan, Suzanna E.",
    title = "{Solar System constraints on light propagation from higher derivative corrections to general relativity and implications for fundamental physics}",
    eprint = "2503.19236",
    archivePrefix = "arXiv",
    primaryClass = "gr-qc",
    doi = "10.1103/r9s4-p74z",
    journal = "Phys. Rev. D",
    volume = "112",
    number = "6",
    pages = "064078",
    year = "2025"
}

@article{DiRusso:2025qpf,
    author = "Di Russo, Giorgio and Tokareva, Anna",
    title = {{Scalar quasinormal modes in Reissner-Nordstr{\"o}m black holes: implications for Weak Gravity Conjecture}},
    eprint = "2510.06813",
    archivePrefix = "arXiv",
    primaryClass = "hep-th",
    doi = "10.1007/JHEP07(2026)202",
    journal = "JHEP",
    volume = "07",
    pages = "202",
    year = "2026"
}

@article{Prasanna:2003ix,
    author = "Prasanna, A. R. and Mohanty, Subhendra",
    title = "{Constraints on nonminimally coupled curved space electrodynamics from astrophysical observations}",
    eprint = "gr-qc/0306021",
    archivePrefix = "arXiv",
    doi = "10.1088/0264-9381/20/14/304",
    journal = "Class. Quant. Grav.",
    volume = "20",
    pages = "3023--3028",
    year = "2003"
}

@article{Guth:1979bh,
    author = "Guth, Alan H. and Tye, S. H. H.",
    title = "{Phase Transitions and Magnetic Monopole Production in the Very Early Universe}",
    reportNumber = "SLAC-PUB-2448, CLNS-79-441",
    doi = "10.1103/PhysRevLett.44.631",
    journal = "Phys. Rev. Lett.",
    volume = "44",
    pages = "631",
    year = "1980",
    note = "[Erratum: Phys.Rev.Lett. 44, 963 (1980)]"
}

@article{10.1093/mnras/stz1904,
    author = {Gong, Yi and Cao, Zhoujian and Gao, He and Zhang, Bing},
    title = {On neutralization of charged black holes},
    journal = {Monthly Notices of the Royal Astronomical Society},
    volume = {488},
    number = {2},
    pages = {2722-2731},
    year = {2019},
    month = {09},
    issn = {0035-8711},
    doi = {10.1093/mnras/stz1904},
    url = {https://doi.org/10.1093/mnras/stz1904},
    eprint = {https://academic.oup.com/mnras/article-pdf/488/2/2722/29002740/stz1904.pdf},
}

@article{PhysRevLett.123.051601,
  title = {Weak Gravity Conjecture from Unitarity and Causality},
  author = {Hamada, Yuta and Noumi, Toshifumi and Shiu, Gary},
  journal = {Phys. Rev. Lett.},
  volume = {123},
  issue = {5},
  pages = {051601},
  numpages = {6},
  year = {2019},
  month = {Jul},
  publisher = {American Physical Society},
  doi = {10.1103/PhysRevLett.123.051601},
  url = {https://link.aps.org/doi/10.1103/PhysRevLett.123.051601}
}

@article{PhysRevLett.43.1365,
  title = {Cosmological Production of Superheavy Magnetic Monopoles},
  author = {Preskill, John P.},
  journal = {Phys. Rev. Lett.},
  volume = {43},
  issue = {19},
  pages = {1365--1368},
  numpages = {0},
  year = {1979},
  month = {Nov},
  publisher = {American Physical Society},
  doi = {10.1103/PhysRevLett.43.1365},
  url = {https://link.aps.org/doi/10.1103/PhysRevLett.43.1365}
}

@article{Arkani-Hamed:2006emk,
    author = "Arkani-Hamed, Nima and Motl, Lubos and Nicolis, Alberto and Vafa, Cumrun",
    title = "{The String landscape, black holes and gravity as the weakest force}",
    eprint = "hep-th/0601001",
    archivePrefix = "arXiv",
    reportNumber = "HUTP-05-A0057",
    doi = "10.1088/1126-6708/2007/06/060",
    journal = "JHEP",
    volume = "06",
    pages = "060",
    year = "2007"
}

@article{Kats:2006xp,
    author = "Kats, Yevgeny and Motl, Lubos and Padi, Megha",
    title = "{Higher-order corrections to mass-charge relation of extremal black holes}",
    eprint = "hep-th/0606100",
    archivePrefix = "arXiv",
    reportNumber = "HUTP-06-A0023",
    doi = "10.1088/1126-6708/2007/12/068",
    journal = "JHEP",
    volume = "12",
    pages = "068",
    year = "2007"
}

\end{document}